\documentclass[letterpaper,12pt]{article}
\usepackage[utf8]{inputenc} 
\usepackage{amsmath}
      \allowdisplaybreaks 
      \everymath{\displaystyle} 
\usepackage{mathtools} %
\usepackage{amsfonts}
\usepackage{mathrsfs}
\usepackage{amssymb}
\usepackage{multicol} 
\usepackage{geometry}
\usepackage{graphicx} 
\usepackage{float} 
\usepackage{subcaption} 
\usepackage{verbatim} 
\usepackage{hyperref} 
\usepackage{cancel} 
\usepackage{authblk} 
\usepackage{lscape} 
\usepackage[all,cmtip]{xy} 
\usepackage{float} 
\usepackage[toc,page]{appendix} 

\usepackage{empheq}

\usepackage{tikz}
\usepackage{tikzscale}
\usetikzlibrary{positioning}
\usetikzlibrary{arrows}
\usetikzlibrary{decorations.pathreplacing,decorations.markings}
            \tikzset{arrow data/.style 2 args={%
            decoration={%
            markings,
            mark=at position #1 with \arrow{#2}},
            postaction=decorate}
            }%
\usetikzlibrary{intersections, backgrounds} 
\usetikzlibrary{calc}

\usepackage{array,longtable}
      \newcolumntype{C}{>{$}c<{$}}  
      \newcolumntype{R}{>{$}r<{$}}  
      \newcolumntype{L}{>{$}l<{$}}  
\newcommand{\pullbackcorner}[1][dr]{\save*!/#1+3pc/#1:(1,-1)@^{|-}\restore} 

\usepackage[nottoc,notlot,notlof]{tocbibind} 
\usepackage[titles]{tocloft} 

\usepackage{titlesec}
\titleformat{\chapter}[display]
{\normalfont\huge\bfseries}{\chaptertitlename\ \thechapter}{20pt}{\Huge}   
\titlespacing*{\chapter}{0pt}{-15pt}{15pt} 

\makeatletter
\let\@fnsymbol\@alph
\makeatother

\makeatletter
      \ams@newcommand{\multiint}[1]{\DOTSI\protect\MultiIntegral{#1}}
      \renewcommand{\MultiIntegral}[1]{%
      \edef\ints@c{\noexpand\intop
	    \ifnum#1=\z@\noexpand\indots@else\noexpand\intkern@\fi
	    \replicate{#1-2}{\noexpand\intop\noexpand\intkern@}%
	    \noexpand\intop
	    \noexpand\ilimits@
      }%
      \futurelet\@let@token\ints@a
      }
\makeatother
\ExplSyntaxOn
\cs_new:Npn \replicate #1 #2 { \prg_replicate:nn { #1 } { #2 } }
\ExplSyntaxOff

\newcommand\numberthis{\addtocounter{equation}{1}\tag{\theequation}} 

\title{ Vassiliev Invariants for Flows Via Chern-Simons Perturbation Theory}
\author[]{\normalsize J. de-la-Cruz-Moreno\thanks{E-mail: jdlcruz@fis.cinvestav.mx}}
\author[]{\normalsize H. Garc\'{\i}a-Compe\'an\thanks{E-mail: compean@fis.cinvestav.mx}}
\author[]{\normalsize E. L\'opez\thanks{E-mail: elopezg@fis.cinvestav.mx}}
\affil[]{\normalsize \it Physics Department, Centro de Investigaci\'on y de Estudios Avanzados del Instituto Polit\'ecnico Nacional, Box 14-740, CP. 07000, Mexico City, Mexico}
\date{}
\begin{document}
\maketitle
      \begin{abstract}
The perturbative expansion of Chern-Simons gauge theory leads
to invariants of knots and links, the finite type invariants
or Vassiliev invariants. It has been proven that at any
order in perturbation theory the resulting expression is an invariant
of that order. Bott-Taubes integrals on configuration spaces are introduced 
in the present context to write Feynman diagrams at a given order in perturbation theory 
in a geometrical and topological setting.  
One of the consequences of the configuration space formalism is that the
resulting amplitudes are given in cohomological terms. 
This cohomological structure can be used to 
translate Bott-Taubes integrals into Chern-Simons perturbative amplitudes and vice versa.
In this article this program is performed up to third order in the coupling constant.  
This expands some work previously worked out by Thurston. 
Finally we take advantage of these results to incorporate in the formalism a smooth and divergenceless vector
field on the $3$-manifold. The Bott-Taubes integrals obtained are used for constructing higher-order asymptotic Vassiliev invariants extending the work of Komendarczyk and Voli\'c.
      \end{abstract}

\newpage
\tableofcontents

\newpage
\section{Introduction}

It is well known that Chern-Simons gauge theory is the appropriate physical framework to describe topological invariants of 3-manifolds such as Ray-Singer torsion \cite{ASchwarz,RST}. In particular, knot and link invariants were described in this context through a non-perturbatibe treatment, which give rise to the Jones polynomial \cite{Wit89}.
The perturbative analysis of Chern-Simons action also leads to invariants of knots and links
\cite{Guadagnini:1989am,Altschuler:1996qe} (for a review see, for instance, Refs. \cite{guadagnini,senhu,Polyak:2004nn, Saw06}). In particular the so called {\it finite type invariants} from which Vassiliev invariants are an example \cite{Bar91,Bar95,Axelrod:1991vq,Axelrod:1993wr} (for an introduction to  Vassiliev invariants see, for instance, \cite{Mostovoy}). In the context of quantum field theory many developments were 
given in more physical terms in Refs. \cite{AL95,ALP97,AF97}. To be precise, in Ref. \cite{AF97} it is proved that at any order in perturbation theory the resulting expression
is a Vassiliev invariant of that order, i.e., a Vassiliev invariant is a sum of certain Feynman diagrams..

On the other hand, in the study of dynamical systems, quantum field theory also has produced a significant contribution.  In Ref. \cite{VV94}, Verjovsky and Vila-Freyer used the Chern-Simons theory and the idea of asymptotic homology cycles of foliations previously proposed by Schwartzman \cite{SA} for these purposes. 
Some further developments on the interface between dynamical systems and algebraic topology were discussed, for instance, in Refs. \cite{RS,S,S2,munoz}. One of the goals of the
article \cite{VV94} was to construct topological invariants of triplets $(M_3,{\cal F}, \mu)$, where $M_3$ is the underlying $3$-manifold, ${\cal F}$ is the foliation in $M_3$ generated by a non-singular global volume-preserving vector field $X$ and $\mu$ is the transverse measure invariant under such a flow. In Ref. \cite{VV94}, for the Abelian Chern-Simons action on $\mathbb{R}^3$ or ${\bf S}^3$ with a vector field $X$, it was shown that the evaluation of Chern-Simons integral functional in Witten's theory \cite{Wit89} leads to the link invariant for a pair of orbits of one non-singular vector field $X$. The result is precisely the {\it helicity} invariant or {\it Hopf invariant}, obtained previously by a series of authors in different contexts of (astro)physics and mathematics \cite{Arnold,BottandTu,moffat,woltjer,Vogel,EPT16}. This is called the asymptotic linking number and it can be regarded to be a topological invariant of the dynamical system defined by the triplet $(M_3,{\cal F}, \mu)$. For a review on these topics see, for instance, \cite{bookAK,K05,Dehornoy}. Thus, in this context it would be possible that the Jones-Witten invariants of manifolds $M_1$ and $M_2$ are equivalent but the invariants  of the triplet $(M_3,{\cal F}, \mu)$ are inequivalent as invariants of dynamical systems.   

In \cite{VV94} it was also discussed the non-Abelian case. This is quite more complicated than the Abelian one. Moreover in this reference it is also found the formal definition of asymptotic Jones-Witten invariant in terms of the average asymptotic Wilson loop functional. In the process the definition it requires to consider the holonomy of the connection and the Ergodic theorem. This is a suitable definition however, it makes very hard the possibility to make explicit computations.  

Helicity or linking numbers can be extended in different directions, one of them is the generalization to higher dimensions. In Ref. \cite{GS10} starting from a BF theory in $n$ dimensions on a homologically trivial manifold, we obtained a generalization of the helicity (or Jones-Witten invariant) found in  \cite{VV94}. Moreover in Ref. \cite{GSV14},  with the use of results from \cite{RS,S,S2}, we were able to find invariants for triplets $(M_4,{\cal F}, \mu)$ in the cases of the Donaldson-Witten and Seiberg-Witten invariants.
These mentioned invariants of four-dimensional dynamical systems involve the use of non-Abelian groups, however, as it was discussed in Ref. \cite{GSV14}, the observables are Lie algebra-valued functionals. Consequently, the mentioned complication arising from non-Abelian features does not appear there. As we mentioned above, in the perturbative Chern-Simons theory the relevant invariants are the Vassiliev polynomials. These objects can be obtained from the expansion of the Wilson loops and therefore they are also Lie algebra valued. Thus similarly to the situation of Ref \cite{GSV14}, the complication does not appear in this case. 

On the perturbative theory, where Vassiliev invariants are defined, the work has been not so extensive.  Configuration spaces were introduced into this context in Refs. \cite{Guadagnini:1989am,Bar91,Bar95,Axelrod:1991vq,Axelrod:1993wr,Kontsevich,Cattaneo:1999ch} to compute Feynman diagrams in Chern-Simons theory. Another important development is the proposal of integration in the configuration space, known as Bott-Taubes integrals. These integrals were introduced in the seminal paper \cite{BT94} in order to study the Feynman diagrams with $3$ points on the knot and $1$ 
point outside from it (for a recent overview on this subject see Ref. \cite{Vol07}). Later Thurston \cite{Thu99} generalized the work of Ref. \cite{BT94} to the case of integration on the configuration space constructed from Feynman diagrams with $p$ points lying on the knot and $q$ points lying outside of it. Moreover Thurston's work also provides a guide to translate Bott-Taubes integrals into Chern-Simons expressions. One of the advantages of the configuration space formalism is that the Feynman amplitudes can be expressed as integrals of differential forms in configuration spece. As a consequence of this fact, in Ref. \cite{KV16}, Komendarczyk and Voli\'c introduced a vector field $X$ into this context with the aim of proposing a manner to obtain {\it asymptotic Vassiliev invariants}. 
Asymptotic Vassiliev invariants were studied also in the context of Kontsevich's integrals in Ref. \cite{BM12}.
In the present article we extend the work done in Refs. \cite{Thu99,KV16}, which uses Bott-Taubes integration. We used systematically the perturbative expansion of Chern-Simons theory to find Bott-Taubes integrals associated to higher-order terms of the perturbative expansion of Chern-Simons theory. The Vassiliev invariants are computed explicitly up to third order in the coupling constant. Furthermore, the results obtained here are used to find their corresponding asymptotic Vassiliev invariants. In order to do all this work, we compile information from various authors into mathematical diagrams. This is highly convenient because some results and constructions are spread in the literature.

This article is organized as follows. In section $2$ we briefly overview the emergence of Vassiliev invariants from the perturbative Chern-Simons theory in the Lorentzian signature. Section $3$ is devoted to review the Bott-Taubes integrals. In section $4$ we obtain the Bott-Taubes integrals for perturbative diagrams of first, second and third orders. In section $5$ we introduce vector fields in the description found in section $4$ to obtain the asymptotic Vassiliev invariants corresponding to second and third order Feynman amplitudes. In section $6$ we give our final remarks. Finally, at the end of the article, appendices $A$, $B$ and $C$ are included to give some mathematical technicalities needed in the bulk of the paper and to provide a derivation of the propagator.

\section{Vassiliev invariants from perturbative Chern-Simons theory}

In this section we briefly overview the perturbative expansion of Chern-Simons theory. We are not intending to be exhaustive but to give the background material to introduce the notation and conventions we will follow in further sections. The Chern-Simons action (or functional) is  written as
\begin{equation}
I_{CS}(A) = \frac{k}{4 \pi} \int_{M} {\rm Tr} \left( A \wedge dA + \frac{2}{3} A \wedge A \wedge A \right),
\label{actioncs}
\end{equation}
where $A$ is a ${\cal G}$-valued connection on a trivial $G$-principal bundle over a $3$-dimensional manifold $M$ which we will take from now on as $\mathbb{R}^3$ or ${\bf S}^3$. Here $G$ is any compact and semi-simple Lie group and ${\cal G}=Lie(G)$ is its associated Lie algebra. 
Moreover Tr$: {\cal G} \times {\cal G} \to \mathbb{R}$ is the Killing quadratic form on ${\cal G}$. In Eq. (\ref{actioncs}) $k$ is the inverse of the coupling constant of the theory and it is an integer number. In this theory the unnormalized correlation functions are given by 
      \begin{equation}
      \big{\langle} W_R^K(A) \big{\rangle} = \int DA e^{\displaystyle i I_{CS}(A)} W_R^K(A),
      \label{vevknot}
      \end{equation}
where
      \begin{equation}
      W_R^K(A) = {\rm Tr}_{R} \left[ {\rm P} \exp \left( \oint_{K} A^a_{\mu}(x) t_a dx^{\mu} \right) \right]
      \label{wilsonloop}
      \end{equation}
is the Wilson loop operator and Tr$_R$ is the Killing form in the representation $R$ of $G$, $t_a$ are the generators of the Lie algebra at representation $R$ and $K: {\bf S}^1 \to M$ is the knot, {\it i.e.}, a smooth embedding.  A nonperturbative analysis \cite{Wit89} of correlation functions (\ref{vevknot}) reveals that these functions coincide with the unnormalized Jones polynomial 
        \begin{equation}
        J(q, K) =  \big{\langle} W_R^K(A) \big{\rangle}.
        \end{equation}
        
These objects are polynomials in the variable $q = \exp \left(\frac{2 \pi i}{k + h^{\vee}} \right)$, where
$h^{\vee}$ stands for the dual Coxeter number of $G$ (for $SU(N)$ it is $N$). It depends on the knot $K$, the Lie group $G$ and its representation $R$. For example, 
$SU(N)$  in the fundamental representation gives the HOMFLY-PT polynomial, $SO(N)$ in the fundamental representation gives the 
Kauffman polynomial and $SU(2)$ in the $(n+1)$-dimensional representation\footnote{In physics literature this is called the $j$ spin representation. Here $2j = n$, i.e., its dimension is $2j + 1$.}
gives the $n$-colored Jones polynomial\footnote{The case $n=1$ (or $j = 1/2$) is the famous Jones polynomial.}. The Jones polynomial can be written as finite $q$-series
        \begin{equation}
        J(q, K) = \sum_n a_n q^n,
        \end{equation}
where $a_n$ are integer numbers. 

Chern-Simons gauge theory can be quantized via BRST method and the resulting quantum action in components in $\mathbb{R}^3$ looks like \cite{Bar95}
      \begin{equation}
      I = \frac{k}{4 \pi} \int_{\mathbb{R}^3} {\rm Tr} \left( \varepsilon^{ijk} A_i \partial_j A_k + 2 \bar{c} \partial_i \partial^i c + 2 \phi \partial^i A_i 
      + \frac{1}{3} \varepsilon^{ijk} A_i [A_j, A_k] + 2 \bar{c} \partial_i [A^i,c] \right),
      \label{actioncstotal}
      \end{equation}
where $c$ and $\bar{c}$ are the ghost and anti-ghost fields, which are Grassmann valued and ${\cal G}$-valued scalar fields coupled to the gauge fields, and $\phi$ is a ${\cal G}$-valued scalar field. This action is composed by the following three parts
 \begin{equation}
      I = I_{0} + I_{g} + I',
      \end{equation}
where  $I_0$ is the kinetic (or free) part of $I$,  
      \begin{equation}
      I_{0} = \frac{k}{4 \pi} \int_{\mathbb{R}^3} {\rm Tr} \left( \varepsilon^{ijk} A_i \partial_j A_k  + 2 \bar{c} \partial_i \partial^i c \right), \numberthis \label{actionfree}
      \end{equation}     
 $I_g$ is the gauge fixing action,
        \begin{equation}     
        I_{g} = \frac{k}{4 \pi} \int_{\mathbb{R}^3} {\rm Tr} \left( 2 \phi \partial^i A_i  \right), 
        \end{equation} 
and  $I'$ is the interaction action,
     \begin{equation}
      I' = \frac{k}{4 \pi} \int_{\mathbb{R}^3} {\rm Tr} \left( \frac{1}{3} \varepsilon^{ijk} A_i [A_j, A_k] + 2 \bar{c} \partial_i [A^i,c] \right). 
      \end{equation}
     
The correlation function (\ref{vevknot}) is then replaced by
      \begin{equation}
      \big{\langle} W_R^K(A) \big{\rangle} = \int DA D\phi Dc D\bar{c} \ e^{\displaystyle i I} W_R^K(A),
      \end{equation}
where the Wilson loop operator $W_R^K(A)$ is given as in (\ref{wilsonloop}). The integral corresponding to $I_{g}$ is a constraint in the space
of connections such that integration should be performed on a submanifold ${\cal U}$ of the space of all connections ${\cal A}$. All additional fields such as the ghost fields are introduced in a gauge invariant way in ${\cal U}$. Thus the above vacuum expectation value can be written as
      \begin{equation}
      \big{\langle} W_R^K(A) \big{\rangle} 
      = \int DA D\phi Dc D\bar{c} \exp \big\{{i (I_{0}+I_g)}\big\} \exp\big({i I'}\big) W_R^K(A).
      \label{vevknottotal}
      \end{equation}
      
From now on the expression $\langle W_R^K(A) \rangle$ means this vacuum expectation value. Of course for the case of $n$-component links 
$K = K_1 \cup \cdots \cup K_n$ the required expression is
      \begin{equation}
      \big{\langle} W_{R_1}^{K_1}(A) \cdots W_{R_n}^{K_n}(A) \big{\rangle} 
      = \int DA D\phi Dc D\bar{c} \exp\big\{{i (I_{0} + I_g)}\big\} \exp\big({i \textit{I'}}\big)
      W_{R_1}^{K_1}(A) \cdots W_{R_n}^{K_n}(A),
      \label{vevlinktotal}
      \end{equation}
where the component $K_i$ has a representation $R_i$ of the gauge group.
      
The perturbative analysis is performed over the interacting terms in $I'$ and in the Wilson loop functional $W_R^K(A)$, while that corresponding to the free part of the action will remain the same. The perturbative expression for a Wilson loop in the fundamental representation ${\bf R}$ at order two  in $1/k$ is
      \begin{alignat*}{7}
      W_R^K(A) 
      &=& & {\rm Tr}_{\bf R} &\biggl[& 1 &+& \oint_{K} ds A_i^a(K(s)) \dot{K}^i(s) t_a \\
      &&&&& &+& \iint\limits_{s_1 < s_2} ds_1 ds_2 A_{i_1}^{a_1}(K(s_1)) A_{i_2}^{a_2}(K(s_2)) \dot{K}^{i_1}(s_1) \dot{K}^{i_2}(s_2) t_{a_1} t_{a_2} + \cdots \biggr] \\ \\
      &=& & {\rm Tr} &\biggl[& I &+& \oint_{K} ds A_i^a(K(s)) \dot{K}^i(s) {\bf R}(t_a) \\
      &&&&& &+& \iint\limits_{s_1 < s_2} ds_1 ds_2 A_{i_1}^{a_1}(K(s_1)) A_{i_2}^{a_2}(K(s_2)) \dot{K}^{i_1}(s_1) \dot{K}^{i_2}(s_2) {\bf R}(t_{a_1}) {\bf R}(t_{a_2}) + \cdots \biggr] \\ \\                       
      &=& & \mathrlap{\rm dim(\bf R)} \\
      & &+& \mathrlap{\oint_{K} ds A_i^a(K(s)) \dot{K}^i(s) [{\bf R}(t_a)]_{\alpha \alpha}} \\
      & &+& \mathrlap{\iint\limits_{s_1 < s_2} ds_1 ds_2 A_{i_1}^{a_1}(K(s_1)) A_{i_2}^{a_2}(K(s_2)) \dot{K}^{i_1}(s_1) \dot{K}^{i_2}(s_2) [{\bf R}(t_{a_1})]_{\alpha_1 \alpha_2} [{\bf R}(t_{a_2})]_{\alpha_2 \alpha_1}} \\ 
      & &+& \mathrlap{\cdots.} \  \  \ 
      \numberthis
      \label{wilsonloopexpansion}
      \end{alignat*}
      
The interaction part at the same order is given by
      \begin{alignat*}{2}
      e^{\displaystyle i I'} 
      = 1 &+ \frac{1}{1!} \left[ \frac{ik}{4 \pi} \int_{\mathbb{R}^3} {\rm Tr} \left( \frac{1}{3} \varepsilon^{ijk} A_i [A_j, A_k] + 2 \bar{c} \partial_i [A^i,c] \right) \right] \\
      &+ \frac{1}{2!} \left[ \frac{ik}{4 \pi} \int_{\mathbb{R}^3} {\rm Tr} \left( \frac{1}{3} \varepsilon^{ijk} A_i [A_j, A_k] + 2 \bar{c} \partial_i [A^i,c] \right) \right]^2 + \cdots. \  \  \
      \numberthis
      \label{actioninteractionexpansion}
      \end{alignat*}
      
In Ref. \cite{AL95} the perturbative analysis is used to give integral expressions for Vassiliev invariants for all prime knots up to six crossings up to order six.  In Ref. \cite{ALP97} the same ideas are applied to all two-component links up to six crossings up to order four. In both works it is used a semi-simple gauge group $G$ because a simple one is not enough to capture all invariants.

\subsection{Feynman diagrams for knots} \label{Feynman diagrams for knots}

The vacuum expectation value (\ref{vevknottotal}) can be written as the following perturbative series expansion \cite{AL95}
      \begin{equation}
      \big{\langle} W_R^K(A) \big{\rangle} = d(R) \sum_{i=0}^{\infty} \sum_{j=1}^{d_i} \alpha_{ij}(K) r_{ij}(G,R) x^i,
      \label{perturbativeseriesknot}
      \end{equation}
where $x = 2 \pi i /k$, $d(R)$ is the dimension of the representation $R$, $d_0 = \alpha_{01} = r_{01} = 1$ and $d_1 = 0$. There are many features of 
(\ref{perturbativeseriesknot}) to discuss:

      \begin{itemize}
      \item The factor $\alpha_{ij}(K)$ is called the geometrical factor and it depends only on the knot $K$ while being independent 
	    of the group and the representation chosen. The factor $r_{ij}(G,R)$ is called the group factor and it depends only on the 
	    group and representation but it is independent of the geometry of the knot.
      \item The index $i$ is the order of the perturbation while the index $j$ accounts for the contributions of the group factors
	    at that order. Actually, there are $d_i$ independent group factors at order $i$ and this quantity is called the dimension
	    of the space of invariants at that order.
      \item Vacuum diagrams are not included.
      \item Diagrams with collapsible propagators are not considered in this expression because they all contribute to the framing
	    and this is not an intrinsic property of the knot. For example, there is no linear term in (\ref{perturbativeseriesknot})
	    (condition $d_1 = 0$) because at this order the only contribution is a diagram with one collapsible propagator.
      \item Diagrams that include loops in the two-point or three-point subdiagrams are also excluded because they only contribute to the 
	    shift $k \longrightarrow k + h^{\vee}$, a quantum correction in the non-perturbative analysis.
      \item It is important to describe how the independent  factors arise. First, all Feynman diagrams at a given order must be written. 
	    Second, ignore the diagrams that contain the structure described in the latter two points above. Third, write the group factors
	    corresponding to the remaining diagrams (this is done by finding the Casimirs of the gauge group). Fourth, use commutator
	    relations and Jacobi identity to relate them and to find the independent true group factors.
      \item Independent group factors are important because they give rise to a classification of Vassiliev invariants into those that
	    are products of lower order ones and those that are not, called primitive Vassiliev invariants.
      \end{itemize}
      
The perturbative expansion (\ref{perturbativeseriesknot}) can be normalized by dividing by the vacuum expectation value of the unknot $K_0$ to give
      \begin{equation}
      \frac{\big{\langle} W_R^K(A) \big{\rangle}}{\big{\langle} W_R^{K_0}(A) \big\rangle} 
      = \sum_{i=0}^{\infty} \sum_{j=1}^{d_i} \widetilde{\alpha}_{ij}(K) r_{ij}(G,R) x^i,
      \label{perturbativeseriesknotnormalized}
      \end{equation}
where $d(R)$ does not longer appear. As $r_{ij}(G,R)$ does not depend on the knot but only on the group and its representation it can be calculated by
means of group theory. The factors $\widetilde{\alpha}_{ij}(K)$ are the Vassiliev invariants; once all the Feynman diagrams at certain order are given the 
integral expression of these factors can be built.

From Ref. \cite{AL95}, expression (\ref{perturbativeseriesknotnormalized}) can be written as
      \begin{alignat*}{6}
      \frac{\big\langle W_R^K(A) \big\rangle}{\big\langle W_R^{K_0}(A) \big\rangle} = 1 
      &+& & \widetilde{\alpha}_{21} r_{21} x^2 + \widetilde{\alpha}_{31} r_{31} x^3 \\
      &+&\bigl[& \widetilde{\alpha}_{41} (r_{21})^2 + \widetilde{\alpha}_{42} r_{42} + \widetilde{\alpha}_{43} r_{43} \bigr] x^4 \\
      &+&\bigl[& \widetilde{\alpha}_{51} r_{21} r_{31} + \widetilde{\alpha}_{52} r_{52} + \widetilde{\alpha}_{53} r_{53} + \widetilde{\alpha}_{54} r_{54} \bigr] x^5 \\
      &+&\bigl[& \widetilde{\alpha}_{61} (r_{21})^3 + \widetilde{\alpha}_{62} (r_{31})^3 + \widetilde{\alpha}_{63} r_{21} r_{42} + \widetilde{\alpha}_{64} r_{21} r_{43}
      + \widetilde{\alpha}_{65} r_{65} + \widetilde{\alpha}_{66} r_{66}\\
      &+& & \widetilde{\alpha}_{67} r_{67} + \widetilde{\alpha}_{68} r_{68} + \widetilde{\alpha}_{69} r_{69} \bigr] x^6 + O(x^7). \numberthis
      \label{ordersixknot}
      \end{alignat*}

\begin{figure}[H]
      \centering
		  \begin{tikzpicture}
		  \draw[thick,dashed] (45:2) -- (225:2); 
		  \filldraw (45:2) circle (2pt);
		  \filldraw (225:2) circle (2pt);
		  \draw [->] [domain=-45:135] plot ({2*cos(\x)}, {2*sin(\x)});
		  \draw [->] [domain=135:315] plot ({2*cos(\x)}, {2*sin(\x)});
		  \end{tikzpicture}
      \caption{Feynman diagram for $\widetilde{\alpha}_{11}(K)$.}
      \label{order1labastida95}
      \end{figure}
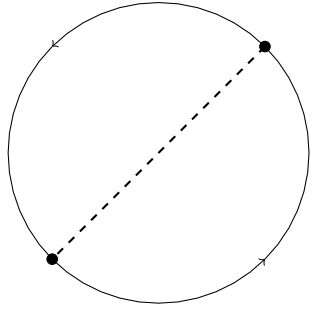

This work will be focused in orders up to three of (\ref{ordersixknot}). As stated at the beginning of this section diagrams with 
collapsible propagators are not considered in (\ref{perturbativeseriesknot}) and that is why there is no linear term there. However this work will also analyse that corresponding term $\widetilde{\alpha}_{11}(K)$ given by the Feynman diagram of figure \ref{order1labastida95} in subsequent sections.

\begin{figure}[H]
      \begin{subfigure}{.5\textwidth}
      \centering
		  \begin{tikzpicture}
		  \draw[thick,dashed] (30:2) -- (0,0)   node[pos=-0.2] {$x$};
		  \draw[thick,dashed] (150:2) -- (0,0)  node[pos=-0.2] {$y$} node[pos=1.25] {$w$};
		  \draw[thick,dashed] (270:2) -- (0,0)  node[pos=-0.2] {$z$};
		  \filldraw (30:2) circle (2pt);
		  \filldraw (150:2) circle (2pt);
		  \filldraw (270:2) circle (2pt);
		  \filldraw (0,0) circle (2pt);
		  \draw [->] [domain=-30:90] plot ({2*cos(\x)}, {2*sin(\x)});
		  \draw [->] [domain=90:210] plot ({2*cos(\x)}, {2*sin(\x)});
		  \draw [->] [domain=210:330] plot ({2*cos(\x)}, {2*sin(\x)});
		  \end{tikzpicture}
      \end{subfigure}%
      \begin{subfigure}{.5\textwidth}
      \centering
		  \begin{tikzpicture}
		  \draw[thick, dashed] (45:2) -- (225:2)     node[pos=-0.1] {$x$} node[pos=1.1] {$z$};
		  \draw[thick, dashed] (135:2) -- (135:0.1)  node[pos=-0.2] {$y$};
		  \draw[thick, dashed] (315:2) -- (315:0.1)  node[pos=-0.2] {$w$};
		  \filldraw (45:2) circle (2pt);
		  \filldraw (135:2) circle (2pt);
		  \filldraw (225:2) circle (2pt);
		  \filldraw (315:2) circle (2pt);
		  \draw [->] [domain=0:90] plot ({2*cos(\x)}, {2*sin(\x)});
		  \draw [->] [domain=90:180] plot ({2*cos(\x)}, {2*sin(\x)});
		  \draw [->] [domain=180:270] plot ({2*cos(\x)}, {2*sin(\x)});
		  \draw [->] [domain=270:360] plot ({2*cos(\x)}, {2*sin(\x)});
		  \draw[draw=none] (270:2) -- (0,0) node[pos=-0.25] {};
		  \end{tikzpicture}
      \end{subfigure}%
      \caption{Effective Feynman diagrams for $\widetilde{\alpha}_{21}(K)$.}
      \label{order2labastida95}
      \end{figure}
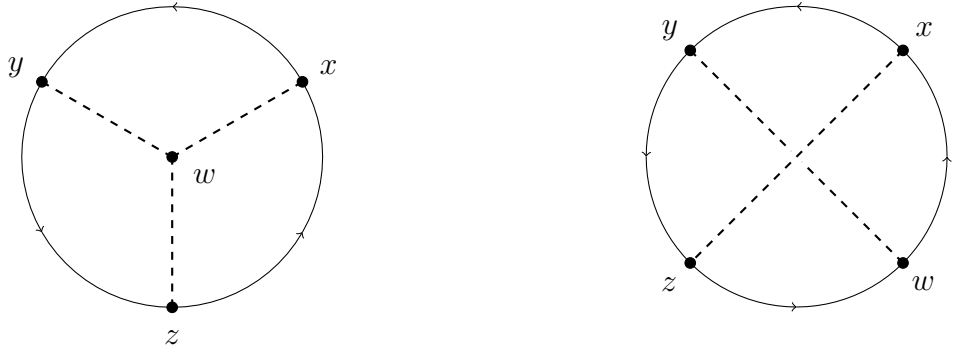

The following geometrical factor to be studied is $\widetilde{\alpha}_{21}(K)$ given by diagrams of figure \ref{order2natan}. Contributions of diagrams $(c)$ and $(d)$ are proved to cancel each other \cite{Bar95} while diagram $(e)$ is a pure-framing one. It is in this sense that the effective contribution to $\widetilde{\alpha}_{21}(K)$ is that coming from figure \ref{order2labastida95}.
The integral expression for many geometrical factors of (\ref{ordersixknot}) are given in \cite{AL95},
where the information of the connection is given through the propagator
      \begin{align}
      \big\langle A_i^a(x) A_j^b(y) \big\rangle = \varepsilon^{ijk} \delta_{ab} \left( \frac{i}{4\pi} \right) \frac{(x-y)_k}{|x-y|^3}.
      \label{propagatorBar91}
      \end{align}
      
      \begin{figure}[H]
      \begin{subfigure}[t]{.5\textwidth}
      \centering
		  \begin{tikzpicture}
		  \draw[thick,dashed] (30:2) -- (0,0); 
		  \draw[thick,dashed] (150:2) -- (0,0); 
		  \draw[thick,dashed] (270:2) -- (0,0); 
		  \filldraw (30:2) circle (2pt);
		  \filldraw (150:2) circle (2pt);
		  \filldraw (270:2) circle (2pt);
		  \filldraw (0,0) circle (2pt);
		  \draw [->] [domain=-30:90] plot ({2*cos(\x)}, {2*sin(\x)});
		  \draw [->] [domain=90:210] plot ({2*cos(\x)}, {2*sin(\x)});
		  \draw [->] [domain=210:330] plot ({2*cos(\x)}, {2*sin(\x)});
		  \end{tikzpicture}
      \caption{Diagram with one internal point.}
      \end{subfigure}%
      \begin{subfigure}[t]{.5\textwidth}
      \centering
		  \begin{tikzpicture}
		  \draw[thick, dashed] (45:2) -- (225:2); 
		  \draw[thick, dashed] (135:2) -- (135:0.1); 
		  \draw[thick, dashed] (315:2) -- (315:0.1); 
		  \filldraw (45:2) circle (2pt);
		  \filldraw (135:2) circle (2pt);
		  \filldraw (225:2) circle (2pt);
		  \filldraw (315:2) circle (2pt);
		  \draw [->] [domain=0:90] plot ({2*cos(\x)}, {2*sin(\x)});
		  \draw [->] [domain=90:180] plot ({2*cos(\x)}, {2*sin(\x)});
		  \draw [->] [domain=180:270] plot ({2*cos(\x)}, {2*sin(\x)});
		  \draw [->] [domain=270:360] plot ({2*cos(\x)}, {2*sin(\x)});
		  \end{tikzpicture}
      \caption{Diagram with non-collapsible propagators.}
      \end{subfigure}%
      \par\bigskip
      \par\bigskip
      \begin{subfigure}[t]{.5\textwidth}
      \centering
		  \begin{tikzpicture}
		  \draw[thick, dashed] (0:2) -- (0:0.75); 
		  \draw[thick, dashed] (180:2) -- (180:0.75); 
		  \filldraw (0:2) circle (2pt);
		  \filldraw (180:2) circle (2pt);
		  \filldraw (0:0.75) circle (2pt);
		  \filldraw (180:0.75) circle (2pt);
		  \draw [->] [domain=-90:90] plot ({2*cos(\x)}, {2*sin(\x)});
		  \draw [->] [domain=90:270] plot ({2*cos(\x)}, {2*sin(\x)});
		  \draw [thick, dashed] [domain=0:360] plot ({0.75*cos(\x)}, {0.75*sin(\x)});
		  \end{tikzpicture}
      \caption{Diagram with a gauge loop.}
      \end{subfigure}%
      \begin{subfigure}[t]{.5\textwidth}
      \centering
		  \begin{tikzpicture}
		  \draw[thick, dashed] (0:2) -- (0:0.75); 
		  \draw[thick, dashed] (180:2) -- (180:0.75); 
		  \filldraw (0:2) circle (2pt);
		  \filldraw (180:2) circle (2pt);
		  \draw [->] [domain=-90:90] plot ({2*cos(\x)}, {2*sin(\x)});
		  \draw [->] [domain=90:270] plot ({2*cos(\x)}, {2*sin(\x)});
		  \draw [->, double] [domain=-90:90] plot ({0.75*cos(\x)}, {0.75*sin(\x)});
		  \draw [->, double] [domain=90:270] plot ({0.75*cos(\x)}, {0.75*sin(\x)});
		  \filldraw (0:0.75) circle (2pt);
		  \filldraw (180:0.75) circle (2pt);
		  \end{tikzpicture}
      \caption{Diagram with a ghost loop.}
      \end{subfigure}%
      \par\bigskip
      \par\bigskip
      \end{figure}%
      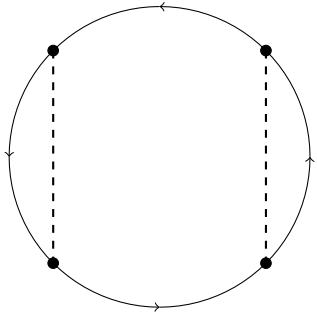
\begin{figure}[H]\ContinuedFloat
      \begin{subfigure}[t]{1\textwidth}
      \centering
		  \begin{tikzpicture}
		  \draw[thick, dashed] (45:2) -- (-45:2);
		  \draw[thick, dashed] (135:2) -- (225:2);
		  \path (45:2) -- (225:2); 
		  \path (135:2) -- (315:2); 
		  \filldraw (45:2) circle (2pt);
		  \filldraw (135:2) circle (2pt);
		  \filldraw (225:2) circle (2pt);
		  \filldraw (315:2) circle (2pt);
		  \draw [->] [domain=0:90] plot ({2*cos(\x)}, {2*sin(\x)});
		  \draw [->] [domain=90:180] plot ({2*cos(\x)}, {2*sin(\x)});
		  \draw [->] [domain=180:270] plot ({2*cos(\x)}, {2*sin(\x)});
		  \draw [->] [domain=270:360] plot ({2*cos(\x)}, {2*sin(\x)});
		  \end{tikzpicture}
      \caption{Diagram with pure framing information.}
      \end{subfigure}%
      \caption{Feynman diagrams for $\widetilde{\alpha}_{21}(K)$ according to \cite{Bar95}.}
      \label{order2natan}
      \end{figure}

\begin{figure}[H]
      \begin{subfigure}[b]{0.5\textwidth}
      \centering
		  \begin{tikzpicture}
		  \draw[thick,dashed] (30:2) -- (0:1)      node[pos=-0.3] {$x$};
		  \draw[thick,dashed] (150:2) -- (180:1)   node[pos=-0.3] {$z$};
		  \draw[thick,dashed] (210:2) -- (180:1)   node[pos=-0.3] {$t$} node[pos=1.3] {$w_2$};
		  \draw[thick,dashed] (330:2) -- (0:1)     node[pos=-0.3] {$y$} node[pos=1.3] {$w_1$};
		  \draw[thick,dashed] (180:1) -- (0:1);
		  \filldraw (30:2) circle (2pt);
		  \filldraw (150:2) circle (2pt);
		  \filldraw (210:2) circle (2pt);
		  \filldraw (-30:2) circle (2pt);
		  \filldraw (0:1) circle (2pt);
		  \filldraw (180:1) circle (2pt);
		  \draw [->] [domain=-90:0] plot ({2*cos(\x)}, {2*sin(\x)});
		  \draw [->] [domain=0:90] plot ({2*cos(\x)}, {2*sin(\x)});
		  \draw [->] [domain=90:180] plot ({2*cos(\x)}, {2*sin(\x)});
		  \draw [->] [domain=180:270] plot ({2*cos(\x)}, {2*sin(\x)});
		  \draw[draw=none] (0,0) -- (0,-2.5);
		  \end{tikzpicture}
      \end{subfigure}%
      \begin{subfigure}[b]{0.5\textwidth}
      \centering
		  \begin{tikzpicture}
		  \draw[thick,dashed] (144:2) -- (0:0)       node[pos=-0.3] {$z$};
		  \draw[thick,dashed] (216:2) -- (0:0)       node[pos=-0.3] {$t$};
			\draw[name path=a, opacity=0] (72:2) -- (288:2);
			\draw[name path=b, opacity=0] (0:0) -- (0:2);
			\path[name intersections={of = a and b, by = ab}];
		  \draw[thick,dashed] (0:0) -- (0:2)         node[pos=-0.3] {$w$} node[pos=1.3] {$x$};
		  \filldraw[white] (ab) circle (4pt);
		  \draw[thick,dashed] (72:2) -- (288:2)      node[pos=-0.1] {$v$} node[pos=1.1] {$y$};
		  \filldraw (0:2) circle (2pt);
		  \filldraw (72:2) circle (2pt);
		  \filldraw (144:2) circle (2pt);
		  \filldraw (216:2) circle (2pt);
		  \filldraw (288:2) circle (2pt);
		  \filldraw (0:0) circle (2pt);
		  \draw [->] [domain=-36:36] plot ({2*cos(\x)}, {2*sin(\x)});
		  \draw [->] [domain=36:108] plot ({2*cos(\x)}, {2*sin(\x)});
		  \draw [->] [domain=108:180] plot ({2*cos(\x)}, {2*sin(\x)});
		  \draw [->] [domain=180:252] plot ({2*cos(\x)}, {2*sin(\x)});
		  \draw [->] [domain=252:324] plot ({2*cos(\x)}, {2*sin(\x)});
		  \draw[draw=none] (0,0) -- (0,-2.5);
		  \end{tikzpicture} 
      \end{subfigure}%
      \\
      \begin{subfigure}[b]{0.5\textwidth}
      \centering
		  \begin{tikzpicture}
		  \draw[thick,dashed] (180:2) -- (0:2)    node[pos=-0.1] {$z$} node[pos=1.1] {$x$};
		  \filldraw[white] (0:1) circle (4pt);
		  \filldraw[white] (180:1) circle (4pt);
		  \draw[thick,dashed] (120:2) -- (240:2)  node[pos=-0.1] {$v$} node[pos=1.1] {$t$};
		  \draw[thick,dashed] (60:2) -- (300:2)   node[pos=-0.1] {$w$} node[pos=1.1] {$y$};
		  \filldraw (0:2) circle (2pt);
		  \filldraw (60:2) circle (2pt);
		  \filldraw (120:2) circle (2pt);
		  \filldraw (180:2) circle (2pt);
		  \filldraw (240:2) circle (2pt);
		  \filldraw (300:2) circle (2pt);
		  \draw [->] [domain=-30:30] plot ({2*cos(\x)}, {2*sin(\x)});
		  \draw [->] [domain=30:90] plot ({2*cos(\x)}, {2*sin(\x)});
		  \draw [->] [domain=90:150] plot ({2*cos(\x)}, {2*sin(\x)});
		  \draw [->] [domain=150:210] plot ({2*cos(\x)}, {2*sin(\x)});
		  \draw [->] [domain=210:270] plot ({2*cos(\x)}, {2*sin(\x)});
		  \draw [->] [domain=270:330] plot ({2*cos(\x)}, {2*sin(\x)});
		  \end{tikzpicture}  
      \end{subfigure}%
      \begin{subfigure}[b]{0.5\textwidth}
      \centering
		  \begin{tikzpicture}
		  \draw[thick,dashed] (120:2) -- (300:2)  node[pos=-0.1] {$v$} node[pos=1.1] {$y$};
		  \draw[thick,dashed] (60:2) -- (240:2)   node[pos=-0.1] {$w$} node[pos=1.1] {$t$};
		  \filldraw[white] (0:0) circle (4pt);
		  \draw[thick,dashed] (180:2) -- (0:2)    node[pos=-0.1] {$z$} node[pos=1.1] {$x$};
		  \filldraw (0:2) circle (2pt);
		  \filldraw (60:2) circle (2pt);
		  \filldraw (120:2) circle (2pt);
		  \filldraw (180:2) circle (2pt);
		  \filldraw (240:2) circle (2pt);
		  \filldraw (300:2) circle (2pt);
		  \draw [->] [domain=-30:30] plot ({2*cos(\x)}, {2*sin(\x)});
		  \draw [->] [domain=30:90] plot ({2*cos(\x)}, {2*sin(\x)});
		  \draw [->] [domain=90:150] plot ({2*cos(\x)}, {2*sin(\x)});
		  \draw [->] [domain=150:210] plot ({2*cos(\x)}, {2*sin(\x)});
		  \draw [->] [domain=210:270] plot ({2*cos(\x)}, {2*sin(\x)});
		  \draw [->] [domain=270:330] plot ({2*cos(\x)}, {2*sin(\x)});
		  \end{tikzpicture}  
      \end{subfigure}%
      \caption{Effective Feynman diagrams for $\widetilde{\alpha}_{31}(K)$.}
      \label{order3labastida95}
      \end{figure}
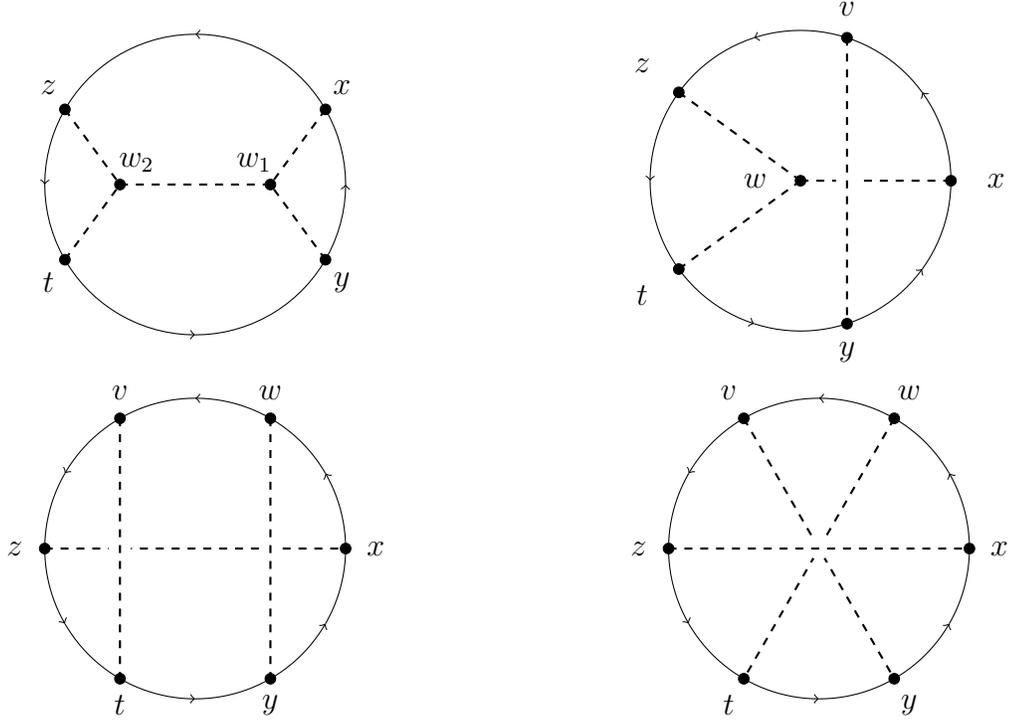

For example for $\widetilde{\alpha}_{21}(K)$ one has
      \begin{alignat*}{6}
      \widetilde{\alpha}_{21}(K) &=& & \frac{\alpha_{21}(K)}{\big\langle W_R^{K_0}(A) \big\rangle} \\
      &=& & \frac{1}{\big\langle W_R^{K_0}(A) \big\rangle} \frac{1}{4 \pi^2} & & \oint_K dx_{\mu} & \int^x dy_{\nu} \int^y dz_{\rho} \int^z dw_{\tau} 
            \biggl[ \varepsilon^{\mu \sigma_1 \rho} \varepsilon^{\nu \sigma_2 \tau}
            \frac{(x-z)_{\sigma_1}}{|x-z|^3} \frac{(y-w)_{\sigma_2}}{|y-w|^3} \biggr] \\
      & &-& \frac{1}{\big\langle W_R^{K_0}(A) \big\rangle} \frac{1}{16 \pi^3} & & 
	    \mathrlap{ \oint_K dx_{\mu} \int^x dy_{\nu} \int^y dz_{\rho} \int_{\mathbb{R}^3} d^3 w 
            \biggl[ \varepsilon^{\mu \rho_1 \sigma_1} \varepsilon^{\nu \rho_2 \sigma_2} \varepsilon^{\rho \rho_3 \sigma_3} 
            \varepsilon_{\sigma_1 \sigma_2 \sigma_3}} \\
      & & & & & & \times \frac{(x-w)_{\rho_1}}{|x-w|^3} \frac{(y-w)_{\rho_2}}{|y-w|^3} \frac{(z-w)_{\rho_3}}{|z-w|^3} \biggr],
      \numberthis
      \end{alignat*}
while for $\widetilde{\alpha}_{31}(K)$ the expression is, from diagrams in figure \ref{order3labastida95},
      \begin{alignat*}{6}
      \widetilde{\alpha}_{31}(K) 
      &=& & \mathrlap{ \frac{1}{\big{\langle} W_R^{K_0}(A) \big{\rangle}} \frac{1}{64 \pi^5} \oint_K dx_{\mu} \int^x dy_{\nu} \int^y dt_{\rho} \int^t dz_{\tau} \int_{\mathbb{R}^3} d^3 w_1 \int_{\mathbb{R}^3} d^3 w_2 \biggl[ \varepsilon_{\alpha \beta \gamma} \varepsilon_{\eta \xi \zeta}  } \\
      & &\times && \varepsilon^{\mu \sigma_1 \alpha} \varepsilon^{\nu \sigma_2 \beta} \varepsilon^{\gamma \sigma_3 \zeta} \varepsilon^{\rho \sigma_4 \eta} \varepsilon^{\tau \sigma_5 \xi}
            \frac{(x-w_1)_{\sigma_1}}{|x-w_1|^3} \frac{(y-w_1)_{\sigma_2}}{|y-w_1|^3} \frac{(w_1-w_2)_{\sigma_3}}{|w_1-w_2|^3} \frac{(t-w_2)_{\sigma_4}}{|t-w_2|^3} \frac{(z-w_2)_{\sigma_5}}{|z-w_2|^3} \biggr] \\ \\
      & &+& \mathrlap{ \frac{1}{\big{\langle} W_R^{K_0}(A) \big\rangle} \frac{5}{32 \pi^4} \oint_K dx_{\mu} \int^x dy_{\nu} \int^y dt_{\rho} \int^t dz_{\tau} \int^z dv_{\eta} \int_{\mathbb{R}^3} d^3 w  } \\
      & & && \times \biggl[ \varepsilon^{\nu \sigma \eta} \varepsilon_{\alpha \beta \gamma} \varepsilon^{\mu \sigma_1 \alpha} \varepsilon^{\rho \sigma_2 \beta} \varepsilon^{\tau \sigma_3 \gamma}
            \frac{(y-v)_{\sigma}}{|y-v|^3} \frac{(x-w)_{\sigma_1}}{|x-w|^3} \frac{(t-w)_{\sigma_2}}{|t-w|^3} \frac{(z-w)_{\sigma_3}}{|z-w|^3} \biggr] \\ \\
      & &+& \mathrlap{ \frac{1}{\big\langle W_R^{K_0}(A) \big\rangle} \frac{3}{8 \pi^3} \oint_K dx_{\mu} \int^x dy_{\nu} \int^y dt_{\rho} \int^t dz_{\tau} \int^z dv_{\eta} \int^v dw_{\zeta} } \\
      & & && \times \biggl[ \varepsilon^{\mu \sigma_1 \tau} \varepsilon^{\nu \sigma_2 \zeta} \varepsilon^{\rho \sigma_3 \eta}
            \frac{(x-z)_{\sigma_1}}{|x-z|^3} \frac{(y-w)_{\sigma_2}}{|y-w|^3} \frac{(t-v)_{\sigma_3}}{|t-v|^3} \biggr] \\ \\
      & &+& \mathrlap{ \frac{1}{\big\langle W_R^{K_0}(A) \big\rangle} \frac{1}{4 \pi^3} \oint_K dx_{\mu} \int^x dy_{\nu} \int^y dt_{\rho} \int^t dz_{\tau} \int^z dv_{\eta} \int^v dw_{\zeta}  } \\
      & & && \times \biggl[ \varepsilon^{\mu \sigma_1 \tau} \varepsilon^{\nu \sigma_2 \eta} \varepsilon^{\rho \sigma_3 \zeta}
            \frac{(x-z)_{\sigma_1}}{|x-z|^3} \frac{(y-v)_{\sigma_2}}{|y-v|^3} \frac{(t-w)_{\sigma_3}}{|t-w|^3} \biggr].
      \numberthis
      \end{alignat*}

\subsection{Feynman diagrams for links} \label{Feynman diagrams for links}

Let $L = K_1 \cup K_2$ be a link with components $K_1$ and $K_2$ and representations $R_1$ and $R_2$ of the gauge group, respectively. The 
important vacuum expectation value will be, from (\ref{vevlinktotal}),
      \begin{equation}
      \big\langle W_{R_1}^{K_1}(A) W_{R_2}^{K_2}(A) \big\rangle 
      = \int DA D\phi Dc D\bar{c} \exp\big\{i (I_{0} + I_g)\big\}  e^{i I'}
      W_{R_1}^{K_1}(A) W_{R_2}^{K_2}(A). 
      \label{vevlinktotaltwocomponent}
      \end{equation}
      
The generalization of (\ref{perturbativeseriesknot}) from knots to links is not trivial because the group factors in the latter case have a more complicated structure. In \cite{ALP97} via a factorization theorem for Wilson lines it is found that such a generalization is given by
      \begin{equation}
      \big\langle W_{R_1}^{K_1}(A) W_{R_2}^{K_2}(A) \big\rangle = \big\langle W_{R_1}^{K_1}(A) \big\rangle \big\langle W_{R_2}^{K_2}(A) \big\rangle 
      \big\langle \mathcal{Z}_{R_1, R_2}^{K_1,K_2}(A) \big\rangle,
      \end{equation}
where $\big\langle W_{R_1}^{K_1}(A) \big\rangle$ and $\big\langle W_{R_2}^{K_2}(A) \big\rangle$ are written as in (\ref{perturbativeseriesknot}) and
      \begin{equation}
      \big\langle \mathcal{Z}_{R_1, R_2}^{K_1,K_2}(A) \big\rangle = \sum_{i=0}^{\infty} \sum_{j=1}^{\delta_i} \gamma_i^{\ j}(K_1, K_2) s_{ij}(G, R_1, R_2) x^i
      \label{interactionpartloop}
      \end{equation}
is the pure link contribution. The objects $\gamma_i^{\ j}(K_1, K_2)$ are called the Vassiliev link invariants and they depend only on the knots 
$K_1$ and $K_2$, the objects $s_{ij}(G, R_1, R_2)$ are the new group factors that depend on the gauge group and its representations $R_1$ and 
$R_2$. Here again the index $i$ is the order of the perturbative expansion while $j$ stands for the contributions of the group factors at that order, 
$x = 2 \pi i /k$ and $\delta_i$ is the number of independent group factors at order $i$ or the dimension of the space of invariants at that order.
Similar considerations to those given in Eq. (\ref{perturbativeseriesknot}), concerning the type of diagrams appearing in the expansion as
well as the independence of the group factors, also apply in this case. Expression (\ref{interactionpartloop}) at order four can be written as
      \begin{align*}
      \big\langle \mathcal{Z}_{R_1, R_2}^{K_1,K_2}(A) \big\rangle = 
      1 &+ \left[ \frac{\left( \gamma_1^{\ 1} \right)^2}{2!} s_{21} \right] x^2
	  + \left[ \frac{\left( \gamma_1^{\ 1} \right)^3}{3!} s_{31} + \gamma_3^{\ 2} s_{32} \right] x^3 \\
	  &+ \left[ \frac{\left( \gamma_1^{\ 1} \right)^4}{4!} s_{41} + \frac{\gamma_1^{\ 1} \gamma_3^{\ 2}}{2} s_{42} + \gamma_4^{\ 3} s_{43}   \right] x^4 + O(x^5). 
	  \numberthis
      \label{orderfourloop}
      \end{align*}

At this order the primitive Vassiliev invariants are then $\gamma_1^{\ 1}$, $\gamma_3^{\ 2}$ and $\gamma_4^{\ 3}$. In Ref. \cite{ALP97} the explicit integral expressions for these three $\gamma$'s are given. For example for $\gamma_1^{\ 1}$ one has 
      \begin{equation}
      \gamma_1^{\ 1} = \frac{1}{2} \oint dx \oint dy \ p(x,y),
      \label{primitivevassiliev11}
      \end{equation}
where
      \begin{equation}
      p(x,y) = \Delta_{\mu \nu}(x-y) = \frac{1}{\pi} \varepsilon_{\mu \rho \nu} \frac{(x-y)^{\rho}}{|x-y|^3}.
      \end{equation}
      
Expression (\ref{primitivevassiliev11}) is twice the linking number of the link while $\gamma_3^{\ 2}$ and $\gamma_4^{\ 3}$ are new invariants found in Ref. \cite{ALP97} that are not clear to be related with known numerical link invariants.

\begin{figure}[H]
      \centering
	    \begin{tikzpicture}[scale=0.8]
	    \draw      [domain=45:180] plot ({-3+1*cos(\x)}, {0+2*sin(\x)}) node[circle, fill, inner sep=2pt] (a1) {};
	    \draw [->] [domain=-180:45] plot ({3+1*cos(\x)}, {0+2*sin(\x)});
	    \draw      [domain=45:180] plot ({3+1*cos(\x)}, {0+2*sin(\x)}) node[circle, fill, inner sep=2pt] (a2) {};
	    \draw [name path=a, opacity=0] (a1) -- (a2);
	    \draw [->] [name path=arc1, domain=-180:45, opacity = 0] plot ({-3+1*cos(\x)}, {0+2*sin(\x)});
		  \path [name intersections={of = arc1 and a, by = arc1a}];
	    \draw [->] [domain=-180:45] plot ({-3+1*cos(\x)}, {0+2*sin(\x)});
		  \filldraw [white] (arc1a) circle (3pt);
			\draw [thick, dashed] (a1) -- (a2)                            node[pos=-0.1] {$y$} node[pos=1.1] {$x$};
	    \end{tikzpicture}
      \caption{Feynman diagram for $\gamma_1^{\ 1}$.}
      \label{figuradellinkingnumber}
      \end{figure}
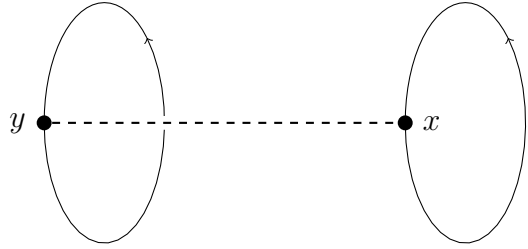

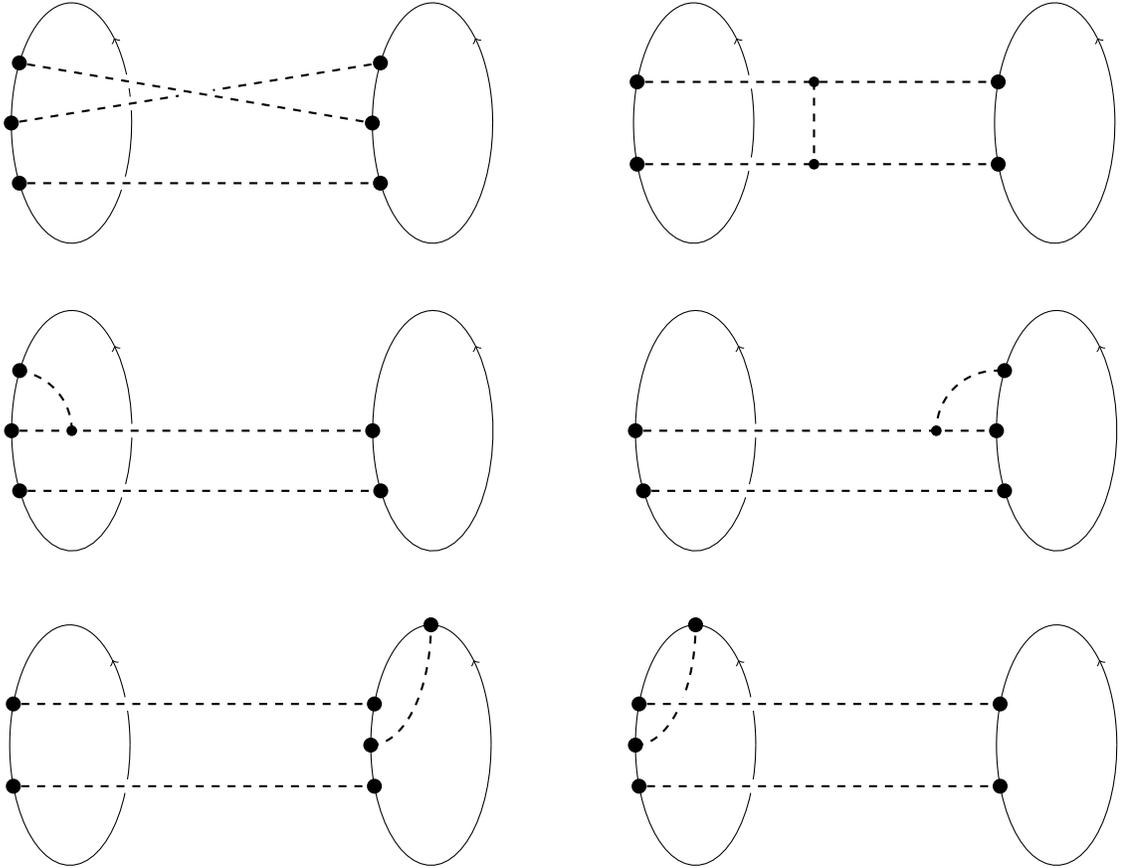
\begin{figure}[ht]
      \begin{subfigure}{0.5\textwidth}
      \centering
		  \begin{tikzpicture}[scale=0.8]
		  \draw      [domain=45:150] plot ({-3+1*cos(\x)}, {0+2*sin(\x)}) node[circle, fill, inner sep=2pt] (a1) {};
		  \draw      [domain=150:180] plot ({-3+1*cos(\x)}, {0+2*sin(\x)}) node[circle, fill, inner sep=2pt] (b1) {};
		  \draw      [domain=180:210] plot ({-3+1*cos(\x)}, {0+2*sin(\x)}) node[circle, fill, inner sep=2pt] (c1) {};
		  \draw [->] [domain=0:45] plot ({3+1*cos(\x)}, {0+2*sin(\x)});
		  \draw      [domain=45:150] plot ({3+1*cos(\x)}, {0+2*sin(\x)}) node[circle, fill, inner sep=2pt] (a2) {};
		  \draw      [domain=150:180] plot ({3+1*cos(\x)}, {0+2*sin(\x)}) node[circle, fill, inner sep=2pt] (b2) {};
		  \draw      [domain=180:210] plot ({3+1*cos(\x)}, {0+2*sin(\x)}) node[circle, fill, inner sep=2pt] (c2) {};
		  \draw      [domain=210:360] plot ({3+1*cos(\x)}, {0+2*sin(\x)});
		  \draw [name path=a, opacity=0] (a1) -- (b2);
		  \draw [name path=b, opacity=0] (b1) -- (a2);
		  \draw [name path=c, opacity=0] (c1) -- (c2);
		  \draw [->] [name path=arc1, domain=0:45, opacity = 0] plot ({-3+1*cos(\x)}, {0+2*sin(\x)});
		  \draw [->] [name path=arc1phant, domain=0:40, opacity = 0] plot ({-3+1*cos(\x)}, {0+2*sin(\x)});
		  \draw      [name path=arc2, domain=210:360, opacity=0] plot ({-3+1*cos(\x)}, {0+2*sin(\x)});
			\path [name intersections={of = arc1 and a, by = arc1a}];
			\path [name intersections={of = arc1phant and b, by = arc1phantb}];
			\path [name intersections={of = arc2 and c, by = arc2c}];
			\path [name intersections={of = a and b, by = ab}];
			      \filldraw [blue, opacity = 0] (arc1a) circle (3pt);
			      \filldraw [blue, opacity = 0] (arc1phantb) circle (3pt);
			      \filldraw [blue, opacity = 0] (arc2c) circle (3pt);
			      \filldraw [blue, opacity = 0] (ab) circle (3pt);
		  \draw [->] [domain=0:45] plot ({-3+1*cos(\x)}, {0+2*sin(\x)});
		  \draw      [domain=210:360] plot ({-3+1*cos(\x)}, {0+2*sin(\x)});
			\filldraw [white] (arc1a) circle (3pt);
			\filldraw [white] (arc1phantb) circle (3pt);
			\filldraw [white] (arc2c) circle (3pt);
			      \draw [thick, dashed] (b1) -- (a2);
			      \filldraw [white] (ab) circle (8pt);
			      \draw [thick, dashed] (a1) -- (b2);
			      \draw [thick, dashed] (c1) -- (c2);
		  \end{tikzpicture}
      \end{subfigure}%
      \begin{subfigure}{0.5\textwidth}
      \centering
		  \begin{tikzpicture}[scale=0.8]
		  \draw      [domain=45:160] plot ({-3+1*cos(\x)}, {0+2*sin(\x)}) node[circle, fill, inner sep=2pt] (a1) {};
		  \draw      [domain=160:200] plot ({-3+1*cos(\x)}, {0+2*sin(\x)}) node[circle, fill, inner sep=2pt] (b1) {};
		  \draw [->] [domain=0:45] plot ({3+1*cos(\x)}, {0+2*sin(\x)});
		  \draw      [domain=45:160] plot ({3+1*cos(\x)}, {0+2*sin(\x)}) node[circle, fill, inner sep=2pt] (a2) {};
		  \draw      [domain=160:200] plot ({3+1*cos(\x)}, {0+2*sin(\x)}) node[circle, fill, inner sep=2pt] (b2) {};
		  \draw      [domain=200:360] plot ({3+1*cos(\x)}, {0+2*sin(\x)});
		  \draw [name path=a, opacity=0] (a1) -- (a2);
		  \draw [name path=b, opacity=0] (b1) -- (b2);
		  \draw [->] [name path=arc1, domain=0:45, opacity = 0] plot ({-3+1*cos(\x)}, {0+2*sin(\x)});
		  \draw      [name path=arc2, domain=200:360, opacity=0] plot ({-3+1*cos(\x)}, {0+2*sin(\x)});
		  \draw [name path=c, opacity=0] (-1,0) -- (-1,1);
		  \draw [name path=d, opacity=0] (-1,0) -- (-1,-1);
			\path [name intersections={of = arc1 and a, by = arc1a}];
			\path [name intersections={of = arc2 and b, by = arc2b}];
			\path [name intersections={of = a and c, by = ac}];
			\path [name intersections={of = b and d, by = bd}];
		  \draw [->] [domain=0:45] plot ({-3+1*cos(\x)}, {0+2*sin(\x)});
		  \draw      [domain=200:360] plot ({-3+1*cos(\x)}, {0+2*sin(\x)});
			\filldraw [white] (arc1a) circle (3pt);
			\filldraw [white] (arc2b) circle (3pt);
			      \draw [thick, dashed] (a1) -- (a2);
			      \draw [thick, dashed] (b1) -- (b2);
			      \draw [thick, dashed] (ac) -- (bd);
			\filldraw (ac) circle (2.25pt);
			\filldraw (bd) circle (2.25pt);
		  \end{tikzpicture}
      \end{subfigure}%
      \par\bigskip
      \par\bigskip
      \begin{subfigure}{0.5\textwidth}
      \centering
		  \begin{tikzpicture}[scale=0.8]
		  \draw      [domain=45:150] plot ({-3+1*cos(\x)}, {0+2*sin(\x)}) node[circle, fill, inner sep=2pt] (a1) {};
		  \draw      [domain=150:180] plot ({-3+1*cos(\x)}, {0+2*sin(\x)}) node[circle, fill, inner sep=2pt] (b1) {};
		  \draw      [domain=180:210] plot ({-3+1*cos(\x)}, {0+2*sin(\x)}) node[circle, fill, inner sep=2pt] (c1) {};
		  \draw [->] [domain=-150:45] plot ({3+1*cos(\x)}, {0+2*sin(\x)});
		  \draw      [domain=45:180] plot ({3+1*cos(\x)}, {0+2*sin(\x)}) node[circle, fill, inner sep=2pt] (b2) {};
		  \draw      [domain=180:210] plot ({3+1*cos(\x)}, {0+2*sin(\x)}) node[circle, fill, inner sep=2pt] (c2) {};
		  \draw [name path=a, opacity=0] (b1) -- (b2);
		  \draw [name path=b, opacity=0] (c1) -- (c2);
		  \draw [->] [name path=arc1, domain=-150:45, opacity = 0] plot ({-3+1*cos(\x)}, {0+2*sin(\x)});
			\path [name intersections={of = arc1 and a, by = arc1a}];
			\path [name intersections={of = arc1 and b, by = arc1b}];
		  \draw [->] [domain=-150:45] plot ({-3+1*cos(\x)}, {0+2*sin(\x)});
			\filldraw [white] (arc1a) circle (3pt);
			\filldraw [white] (arc1b) circle (3pt);
			      \draw [thick, dashed] (b1) -- (b2);
			      \draw [thick, dashed] (c1) -- (c2);
			\filldraw (-3,0) circle (2.25pt);
		  \draw [thick, dashed, domain=0:85] plot ({-4+1*cos(\x)}, {0+1*sin(\x)});
		  \end{tikzpicture}
      \end{subfigure}%
      \begin{subfigure}{0.5\textwidth}
      \centering
		  \begin{tikzpicture}[scale=0.8]
		  \draw      [domain=45:180] plot ({-3+1*cos(\x)}, {0+2*sin(\x)}) node[circle, fill, inner sep=2pt] (b1) {};
		  \draw      [domain=180:210] plot ({-3+1*cos(\x)}, {0+2*sin(\x)}) node[circle, fill, inner sep=2pt] (c1) {};
		  \draw [->] [domain=-150:45] plot ({3+1*cos(\x)}, {0+2*sin(\x)});
		  \draw      [domain=45:150] plot ({3+1*cos(\x)}, {0+2*sin(\x)}) node[circle, fill, inner sep=2pt] (a2) {};
		  \draw      [domain=150:180] plot ({3+1*cos(\x)}, {0+2*sin(\x)}) node[circle, fill, inner sep=2pt] (b2) {};
		  \draw      [domain=180:210] plot ({3+1*cos(\x)}, {0+2*sin(\x)}) node[circle, fill, inner sep=2pt] (c2) {};
		  \draw [name path=a, opacity=0] (b1) -- (b2);
		  \draw [name path=b, opacity=0] (c1) -- (c2);
		  \draw [->] [name path=arc1, domain=-150:45, opacity = 0] plot ({-3+1*cos(\x)}, {0+2*sin(\x)});
			\path [name intersections={of = arc1 and a, by = arc1a}];
			\path [name intersections={of = arc1 and b, by = arc1b}];
		  \draw [->] [domain=-150:45] plot ({-3+1*cos(\x)}, {0+2*sin(\x)});
			\filldraw [white] (arc1a) circle (3pt);
			\filldraw [white] (arc1b) circle (3pt);
			      \draw [thick, dashed] (b1) -- (b2);
			      \draw [thick, dashed] (c1) -- (c2);
			\filldraw (1,0) circle (2.25pt);
		  \draw [thick, dashed, domain=85:180] plot ({2+1*cos(\x)}, {0+1*sin(\x)});
		  \end{tikzpicture}
      \end{subfigure}%
      \par\bigskip
      \par\bigskip
      \begin{subfigure}{0.5\textwidth}
      \centering
		  \begin{tikzpicture}[scale=0.8]
		  \draw      [domain=45:160] plot ({-3+1*cos(\x)}, {0+2*sin(\x)}) node[circle, fill, inner sep=2pt] (a1) {};
		  \draw      [domain=160:200] plot ({-3+1*cos(\x)}, {0+2*sin(\x)}) node[circle, fill, inner sep=2pt] (b1) {};
		  \draw [->] [domain=0:45] plot ({3+1*cos(\x)}, {0+2*sin(\x)});
		  \draw      [domain=45:90] plot ({3+1*cos(\x)}, {0+2*sin(\x)}) node[circle, fill, inner sep=2pt] (a0) {};
		  \draw      [domain=90:160] plot ({3+1*cos(\x)}, {0+2*sin(\x)}) node[circle, fill, inner sep=2pt] (a2) {};
		  \draw      [domain=160:180] plot ({3+1*cos(\x)}, {0+2*sin(\x)}) node[circle, fill, inner sep=2pt] (c2) {};
		  \draw      [domain=180:200] plot ({3+1*cos(\x)}, {0+2*sin(\x)}) node[circle, fill, inner sep=2pt] (b2) {};
		  \draw      [domain=200:360] plot ({3+1*cos(\x)}, {0+2*sin(\x)});
		  \draw [name path=a, opacity=0] (a1) -- (a2);
		  \draw [name path=b, opacity=0] (b1) -- (b2);
		  \draw [->] [name path=arc1, domain=0:45, opacity = 0] plot ({-3+1*cos(\x)}, {0+2*sin(\x)});
		  \draw      [name path=arc2, domain=200:360, opacity=0] plot ({-3+1*cos(\x)}, {0+2*sin(\x)});
			\path [name intersections={of = arc1 and a, by = arc1a}];
			\path [name intersections={of = arc2 and b, by = arc2b}];
		  \draw [->] [domain=0:45] plot ({-3+1*cos(\x)}, {0+2*sin(\x)});
		  \draw      [domain=200:360] plot ({-3+1*cos(\x)}, {0+2*sin(\x)});
			\filldraw [white] (arc1a) circle (3pt);
			\filldraw [white] (arc2b) circle (3pt);
			      \draw [thick, dashed] (a1) -- (a2);
			      \draw [thick, dashed] (b1) -- (b2);
		  \draw      [thick, dashed, domain=-90:0] plot ({2+1*cos(\x)}, {2+2*sin(\x)});
		  \end{tikzpicture}
      \end{subfigure}%
      \begin{subfigure}{0.5\textwidth}
      \centering
		  \begin{tikzpicture}[scale=0.8]
		  \draw      [domain=45:90] plot ({-3+1*cos(\x)}, {0+2*sin(\x)}) node[circle, fill, inner sep=2pt] (a0) {};
		  \draw      [domain=90:160] plot ({-3+1*cos(\x)}, {0+2*sin(\x)}) node[circle, fill, inner sep=2pt] (a1) {};
		  \draw      [domain=160:180] plot ({-3+1*cos(\x)}, {0+2*sin(\x)}) node[circle, fill, inner sep=2pt] (c1) {};
		  \draw      [domain=180:200] plot ({-3+1*cos(\x)}, {0+2*sin(\x)}) node[circle, fill, inner sep=2pt] (b1) {};
		  \draw [->] [domain=0:45] plot ({3+1*cos(\x)}, {0+2*sin(\x)});
		  \draw      [domain=45:160] plot ({3+1*cos(\x)}, {0+2*sin(\x)}) node[circle, fill, inner sep=2pt] (a2) {};
		  \draw      [domain=160:200] plot ({3+1*cos(\x)}, {0+2*sin(\x)}) node[circle, fill, inner sep=2pt] (b2) {};
		  \draw      [domain=200:360] plot ({3+1*cos(\x)}, {0+2*sin(\x)});
		  \draw [name path=a, opacity=0] (a1) -- (a2);
		  \draw [name path=b, opacity=0] (b1) -- (b2);
		  \draw [->] [name path=arc1, domain=0:45, opacity = 0] plot ({-3+1*cos(\x)}, {0+2*sin(\x)});
		  \draw      [name path=arc2, domain=200:360, opacity=0] plot ({-3+1*cos(\x)}, {0+2*sin(\x)});
		  \draw      [name path=circ, domain=-90:0, opacity=0] plot ({-4+1*cos(\x)}, {2+2*sin(\x)});
			\path [name intersections={of = arc1 and a, by = arc1a}];
			\path [name intersections={of = arc2 and b, by = arc2b}];
			\path [name intersections={of = circ and a, by = circa}];
		  \draw [->] [domain=0:45] plot ({-3+1*cos(\x)}, {0+2*sin(\x)});
		  \draw      [domain=200:360] plot ({-3+1*cos(\x)}, {0+2*sin(\x)});
		  \draw      [thick, dashed, domain=-90:0] plot ({-4+1*cos(\x)}, {2+2*sin(\x)});
			\filldraw [white] (arc1a) circle (3pt);
			\filldraw [white] (arc2b) circle (3pt);
			\filldraw [white] (circa) circle (3pt);
			      \draw [thick, dashed] (a1) -- (a2);
			      \draw [thick, dashed] (b1) -- (b2);
		  \end{tikzpicture}
      \end{subfigure}%
      \caption{Feynman diagrams for $\gamma_3^{\ 2}$.}
      \label{figuradelordertwovassilievlinkinvariant}
      \end{figure}

In the present work we will focus in the first three orders of (\ref{orderfourloop}) so $\gamma_1^{\ 1}$ and $\gamma_3^{\ 2}$ need to be analysed.
The Feynman diagram corresponding to $\gamma_1^{\ 1}$ is that of figure \ref{figuradellinkingnumber} (note again that there is no linear 
term in (\ref{interactionpartloop}) but this is exactly the diagram for that order) while figure \ref{figuradelordertwovassilievlinkinvariant} stands for the diagrams corresponding to $\gamma_3^{\ 2}$.
      
\subsection{About numerical calculations}

The invariants $\widetilde{\alpha}_{ij}(K)$ can be calculated by performing the integrals defining them but it is easier to use Eq. (\ref{ordersixknot}) because its 
left-hand-side can be found in the literature, they are the quantum group invariants obtained from the nonperturbative analysis. 

As an example, the values of $\widetilde{\alpha}_{21}(K)$ and $\widetilde{\alpha}_{31}(K)$ are calculated for the right-handed trefoil $3_1$ with gauge group $G=SU(2)$. In the left-hand-side of (\ref{ordersixknot}) the normalized HOMFLY-PT polynomial is used, {\it i.e.},
      \begin{alignat*}{6}
      \lambda \left( 1 + q^2 - \lambda q^2 \right) &=& & q \left( 1 + q^2 - q^3 \right) \\
      &=& & q + q^3 - q^4 \\
      &=& & \exp(x) + \exp(3x) - \exp(4x) \\
      &=& & \left( 1 + x + \frac{x^2}{2} + \frac{x^3}{6} \right) + \left( 1 + 3x + \frac{9x^2}{2} + \frac{27x^3}{6} \right) \\
      & &-& \left( 1 + 4x + \frac{16x^2}{2} + \frac{64x^3}{6} \right) + O(x^4) \\
      &=& & 1 + x^2 \left( \frac{1}{2} + \frac{9}{2} - \frac{16}{2} \right) + x^3 \left( \frac{1}{6} + \frac{27}{6} - \frac{64}{6} \right) + O(x^4) \\
      &=& & 1 - 3x^2 - 6x^3 + O(x^4),
      \numberthis
      \end{alignat*}
where $\lambda = q^{N-1} = q$ because $N=2$. In the right hand side of (\ref{ordersixknot}) the values \cite{AL95}
     \begin{align*}
     r_{2,1} &= C_3  
              = - \frac{1}{4} \left( N^2 - 1 \right) = - \frac{3}{4}, \\
     r_{3,1} &= (C_3)^2 (C_2)^{-1} 
              = \frac{\left( \displaystyle - \frac{1}{4} \left( N^2 - 1 \right) \right)^2}{\displaystyle - \frac{1}{2N} \left( N^2 - 1 \right)} 
              = - \frac{2N}{16} \left( N^2 - 1 \right) = - \frac{3}{4},
     \numberthis
     \end{align*}
for the group factors (the Casimirs are given in the fundamental representation) are used, {\it i.e.}, it looks like
      \begin{equation}
      1 + \widetilde{\alpha}_{21} \left( -\frac{3}{4} \right) x^2 + \widetilde{\alpha}_{31} \left( -\frac{3}{4} \right) x^3 + O(x^4).
      \end{equation}

By equating both sides at order three it is found that
      \begin{equation}
      1 - 3x^2 - 6x^3 = 1 + \widetilde{\alpha}_{21} \left( -\frac{3}{4} \right) x^2 + \widetilde{\alpha}_{31} \left( -\frac{3}{4} \right) x^3,
      \end{equation}
or simply $\widetilde{\alpha}_{21} = 4$ and $\widetilde{\alpha}_{31} = 8$, in accordance with equation $(5.1)$ of \cite{AL95}.

\section[Bott-Taubes integration and volume-preserving vector fields]{Bott-Taubes integration and volume-preserving \\ vector fields}
\label{volumepreserving}

Some efforts to describe Vassiliev invariants (of finite type) in a geometrical framework were made by Kontsevich \cite{Kontsevich} and by Bott and Taubes 
\cite{BT94}, where the identification of the correct spaces in which Feynman integrals could be rewriting was one of the key achievements of such a description.

Since the $19$\textsuperscript{th} century Gauss work on electromagnetic theory showed that there is an integral formula for the linking number of two curves $\gamma_0$ and $\gamma_1$ in $\mathbb{R}^3$, which also represents a homotopy invariant of both curves under some assumptions of transversal intersection. It is
given by 
      \begin{equation}
      \frac{1}{4\pi}\int\limits_{{\bf S}^1\times {\bf S}^1}\Phi^*\omega,
      \end{equation}
where $\Phi^*\omega$ is the pullback of $\omega\in\Omega^2\left({\bf S}^2\right)$, which is the volume 2-form given in \cite{Mos04}. Moreover given a pair of curves $\gamma_0,\gamma_1: {\bf S}^1 \to \mathbb{R}^3$, the map $\Phi: {\bf S}^1 \times {\bf S}^1 \to {\bf S}^2$ is given by $\Phi=\phi\circ\left(\gamma_0 \times\gamma_{1}\right)$, where $\phi: \mathbb{R}^3 \times \mathbb{R}^3 \to {\bf S}^2$ is the map defined via $\phi(x_1,x_2)=\frac{x_1-x_2}{|x_1-x_2|^3}$ with $x_1,x_2 \in \mathbb{R}^3$. This map is known as the Gauss map and in components it is given by
      \begin{equation}
      \omega_{\mu\nu}=\frac{\varepsilon_{\mu\nu\sigma}}{4\pi}\frac{x^{\sigma}}{|x|^3}.
      \label{omegaform}
      \end{equation}
      
This procedure gives a new way to formulate the question of building up a homotopy invariant for a single knot in the same way as the linking number was developed. The answer to that question requires the introduction of {\it configuration spaces}, $C(n,M)$, defined by
	   \begin{equation}
       C(n,M)=\left\lbrace (x_1,\ldots,x_n)\in \prod_{i=1}^n M \; | \; x_i\neq x_j \Leftrightarrow i\neq j \right\rbrace
	   \end{equation}	   	
as the natural framework where such an integral has to be defined (because of the explicit form of $\phi$). To ensure that an integral defined on this kind of spaces converges, some new features such as compactification of the configuration space $C(n,M)$ are required.  Here we will use the well known {\it Fulton-MacPherson compactification} with some refinements worked out in \cite{Sin04}. 

In Ref. \cite{KV16} there is a description of how to obtain a finite type knot invariant via a linear combination of integrals on some bundles (pullback bundles, see Appendix \ref{appendixA}) of {\it compactified configuration spaces} $C[n,M]$. The building blocks for writing this linear combinations depend on some trivalent diagrams $D$ such as those in figure \ref{order2labastida95} and in the diagram of figure \ref{diagramota01}. In this diagram $P(D)$ denotes the set of dashed lines in each Feynman diagram, $\phi$ is a product of the Gauss maps each one associated to a line in $P(D)$ and $\Phi$ is an extension of the function $\phi$ to the corresponding compactified space \cite{Vol13}.

      \begin{figure}[ht]
      \[ \xymatrix{
      & & C\left(p; {\bf S}^1\right) \times \mathcal{K} \ar[dd]^{ev} & & & \\
      & & & & & \\
      & C\left(p; {\bf S}^1\right) \ar[r]^{ev_K} \ar[dd]^{\alpha_p^1} & C\left(p; {\bf S}^3\right) \ar[dd]^{\alpha_p^3} & \ar[l]_{\pi_p} C\left(p+q; {\bf S}^3\right) \ar[rr]^{\phi} \ar[dd]^{\alpha_{p+q}^3} & & \Pi_{e\in P(D)}{\bf S}^2 \\
      & & & & & \\
      & C\left[p; {\bf S}^1\right] \ar[r]^{\overline{ev}_K} & C\left[p; {\bf S}^3\right] & \ar[l]_{\overline{\pi}_p} C\left[p+q; {\bf S}^3\right] \ar[uurr]_{\Phi} & & \\
      & & & & & \\
      \mathcal{K} & & \ar[ll]_{pr} C\left[p; {\bf S}^1\right] \times \mathcal{K} \ar[uu]_{\overline{ev}} & \ar[l]_{pr_1} C\left[p, q; {\bf S}^1, {\bf S}^3\right] \ar[uu]_{pr_2} \pullbackcorner & &
      } \]
      \caption{Mathematical diagram for a trivalent diagram $D$ with $p$ points on the knot and $q$ points out of it.}
      \label{diagramota01} 
      \end{figure}
      
In figure \ref{diagramota01}, $\mathcal{K}=\{\beta:{\bf S}^1 \rightarrow {\bf S}^3\;|\; K \text{ is an smooth embedding}\}$ is the space of all (smooth) knots in ${\bf S}^3$ and $K$ is one of these knots. The map $ev$ is given by
	  \begin{equation}
			ev((s_1,\ldots,s_p),K)=(K(s_1),\ldots,K(s_p))\in C(p,{\bf S}^3),
	  \end{equation}
so that $ev_K$ can be expressed as $ev_K:=ev(\cdot,K)$. Also $\pi_p$ and $pr$ are projections defined via
	  \begin{equation}
      \pi_p(x_1,\ldots,x_p,x_{p+1},\ldots,x_{p+q})=(x_1,\ldots,x_p), \qquad pr(v,K)=K,
 	  \end{equation}
respectively. As stated in Ref. \cite{KV16} the maps $\alpha_m^n$ are the inclusion of configuration spaces into their compactifications while $m$ and $n$ denote, respectively, the number of points in $C(m,{\bf S}^n)$ and the dimension of the underlaying sphere. Finally $\overline{ev}_K$ and $\overline{\pi}_p$ are extensions of the maps $ev_K$ and $\pi_p$ to the corresponding compactifications of their domains and codomains.

The building blocks are actually functions $I_D:\mathcal{K}\rightarrow\mathbb{R}$ on the knots space $\mathcal{K}$ defined via
      \begin{equation}
      I_{D}(K)=(pr\circ pr_1)_*\left(\Phi\circ pr_2\right)^*\overline{\omega}, \; \; \; K\in\mathcal{K},
      \label{Ivolic}
      \end{equation} 
where the mapping $pr: C[p;{\bf S}^1] \times {\cal K} \to {\cal K}$ is the projection in the second entry and 
      \begin{equation}
      \overline{\omega} := \bigwedge_{e \in P(D)} \omega \in \Omega^{2P(D)}\bigg(\prod_{e\in P(D)}{\bf S}^2\bigg)
      \end{equation}
is a product of the unit volume form $\omega$ given as in Eq. (\ref{omegaform}), $\left(pr\circ pr_1\right)_*$ stands for the pushforward (or integration over the fiber, see Appendix \ref{appendixB} for the definition) of the form $\left(\Phi\circ pr_2\right)^*\overline{\omega}$ and the fibers of $\left(pr\circ pr_1\right)$ are compact smooth manifolds with 
corners \cite{Vol07}.

One of the results proved in Ref. \cite{KV16} asserts that for each diagram $D$ the value of these blocks in a specific knot $K$ can also be calculated by integration on the original configuration space $C\left(p,{\bf S}^1\right)$ of only the points that belong to $K$ (proposition (3.7) in \cite{KV16}), {\it i.e.},
 	  \begin{equation}
	  I_{D}(K)=\int\limits_{C(p,{\bf S}^1)}f_{D,K}(\overline{s})d\overline{s},
	  \label{Ieqintf}
	  \end{equation}
where $f_{D,K}$ is defined as
      \begin{alignat*}{5}
      f_{D,K}(\overline{s})&=&&\left(\left(\alpha_p^3\circ ev_K\right)^*\left(\overline{\pi}_p\right)_*\Phi^*\overline{\omega}\right)_{\overline{s}}(\partial_{\overline{s}})\\
      &=&&\left(\left(\alpha_p^3\right)^*\left(\overline{\pi}_p\right)_*\Phi^*\overline{\omega}\right)_{K(\overline{s})}(\dot{K}(s_1),\ldots,\dot{K}(s_p)),
      \numberthis
      \label{fvolic}
      \end{alignat*}
with $\overline{s}\in C\left(p,{\bf S}^1\right)$ and $\partial_{\overline{s}}$ a $p$-tuple where each element is the canonical vector field $\partial_{s_i}$ on ${\bf S}^1$, also $\left(\dot{K}(s_1),\ldots,\dot{K}(s_p)\right)$ is given by the pushforward of $\partial_{\overline{s}}$ by $ev_K$. This way $\dot{K}$ defines a vector field along the curve $K$ in $\mathbb{R}^3$ and also for each point $\overline{s} \in C(p,{\bf S}^1)$ a frame $\left(\dot{K}(s_1),\ldots,\dot{K}(s_p)\right)$ in $C(p,{\bf S}^3)$ which is endowing the knot $K$.

These basic blocks can also be used to define invariants associated with a volume-preserving vector field $X$ on a compact domain 
$\mathcal{S}$ of $\mathbb{R}^3$ and tangent to its boundary by making an appropriate generalization of Eq. (\ref{fvolic}). 

\begin{figure}[H]
            \centering
            \begin{tikzpicture}[scale=0.5]
            \draw [black, arrow data={0.90}{stealth}] plot [smooth] coordinates {(-4.5,-2) (-3.5,-1) (-2,0) (2,1) (4,3)};
            \draw [blue, arrow data={0.75}{stealth}] plot [smooth] coordinates {(2,1) (2.25,0) (2,-0.5) (1,-0.5) (0.5,-1) (-0.25,-3) (-1.5,-3.6) (-2.5,-3) (-3.5,-1)};
            \draw [red, arrow data={0.65}{stealth}] plot [smooth] coordinates {(2,1) (1.5,0.5) (1.5,-1) (0.5,-2) (-1,-2) (-3,-0.5) (-3.5,-0.5) (-3.5,-1)};
            \draw [black] plot [smooth cycle] coordinates {(3,6) (2,5) (0,3) (-5,2) (-7.5,-1) (-7,-4) (-3,-6) (2,-5) (4,-3) (8,-1) (9,2) (8,4) (5,6)};
            \filldraw (-3.5,-1) circle (2pt);
            \filldraw (2,1) circle (2pt);
            \node[label=:] (A) at (-1.5,-4) {\textcolor{blue}{$\gamma$}};
            \node[label=:] (B) at (1.5,-2) {\textcolor{red}{$\gamma'$}};
            \node[label=:] (C) at (-4,-1) {\textcolor{black}{$x$}};
            \node[label=:] (D) at (1.5,1.5) {\textcolor{black}{\small $\theta^x(T)$}};
            \node[label=:] (E) at (-6,3) {\textcolor{black}{{\large $S$}}};
            \draw [black, yshift=1cm] plot [smooth] coordinates {(5,0) (5.25,-0.25) (5.75,-0.4) (6.25,-0.4) (6.75,-0.25) (7,0)};
            \draw [black, yshift=1cm] plot [smooth] coordinates {(5.25,-0.25) (5.5,0) (5.80,0.1) (6.20,0.1) (6.5,0) (6.75,-0.25)};
            \draw [black, yshift=-2.5cm, xshift=-7.5cm] plot [smooth] coordinates {(5,0) (5.25,-0.25) (5.75,-0.4) (6.25,-0.4) (6.75,-0.25) (7,0)};
            \draw [black, yshift=-2.5cm, xshift=-7.5cm] plot [smooth] coordinates {(5.25,-0.25) (5.5,0) (5.80,0.1) (6.20,0.1) (6.5,0) (6.75,-0.25)};
            \end{tikzpicture}
            \caption{Asymptotic cycles for $\gamma$ and $\gamma'$ defined via the flow $\theta$ in the domain $S$.}
            \label{dibujodelosciclos}
            \end{figure}
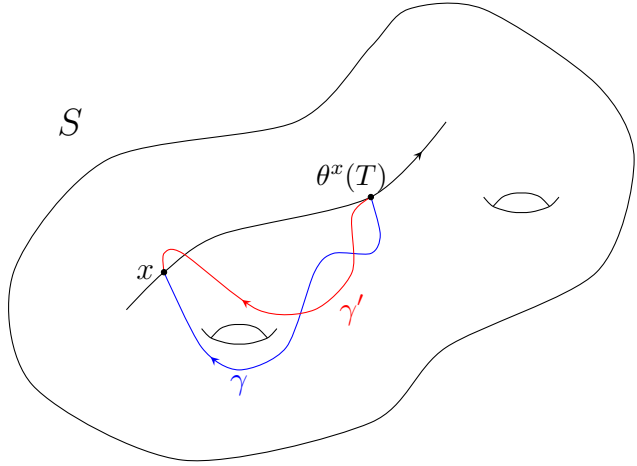

Explicitly, let $X$ be a volume-preserving vector field on a domain $\mathcal{S}$ in $\left(\mathbb{R}^3,\mu\right)$ with flow $\theta:\mathbb{R}\times \mathcal{S}\rightarrow \mathcal{S}$ and let $x\in \mathcal{S}$, then 
$\theta^x:=\theta(\cdot,x)$ defines a curve on $\mathcal{S}$. Moreover, $\mu$ is a Borel probability measure invariant under the flow. Now for every $T\in \mathbb{R}$, by taking $\sigma(x,\theta^x(T))$ to be the set of uniformly bounded curves between $x$ and $\theta^x(T)$ and $\gamma\in\sigma(x,\theta^x(T))$, then 
$\gamma_T^x:{\bf S}^1\rightarrow\theta([0,T],x)\cup\gamma$ is a piecewise smooth closed curve on $\mathcal{S}$ that can be defined in the interval $[0,T+1]$, where $[T,T+1]$ parametrizes the points in $\gamma$. This construction is illustrated in figure \ref{dibujodelosciclos} below and corresponds to the asymptotic cycles of Schwartzman \cite{SA,S2}.

In a similar way, $f_{D,K}$ given in Eq. (\ref{fvolic}) is generalized to $f_{D,X}$ who belongs to $\Omega^{0}\left(C(p,\mathcal{S})\right)$ and it is defined by
      \begin{alignat*}{5}
      f_{D,X}(\overline{x})&:=&&\left(\left(\alpha_p^3\right)^*\left(\overline{\pi}_p\right)_*\Phi^*\overline{\omega}\right)_{\overline{x}}(X_{x_1},\ldots,X_{x_p}).
      \numberthis
      \label{lafdevolicconx}
      \end{alignat*}
      
\noindent      
Then by the key lemma in Ref. \cite{KV16} one gets the asymptotic value of $I_{D}$ of order $p$ along the flow of $X$, $\overline{\lambda}_{D}$, as
      \begin{alignat*}{5}
      \overline{\lambda}_{D}(x)&=&&\lim_{T\rightarrow \infty}\frac{1}{T^p}I_D\left(\gamma^x_T\right)\\
      &=&&\lim_{T\rightarrow \infty}\frac{1}{T^p}\int\limits_0^{T+1}\cdots\int\limits_0^{T+1} f_{D,X}(\gamma^x_T(t_1),\ldots,\gamma^x_T(t_p))dt_1\wedge\cdots\wedge dt_p\\
      &=&&\lim_{T\rightarrow \infty}\frac{1}{T^p}\int\limits_0^T\cdots\int\limits_0^T f_{D,X}(\theta^x(t_1),\ldots,\theta^x(t_p))dt_1\wedge\cdots\wedge dt_p,
      \numberthis
      \label{intlambdaeqintf}
      \end{alignat*}
where the last expression stands for $\lambda_{D}(x)$ as in the same reference.
      
This is a function in $L^1(\mathcal{S},\mu)$ such that when integrated via the invariant measure gives a new kind of
flow-invariant quantity that can be rewriting as
      \begin{alignat*}{5}
      \int\limits_{\mathcal{S}}\lambda_D\mu=\int\limits_{\mathcal{S}^p}f_{D,X}\overline{\mu}_{\Delta},
      \numberthis
      \label{intasyminv}	
      \end{alignat*}
where, by taking $\theta^p$ to be the $p$-fold product of the flow $\theta$, we have 
      \begin{alignat*}{5}
      \overline{\mu}_{\Delta}=&&\lim_{T\rightarrow \infty}\frac{1}{T^p}\int\limits_0^T\cdots\int\limits_0^T \left(\left(\theta^p\right)_*\mu_{\Delta}\right)dt_1\wedge\cdots\wedge dt_p
      \numberthis
      \label{invkfoldmeasure}
      \end{alignat*}	
is a well defined limit measure. 

Finally, theorems $A$ and $B$ from Ref. \cite{KV16} assert that this generalization also works as the basic building blocks for asymptotic invariants of the vector field $X$, {\it i.e.}, quantities that are invariant under the action of their own flow $\theta$ that are going to be calculated through asymptotic values of some functions defined on the space of knots $\mathcal{K}$.

\section[Correspondence between Feynman diagrams and Bott-Taubes integrals]{Correspondence between Feynman diagrams and \\ Bott-Taubes integrals}

The diagrams for a knot to be analysed in this work can be built from Chern-Simons theory by considering the pertubative expansion of its Wilson loop (\ref{wilsonloopexpansion}) at order four (the information relative to the group is not written explicitly), {\it i.e.},
      \begin{alignat*}{4}
      W_R^K(A) \approx & \ 1 &+& \mathrlap{\left( \frac{1}{\sqrt{k}} \right) \oint_{{\bf S}^1} ds A_i(K(s)) \dot{K}^i(s)} \\
      & &+& \mathrlap{\left( \frac{1}{\sqrt{k}} \right)^2 \iint\limits_{s_1 < s_2} ds_1 ds_2 A_{i_1}(K(s_1)) A_{i_2}(K(s_2)) \dot{K}^{i_1}(s_1) \dot{K}^{i_2}(s_2)} \\
      & &+& \mathrlap{\left( \frac{1}{\sqrt{k}} \right)^3 \iiint\limits_{s_1 < s_2 < s_3} ds_1 ds_2 ds_3 \biggl[ A_{i_1}(K(s_1)) A_{i_2}(K(s_2)) A_{i_3}(K(s_3))} \\
      & & & & \times \dot{K}^{i_1}(s_1) \dot{K}^{i_2}(s_2) \dot{K}^{i_3}(s_3) \biggr] \\ 
      & &+& & \left( \frac{1}{\sqrt{k}} \right)^4 \iiiint\limits_{s_1 < s_2 < s_3 < s_4} ds_1 ds_2 ds_3 ds_4 \biggl[ A_{i_1}(K(s_1)) A_{i_2}(K(s_2)) A_{i_3}(K(s_3)) A_{i_4}(K(s_4)) \\
      & & & & \times \dot{K}^{i_1}(s_1) \dot{K}^{i_2}(s_2) \dot{K}^{i_3}(s_3) \dot{K}^{i_4}(s_4) \biggr],
      \numberthis
      \end{alignat*}  
where $K:{\bf S}^1 \to {\bf S}^3$ is the knot embedding, and the interaction term (\ref{actioninteractionexpansion}) at first order, {\it i.e.},
      \begin{align*}
      e^{\displaystyle i I'} 
      &\approx 1 + \left( \frac{1}{\sqrt{k}} \right)^3 \frac{ik}{4 \pi} \int_{\mathbb{R}^3} {\rm Tr} \left( \frac{1}{3} \varepsilon^{ijk} A_i [A_j, A_k] + 2 \bar{c} \partial_i [A^i,c] \right) \\ 
      &= 1 + \left( \frac{1}{\sqrt{k}} \right) \frac{i}{4 \pi} \int_{\mathbb{R}^3} {\rm Tr} \left( \frac{1}{3} \varepsilon^{ijk} A_i [A_j, A_k] 
      +  2 \bar{c} \partial_i [A^i,c] \right). \numberthis
      \end{align*}  

In the procedure to get the previous equations we have imposed the following redefinitions       
      \begin{equation}
      A \longmapsto \frac{A}{\sqrt{k}}, \ \ \ \ \ \ \ \ \ \ c \longmapsto \frac{c}{\sqrt{k}},  \ \ \ \ \ \ \ \ \ \
      \bar{c} \longmapsto \frac{\bar{c}}{\sqrt{k}},
      \end{equation}
for the gauge, ghost and antighost fields, respectively. This way the vacuum expectation value (\ref{vevknottotal}) gives
            \begin{align*}
                    & \int DA D\phi Dc D\bar{c} e^{\displaystyle i(I_{0} + I_g)} e^{\displaystyle i I'} W_R^K(A) \\
            \approx & \int DA D\phi Dc D\bar{c} e^{\displaystyle i(I_{0} + I_g)}  \biggr[
                    1 + \left( \frac{1}{\sqrt{k}} \right) \frac{i}{4 \pi} \int_{\mathbb{R}^3} {\rm Tr} \left( \frac{1}{3} \varepsilon^{ijk} A_i [A_j, A_k] + 2 \bar{c} \partial_i [A^i,c] \right) \biggr] \\
                    & \times \Biggl[ 1 +  \left( \frac{1}{\sqrt{k}} \right) \oint_{{\bf S}^1} ds A_i(K(s)) \dot{K}^i(s) \\
                    & + \left( \frac{1}{\sqrt{k}} \right)^2 \iint\limits_{s_1 < s_2} ds_1 ds_2 A_{i_1}(K(s_1)) A_{i_2}(K(s_2)) \dot{K}^{i_1}(s_1) \dot{K}^{i_2}(s_2) \\
                    & + \left( \frac{1}{\sqrt{k}} \right)^3 \iiint\limits_{s_1 < s_2 < s_3} ds_1 ds_2 ds_3 A_{i_1}(K(s_1)) A_{i_2}(K(s_2)) A_{i_3}(K(s_3))              \dot{K}^{i_1}(s_1) \dot{K}^{i_2}(s_2) \dot{K}^{i_3}(s_3) \\
                    & + \bigg( \frac{1}{\sqrt{k}} \bigg)^4 \iiiint\limits_{s_1 < s_2 < s_3 < s_4} ds_1 ds_2 ds_3 ds_4 \bigg\{ A_{i_1}(K(s_1)) A_{i_2}(K(s_2)) A_{i_3}(K(s_3)) A_{i_4}(K(s_4)) \\
                    & \hspace{8.5cm} \times \dot{K}^{i_1}(s_1) \dot{K}^{i_2}(s_2) \dot{K}^{i_3}(s_3) \dot{K}^{i_4}(s_4) \bigg\} \Biggr].
            \numberthis
            \label{origendiagramas} 
            \end{align*}
            
The interest of this work is focused on some normalized terms coming from Eq. (\ref{origendiagramas}), specifically, that of order $1/k$ coming from the third term of $W_R^K(A)$ and the first term of $e^{i I'}$, {\it i.e.},
      \begin{align}
      V_1 = \frac{1}{N} \int DA D\phi Dc D\bar{c} e^{\displaystyle i (I_{0} + I_g)} 
      \iint\limits_{s_1 < s_2} ds_1 ds_2 A_{i_1}(K(s_1)) A_{i_2}(K(s_2)) \dot{K}^{i_1}(s_1) \dot{K}^{i_2}(s_2),
      \end{align}
where $N = \int DA Dc D\bar{c} e^{\displaystyle i I_{0}}$. The previous equation gives rise to the self-linking invariant or the Vassiliev invariant of first order. Also important are those of order $1/k^2$ coming from the fourth term of
$W_R^K(A)$ and the part without ghosts of the second term of $e^{i I'}$, {\it i.e.},
      \begin{align*}
      V_{21} = \frac{1}{N} \left( \frac{i}{4 \pi} \right) \int DA D\phi Dc D\bar{c} e^{\displaystyle i (I_{0} + I_g)}
      \int_{\mathbb{R}^3} {\rm Tr} \left( \frac{1}{3} \varepsilon^{ijk} A_i [A_j, A_k] \right) \iiint\limits_{s_1 < s_2 < s_3} ds_1 ds_2 ds_3 & \\
      \times \biggl\{ A_{i_1}(K(s_1)) A_{i_2}(K(s_2)) A_{i_3}(K(s_3)) \dot{K}^{i_1}(s_1) \dot{K}^{i_2}(s_2) \dot{K}^{i_3}(s_3) \biggr\}, &
      \numberthis
      \end{align*}
and from the fifth term of $W_R^K(A)$ and the first term of $e^{i I'}$, {\it i.e.},
      \begin{align*}
      V_{22} = \frac{1}{N} \int DA D\phi Dc D\bar{c} e^{\displaystyle i (I_{0} + I_g)}
      \iiiint\limits_{s_1 < s_2 < s_3 < s_4} ds_1 ds_2 ds_3 ds_4 \biggl\{ A_{i_1}(K(s_1)) A_{i_2}(K(s_2)) & \\
      \times A_{i_3}(K(s_3)) A_{i_4}(K(s_4)) \dot{K}^{i_1}(s_1) \dot{K}^{i_2}(s_2) \dot{K}^{i_3}(s_3) \dot{K}^{i_4}(s_4) \biggr\}, &
      \numberthis
      \end{align*}
that form the Vassiliev invariant of second order. There are other three terms of order $1/k^2$ but according to \cite{Bar95} they do not contribute to the 
invariant (see section \ref{Feynman diagrams for knots}).
      
The three integrals above have ghosts dependence only in the $e^{i I_{0}}$ factor (\ref{actionfree}). Thus the factor obtained by
performing integration of $c$ and $\bar{c}$ fields will cancel that from the normalization factor $N$ yielding
      \begin{alignat*}{6}
      & V_1 &=& \mathrlap{\frac{1}{N_A} \int DA e^{\displaystyle i I_A}  \iint\limits_{s_1 < s_2} ds_1 ds_2 A_{i_1}(K(s_1)) A_{i_2}(K(s_2)) \dot{K}^{i_1}(s_1) \dot{K}^{i_2}(s_2),}
      \numberthis \label{selflinkingcs} \\ \\
      & V_{21} &=& \mathrlap{\frac{1}{N_A} \int DA e^{\displaystyle i I_A} 
      \int_{\mathbb{R}^3} {\rm Tr} \left( \frac{i}{12\pi} \varepsilon^{ijk} A_i [A_j, A_k] \right) \iiint\limits_{s_1 < s_2 < s_3} ds_1 ds_2 ds_3 \biggl\{ A_{i_1}(K(s_1))} \\
      &&&& \times A_{i_2}(K(s_2)) A_{i_3}(K(s_3)) \dot{K}^{i_1}(s_1) \dot{K}^{i_2}(s_2) \dot{K}^{i_3}(s_3) \biggr\},
      \numberthis \label{laycs} \\ \\
      & V_{22} &=& & \frac{1}{N_A} \int DA e^{\displaystyle i I_A} 
      \iiiint\limits_{s_1 < s_2 < s_3 < s_4} ds_1 ds_2 ds_3 ds_4 \biggl\{ A_{i_1}(K(s_1)) A_{i_2}(K(s_2)) A_{i_3}(K(s_3)) \\
      &&&& \times A_{i_4}(K(s_4)) \dot{K}^{i_1}(s_1) \dot{K}^{i_2}(s_2) \dot{K}^{i_3}(s_3) \dot{K}^{i_4}(s_4) \biggr\},
      \numberthis \label{laxcs}
      \end{alignat*}
where $N_A = \int DA  e^{\displaystyle i I_A}$ and 
      \begin{equation}      
      I_A = \frac{k}{4 \pi} \int_{\mathbb{R}^3} {\rm Tr} \left( \varepsilon^{ijk} A_i \partial_j A_k \right).
      \end{equation}

Following a similar procedure to get the first and second order expressions it can be shown that the third order expressions can be written as
      \begin{alignat*}{6}
      V_{31} &=
      \mathrlap{ \frac{1}{N_A} \int DA e^{\displaystyle I_A}
      \biggl[ \int_{\mathbb{R}^3} {\rm Tr} \left( \frac{i}{12\pi} \varepsilon^{ijk} A_i [A_j, A_k] \right) \biggr]^2
      \iiiint\limits_{s_1 < s_2 < s_3 < s_4} ds_1 ds_2 ds_3 ds_4 } \\
      && \times \bigg\{ A_{i_1}(K(s_1)) A_{i_2}(K(s_2)) A_{i_3}(K(s_3)) A_{i_4}(K(s_4)) \dot{K}^{i_1}(s_1) \dot{K}^{i_2}(s_2) \dot{K}^{i_3}(s_3) \dot{K}^{i_4}(s_4) \bigg\},
      \numberthis \label{dibujo31cs} \\ \\
      V_{32} &=
      \mathrlap{ \frac{1}{N_A} \int DA e^{\displaystyle i I_A}
      \int_{\mathbb{R}^3} {\rm Tr} \left( \frac{i}{12\pi} \varepsilon^{ijk} A_i [A_j, A_k] \right)
      \multiint{5}\limits_{s_1 < s_2 < s_3 < s_4 < s_5} ds_1 ds_2 ds_3 ds_4 ds_5 } \\
      && \times \biggl\{ A_{i_1}(K(s_1)) A_{i_2}(K(s_2)) A_{i_3}(K(s_3)) A_{i_4}(K(s_4)) A_{i_5}(K(s_5))  \\
      && \times \dot{K}^{i_1}(s_1) \dot{K}^{i_2}(s_2) \dot{K}^{i_3}(s_3) \dot{K}^{i_4}(s_4) \dot{K}^{i_5}(s_5) \biggr\},
      \numberthis \label{dibujo32cs} \\ \\
      V_{33} &=
      \mathrlap{ \frac{1}{N_A} \int DA e^{\displaystyle i I_A} 
      \multiint{6}\limits_{s_1 < s_2 < s_3 < s_4 < s_5 < s_6} ds_1 ds_2 ds_3 ds_4 ds_5 ds_6 } \\
      && \times \biggl\{ A_{i_1}(K(s_1)) A_{i_2}(K(s_2)) A_{i_3}(K(s_3)) A_{i_4}(K(s_4)) A_{i_5}(K(s_5)) A_{i_6}(K(s_6)) \\ 
      && \times \dot{K}^{i_1}(s_1) \dot{K}^{i_2}(s_2) \dot{K}^{i_3}(s_3) \dot{K}^{i_4}(s_4) \dot{K}^{i_5}(s_5) \dot{K}^{i_6}(s_6) \biggr\},
      \numberthis \label{dibujo33cs} \\ \\
      V_{34} &=
      \mathrlap{ \frac{1}{N_A} \int DA e^{\displaystyle i I_A} 
      \multiint{6}\limits_{s_1 < s_2 < s_3 < s_4 < s_5 < s_6} ds_1 ds_2 ds_3 ds_4 ds_5 ds_6 } \\
      && \times \biggl\{ A_{i_1}(K(s_1)) A_{i_2}(K(s_2)) A_{i_3}(K(s_3)) A_{i_4}(K(s_4)) A_{i_5}(K(s_5)) A_{i_6}(K(s_6))  \\ 
      && \times \dot{K}^{i_1}(s_1) \dot{K}^{i_2}(s_2) \dot{K}^{i_3}(s_3) \dot{K}^{i_4}(s_4) \dot{K}^{i_5}(s_5) \dot{K}^{i_6}(s_6) \biggr\}.
      \numberthis \label{dibujo34cs}
      \end{alignat*}
     
Before working with knots in the configuration space formalism it would be useful to rewrite the above expressions by using the propagator
      \begin{align}
      \big\langle A_{i_1}(K(s_1)) A_{i_2}(K(s_2)) \big\rangle = \frac{1}{N_A} \int DA e^{i I_A} A_{i_1}(K(s_1)) A_{i_2}(K(s_2)).
      \label{propagatorcs}
      \end{align}
      
Specifically, expression (\ref{selflinkingcs}) rewrites as
      \begin{alignat*}{6}
      V_1 &=& & \frac{1}{N_A} \int DA e^{i I_A} \iint\limits_{s_1 < s_2} ds_1 ds_2 A_{i_1}(K(s_1)) A_{i_2}(K(s_2)) \dot{K}^{i_1}(s_1) \dot{K}^{i_2}(s_2) \\
      &=& & \iint\limits_{s_1 < s_2} ds_1 ds_2 \biggl[ \frac{1}{N_A} \int DA e^{i I_A} A_{i_1}(K(s_1)) A_{i_2}(K(s_2)) \biggr] \dot{K}^{i_1}(s_1) \dot{K}^{i_2}(s_2) \\
      &=& & \iint\limits_{s_1 < s_2}   ds_1 ds_2 \big\langle A_{i_1}(K(s_1)) A_{i_2}(K(s_2)) \big\rangle \dot{K}^{i_1}(s_1) \dot{K}^{i_2}(s_2).
      \numberthis
      \label{selflinkingcstrabajado}      
      \end{alignat*}
      
Expression (\ref{laycs}) looks like
      \begin{alignat*}{6}
      V_{21} 
      &=& \mathrlap{ \frac{1}{N_A} \int DA e^{i I_A}
      \int_{\mathbb{R}^3} {\rm Tr} \left( \frac{i}{12\pi} \varepsilon^{ijk} A_i [A_j, A_k] \right) \iiint\limits_{s_1 < s_2 < s_3} ds_1 ds_2 ds_3 \biggl\{ A_{i_1}(K(s_1)) } && \\
      && && 
      \times A_{i_2}(K(s_2)) A_{i_3}(K(s_3)) \dot{K}^{i_1}(s_1) \dot{K}^{i_2}(s_2) \dot{K}^{i_3}(s_3) \biggr\} \\
      &=& && 
      \iiint\limits_{s_1 < s_2 < s_3} ds_1 ds_2 ds_3 \int\limits_{\mathbb{R}^3} \varepsilon^{ijk} d^3 x_4
      \biggl\{ \big\langle A_i(x_4) A_j(x_4) A_k(x_4) A_{i_1}(K(s_1)) A_{i_2}(K(s_2)) \\
      && && 
      \times A_{i_3}(K(s_3)) \big\rangle \dot{K}^{i_1}(s_1) \dot{K}^{i_2}(s_2) \dot{K}^{i_3}(s_3) \biggr\} \\
      &=& \mathrlap{
      \iiint\limits_{s_1 < s_2 < s_3} ds_1 ds_2 ds_3 \int\limits_{\mathbb{R}^3} \varepsilon^{ijk} d^3 x_4
      \biggl\{\big\langle A_i(x_4) A_{i_1}(K(s_1)) \big\rangle \big\langle A_j(x_4) A_{i_2}(K(s_2)) \big\rangle } && \\
      && &&
      \times \big\langle A_k(x_4) A_{i_3}(K(s_3)) \big\rangle \dot{K}^{i_1}(s_1) \dot{K}^{i_2}(s_2) \dot{K}^{i_3}(s_3) \biggr\}. 
      \numberthis
      \label{laycstrabajado}
      \end{alignat*}
        
Expression (\ref{laxcs}) can be seen to be
      \begin{alignat*}{6}
      V_{22} 
      &=& \mathrlap{ \frac{1}{N_A} \int DA e^{i I_A} \iiiint\limits_{s_1 < s_2 < s_3 < s_4} ds_1 ds_2 ds_3 ds_4 \biggl\{ A_{i_1}(K(s_1)) A_{i_2}(K(s_2)) A_{i_3}(K(s_3))} && \\
      && && \times A_{i_4}(K(s_4)) \dot{K}^{i_1}(s_1) \dot{K}^{i_2}(s_2) \dot{K}^{i_3}(s_3) \dot{K}^{i_4}(s_4) \biggr\} \\
      &=& && \iiiint\limits_{s_1 < s_2 < s_3 < s_4} ds_1 ds_2 ds_3 ds_4 \biggl\{ \big\langle A_{i_1}(K(s_1)) A_{i_2}(K(s_2)) A_{i_3}(K(s_3))A_{i_4}(K(s_4)) \big\rangle \dot{K}^{i_1}(s_1)  \\
      && && \times \dot{K}^{i_2}(s_2) \dot{K}^{i_3}(s_3) \dot{K}^{i_4}(s_4) \biggr\} \\
      &=& \mathrlap{ \iiiint\limits_{s_1 < s_2 < s_3 < s_4} ds_1 ds_2 ds_3 ds_4 \biggl\{ \big\langle A_{i_1}(K(s_1)) A_{i_3}(K(s_3)) \big\rangle \big\langle A_{i_2}(K(s_2)) A_{i_4}(K(s_4)) \big\rangle } && \\
      && && \times \dot{K}^{i_1}(s_1) \dot{K}^{i_2}(s_2) \dot{K}^{i_3}(s_3) \dot{K}^{i_4}(s_4) \biggr\}.
      \numberthis
      \label{laxcstrabajado}
      \end{alignat*}
      
Analogously, expressions (\ref{dibujo31cs}) - (\ref{dibujo34cs}) rewrite also as
            \begin{alignat*}{5}
            V_{31}
            &=& \mathrlap{ \iiiint\limits_{s_1 < s_2 < s_3 < s_4} ds_1 ds_2 ds_3 ds_4 \int\limits_{\mathbb{R}^3} \varepsilon^{ijk} d^3 x_5 \int\limits_{\mathbb{R}^3} \varepsilon^{lmn} d^3 x_6 \biggl\{ \big\langle A_m(x_5) A_n(x_6) \big\rangle } && \\
            && && \times \big\langle A_{i_1}(K(s_1)) A_i(x_6) \big\rangle \big\langle A_{i_2}(K(s_2)) A_j(x_5) \big\rangle \big\langle A_{i_3}(K(s_3)) A_k(x_5) \big\rangle \\
            && && \times \big\langle A_{i_4}(K(s_4)) A_l(x_6) \big\rangle \dot{K}^{i_1}(s_1) \dot{K}^{i_2}(s_2) \dot{K}^{i_3}(s_3) \dot{K}^{i_4}(s_4) \biggr\},
            \numberthis 
            \label{dibujo31cstrabajado} \\ \\
            V_{32}
            &=& && \multiint{5}\limits_{s_1 < s_2 < s_3 < s_4 < s_5} ds_1 ds_2 ds_3 ds_4 ds_5 \int\limits_{\mathbb{R}^3} \varepsilon^{ijk} d^3 x_6 \biggl\{ \big\langle A_{i_2}(K(s_2)) A_{i_5}(K(s_5)) \big\rangle \\
            && && \times \big\langle A_{i_1}(K(s_1)) A_i(x_6) \big\rangle \big\langle A_{i_3}(K(s_3)) A_j(x_6) \big\rangle \big\langle A_{i_4}(K(s_4)) A_k(x_6) \big\rangle \\
            && && \times \dot{K}^{i_1}(s_1) \dot{K}^{i_2}(s_2) \dot{K}^{i_3}(s_3) \dot{K}^{i_4}(s_4) \dot{K}^{i_5}(s_5) \biggr\},
            \numberthis 
            \label{dibujo32cstrabajado} \\ \\
            V_{33}
            &=& \mathrlap{ \multiint{6}\limits_{s_1 < s_2 < s_3 < s_4 < s_5 < s_6} ds_1 ds_2 ds_3 ds_4 ds_5 ds_6 \biggl\{ \big\langle A_{i_1}(K(s_1)) A_{i_4}(K(s_4)) \big\rangle } && \\
            && && \times \big\langle A_{i_2}(K(s_2)) A_{i_6}(K(s_6)) \big\rangle \big\langle A_{i_3}(K(s_3)) A_{i_5}(K(s_5)) \big\rangle \\
            && && \times \dot{K}^{i_1}(s_1) \dot{K}^{i_2}(s_2) \dot{K}^{i_3}(s_3) \dot{K}^{i_4}(s_4) \dot{K}^{i_5}(s_5) \dot{K}^{i_6}(s_6) \biggr\},
            \numberthis 
            \label{dibujo33cstrabajado} \\ \\
            V_{34}
            &=& \mathrlap{ \multiint{6}\limits_{s_1 < s_2 < s_3 < s_4 < s_5 < s_6} ds_1 ds_2 ds_3 ds_4 ds_5 ds_6 \biggl\{ \big\langle A_{i_1}(K(s_1)) A_{i_4}(K(s_4)) \big\rangle } && \\
            && && \times \big\langle A_{i_2}(K(s_2)) A_{i_5}(K(s_5)) \big\rangle \big\langle A_{i_3}(K(s_3)) A_{i_6}(K(s_6)) \big\rangle \\
            && && \times \dot{K}^{i_1}(s_1) \dot{K}^{i_2}(s_2) \dot{K}^{i_3}(s_3) \dot{K}^{i_4}(s_4) \dot{K}^{i_5}(s_5) \dot{K}^{i_6}(s_6) \biggr\}.
            \numberthis
            \label{dibujo34cstrabajado}
            \end{alignat*}

Expressions (\ref{selflinkingcstrabajado}), (\ref{laycstrabajado}), (\ref{laxcstrabajado}) and (\ref{dibujo31cstrabajado}) - (\ref{dibujo34cstrabajado}) will be the subject of the following subsections.

\subsection{First order Vassiliev knot invariant: self-linking of a knot}

\begin{figure}[H]
      \begin{subfigure}[b]{.6\textwidth}
      \centering
      \[ \xymatrix{
      & & C\left(2; {\bf S}^1\right) \times \mathcal{K} \ar[dd]^{ev} & & \\
      & & & & \\
      & C\left(2; {\bf S}^1\right) \ar[r]^{ev_K} \ar[dd]^{\alpha_2^1} & C\left(2; {\bf S}^3\right) \ar[dd]^{\alpha_2^3} \ar[rr]^{\phi} & & {\bf S}^2 \\
      & & & & \\
      & C\left[2; {\bf S}^1\right] \ar[r]^{\overline{ev}_K} & C\left[2; {\bf S}^3\right] \ar[uurr]_{\Phi} & & \\
      & & & & \\
      \mathcal{K} & & \ar[ll]_{pr} C\left[2; {\bf S}^1\right] \times \mathcal{K} \ar[uu]_{\overline{ev}} & & \\
      } \]
      \caption{Mathematical diagram for $D_1$.}
      \label{diagramota02}
      \end{subfigure}%
      \begin{subfigure}[b]{.4\textwidth}
      \centering
		  \begin{tikzpicture}
		  \draw[thick,dashed] (45:2) -- (225:2) node[pos=-0.1] {$s_1$} node[pos=1.1] {$s_2$};
		  \filldraw (45:2) circle (2pt);
		  \filldraw (225:2) circle (2pt);
		  \draw [->] [domain=-45:135] plot ({2*cos(\x)}, {2*sin(\x)});
		  \draw [->] [domain=135:315] plot ({2*cos(\x)}, {2*sin(\x)});
		  \end{tikzpicture}
		  \vspace{4.2\baselineskip}
      \caption{Feynman diagram $D_{1}$.}
      \label{Feynmanselflinking}
      \end{subfigure}
      \caption{First order Vassiliev knot invariant: self-linking of a knot.}
      \label{figuradelselflinking}
      \end{figure}
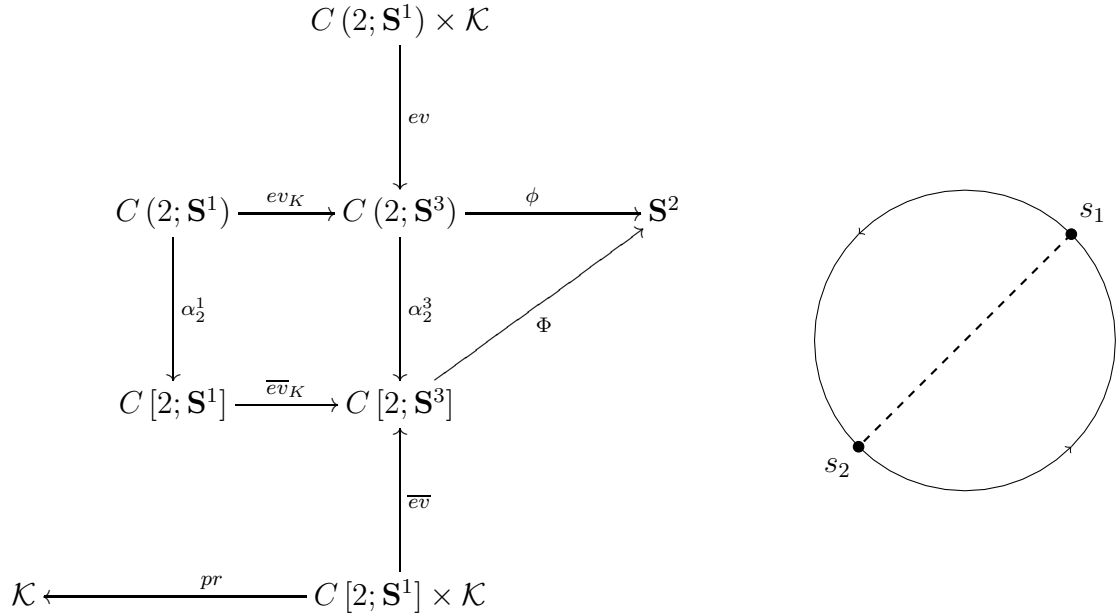

The Feynman diagram $D_1$ corresponding to the term (\ref{selflinkingcstrabajado}) is that of figure (\ref{Feynmanselflinking}) while the mathematical one is, according to the theory of section \ref{volumepreserving}, that of figure (\ref{diagramota02})
\noindent
where the column corresponding to internal points has been supressed (see figure \ref{diagramota01}). The integral corresponding to this diagram in the configuration space formalism is thus
      \begin{alignat*}{6}
      \overline{I}_{D_1} &=& \int\limits_{C[2,{\bf S}^1]} \left( \Phi \circ \overline{ev}_K \right)^* \omega
      &=& & \int\limits_{C(2,{\bf S}^1)} \left( \Phi \circ \overline{ev}_K \right)^* \omega + \int\limits_{\partial C[2,{\bf S}^1]} \left( \Phi \circ \overline{ev}_K \right)^* \omega \\
      &&&=& & \int\limits_{C(2,{\bf S}^1)} \left( \phi \circ ev_K \right)^* \omega + \int\limits_{\partial C[2,{\bf S}^1]} \left( \Phi \circ \overline{ev}_K \right)^* \omega \\
      &&&=& & \int\limits_{C(2,{\bf S}^1)} \left( \frac{K(x_2)-K(x_1)}{|K(x_2)-K(x_1)|} \right)^* \omega 
	  + \int\limits_{\partial C[2,{\bf S}^1]} \left( \pm \frac{\dot{K}(x_1)}{\big| \dot{K}(x_1) \bigr|} \right)^* \omega,
	  \numberthis
      \label{selflinkinconfiguration01}
      \end{alignat*}
where (see Ref. \cite{Vol07})
      \begin{align}
      \Phi \circ \overline{ev}_K = \pm \frac{\dot{K}(x_1)}{\big| \dot{K}(x_1) \bigr|}
      \label{kadot}      
      \end{align}
is the expression for $\Phi \circ \overline{ev}_K$ at the boundary of $C[2,{\bf S}^1]$. The sign in Eq. (\ref{kadot}) depends on whether the collapse of points $x_1$ and $x_2$ is in one direction or the other.

To explicitly make contact of (\ref{selflinkinconfiguration01}) with (\ref{selflinkingcstrabajado}) coming from the Chern-Simons theory it is necessary to calculate the pullback of $\omega$ under $\phi$, $\phi^*\omega$. This result is (see Appendix \ref{appendixC} for the explicit derivation) is given by
      \begin{equation}
      \phi_{a,b}^* \omega
      = \frac{\varepsilon_{\mu \nu \sigma}}{4\pi} \frac{(x_b - x_a)^{\mu}}{|x_b - x_a|^3} 
      \left( \frac{1}{2} dx_a^{\nu} \wedge dx_a^{\sigma} - dx_a^{\nu} \wedge dx_b^{\sigma} + \frac{1}{2} dx_b^{\nu} \wedge dx_b^{\sigma} \right).
      \label{pullbackpreab}
      \end{equation}
      
According to Ref. \cite{Thu99} the integral over the configuration space is non-zero only if there appears exactly one $dx_a$ and one $dx_b$, {\it i.e.},
it is enough to consider
      \begin{equation}
      \phi_{a,b}^* \omega = - \frac{\varepsilon_{\mu \nu \sigma}}{4\pi} \frac{(x_b - x_a)^{\mu}}{|x_b - x_a|^3} dx_a^{\nu} \wedge dx_b^{\sigma}.
      \label{pullbackab}
      \end{equation}
      
Actually the important expression is that for $(\phi \circ ev_K)^* \omega = ev_K^* \phi^* \omega$, to be precise
      \begin{align}
      ev_K^* \phi_{ab}^* \omega = - \frac{\varepsilon_{\mu \nu \sigma}}{4\pi} \frac{(K(x_b) - K(x_a))^{\mu}}{|K(x_b) - K(x_a)|^3} 
                                    \frac{dK^{\nu}(x_a)}{dx_a} \frac{dK^{\sigma}(x_b)}{dx_b} dx_a \wedge dx_b.
      \label{pullbackevkab}
      \end{align}
      
By using this expression in Eq. (\ref{selflinkinconfiguration01}) it is obtained
      \begin{alignat*}{6}
      \overline{I}_{D_1}
      &=& & \int\limits_{C(2,{\bf S}^1)} \biggl[ - \frac{\varepsilon_{\mu \nu \sigma}}{4\pi} \frac{(K(x_2) - K(x_1))^{\mu}}{|K(x_2) - K(x_1)|^3} 
	    \frac{dK^{\nu}(x_1)}{dx_1} \frac{dK^{\sigma}(x_2)}{dx_2} \biggr] dx_1 \wedge dx_2 \\
      & &+& \int\limits_{\partial C[2,{\bf S}^1]} \left( \pm \frac{\dot{K}(x_1)}{\bigl| \dot{K}(x_1) \bigr|} \right)^* \omega \\
      &=& & \int\limits_{C(2,{\bf S}^1)} \Delta_{\nu \sigma} (K(x_1) - K(x_2)) \dot{K}^{\nu}(x_1) \dot{K}^{\sigma}(x_2) dx_1 \wedge dx_2 \\
      & &+& \int\limits_{\partial C[2,{\bf S}^1]} \left( \pm \frac{\dot{K}(x_1)}{\bigl| \dot{K}(x_1) \bigr|} \right)^* \omega,
      \numberthis
      \end{alignat*}
where the last equality used standard notation for the derivatives and the following expression for the propagator \cite{Thu99}
      \begin{align}
      \Delta_{\mu \nu}(\overline{x}) = \frac{\varepsilon_{\mu \nu \sigma}}{4\pi} \frac{x^{\sigma}}{|\overline{x}|^3}.
      \end{align}
      
The boundary term is exactly cancelled with a framing term in order to obtain an invariant of knots with framing, {\it i.e.}, the real topological invariant is
      \begin{align}
      I_{D_1} = \int\limits_{C(2,{\bf S}^1)} \Delta_{\nu \sigma} (K(x_1) - K(x_2)) \dot{K}^{\nu}(x_1) \dot{K}^{\sigma}(x_2) dx_1 \wedge dx_2.
      \label{selflinkinconfiguration02}
      \end{align}
      
The match between expressions (\ref{selflinkingcstrabajado}) coming from Chern-Simons theory and (\ref{selflinkinconfiguration02}) coming from
the configuration space construction, {\it i.e.}, $I_{D_1}=V_1$, establishes a deep correspondence between formalisms.

\subsection{Second order Vassiliev knot invariant}

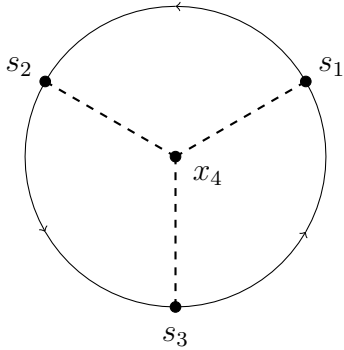
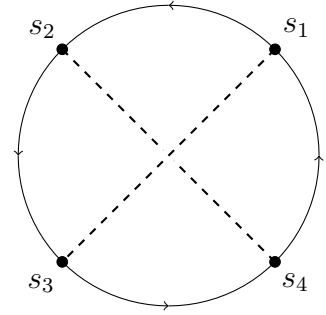
\begin{figure}[H]
      \begin{subfigure}{.5\textwidth}
      \centering
		  \begin{tikzpicture}
		  \draw[thick,dashed] (30:2) -- (0,0) node[pos=-0.2] {$s_1$};
		  \draw[thick,dashed] (150:2) -- (0,0) node[pos=-0.2] {$s_2$} node[pos=1.25] {$x_4$};
		  \draw[thick,dashed] (270:2) -- (0,0) node[pos=-0.2] {$s_3$};
		  \filldraw (30:2) circle (2pt);
		  \filldraw (150:2) circle (2pt);
		  \filldraw (270:2) circle (2pt);
		  \filldraw (0,0) circle (2pt);
		  \draw [->] [domain=-30:90] plot ({2*cos(\x)}, {2*sin(\x)});
		  \draw [->] [domain=90:210] plot ({2*cos(\x)}, {2*sin(\x)});
		  \draw [->] [domain=210:330] plot ({2*cos(\x)}, {2*sin(\x)});
		  \end{tikzpicture}
      \caption{Feynman diagram $D_{21}$.}
      \label{laydibujo}
      \end{subfigure}%
      \begin{subfigure}{.5\textwidth}
      \centering
		  \begin{tikzpicture}
		  \draw[thick, dashed] (45:2) -- (225:2) node[pos=-0.1] {$s_1$} node[pos=1.1] {$s_3$};
		  \draw[thick, dashed] (135:2) -- (135:0.1) node[pos=-0.2] {$s_2$};
		  \draw[thick, dashed] (315:2) -- (315:0.1) node[pos=-0.2] {$s_4$};
		  \filldraw (45:2) circle (2pt);
		  \filldraw (135:2) circle (2pt);
		  \filldraw (225:2) circle (2pt);
		  \filldraw (315:2) circle (2pt);
		  \draw [->] [domain=0:90] plot ({2*cos(\x)}, {2*sin(\x)});
		  \draw [->] [domain=90:180] plot ({2*cos(\x)}, {2*sin(\x)});
		  \draw [->] [domain=180:270] plot ({2*cos(\x)}, {2*sin(\x)});
		  \draw [->] [domain=270:360] plot ({2*cos(\x)}, {2*sin(\x)});
		  \draw[draw=none] (270:2) -- (0,0) node[pos=-0.25] {};
		  \end{tikzpicture}
      \caption{Feynman diagram $D_{22}$.}
      \label{laxdibujo}
      \end{subfigure}%
      \caption{Feynman diagrams for the second order Vassiliev invariant.}
      \label{secondordervassilievdibujo}
      \end{figure}

The second order Vassiliev invariant comes from the contributions of Feynman diagrams $D_{21}$ and $D_{22}$ (figures \ref{laydibujo} and 
\ref{laxdibujo}, respectively) corresponding to terms (\ref{laycstrabajado}) and (\ref{laxcstrabajado}), in that order.

The first step is to analyse diagram $D_{21}$. In this case the map $\phi$ in figure \ref{diagramota03} is given by the restriction of 
      \begin{equation*}
      \phi_{1,4}\times\phi_{2,4}\times\phi_{3,4}:\prod_{i=1}^4 {\bf S}^3\longrightarrow\prod_{i=1}^3 {\bf S}^2 \numberthis
      \end{equation*}     
to $C(3+1,{\bf S}^3)$ where each of these $\phi_{a,b}$ corresponds to a Gauss map. By taking the pullback of 
$\overline{\omega}:=\omega\times\omega\times\omega\in\Omega^6\left({\bf S}^2\times {\bf S}^2\times {\bf S}^2\right)$ under the previous map it is obtained
      \begin{alignat*}{6}
      \phi^*\left(\overline{\omega}\right) 
      &=&& \left(\phi_{1,4}\times\phi_{2,4}\times\phi_{3,4}\right)^*(\overline{\omega}) \\
      &=&&\phi_{1,4}^*\omega\wedge\phi_{2,4}^*\omega\wedge\phi_{3,4}^*\omega,
      \numberthis
      \end{alignat*}
where each of these pullbacks are given in the same way as in Eq. (\ref{pullbackab}).

      \begin{figure}[ht]
      \[ \xymatrix{
      & & C\left(3; {\bf S}^1\right) \times \mathcal{K} \ar[dd]^{ev} & & & \\
      & & & & & \\
      & C\left(3; {\bf S}^1\right) \ar[r]^{ev_K} \ar[dd]^{\alpha_3^1} & C\left(3; {\bf S}^3\right) \ar[dd]^{\alpha_3^3} & \ar[l]_{\pi_3} C\left(3+1; {\bf S}^3\right) \ar[rr]^{\phi} \ar[dd]^{\alpha_{3+1}^3} & & {\bf S}^2 \times {\bf S}^2 \times {\bf S}^2 \\
      & & & & & \\
      & C\left[3; {\bf S}^1\right] \ar[r]^{\overline{ev}_K} & C\left[3; {\bf S}^3\right] & \ar[l]_{\overline{\pi}_3} C\left[3+1; {\bf S}^3\right] \ar[uurr]_{\Phi} & & \\
      & & & & & \\
      \mathcal{K} & & \ar[ll]_{pr} C\left[3; {\bf S}^1\right] \times \mathcal{K} \ar[uu]_{\overline{ev}} & \ar[l]_{pr_1} C\left[3, 1; {\bf S}^1, {\bf S}^3\right] \ar[uu]_{pr_2} \pullbackcorner & &
      } \]
      \caption{Mathematical diagram for $D_{21}$.}
      \label{diagramota03}
      \end{figure}
      
Let $\overline{s}\in C(3,{\bf S}^1)$ and $\overline{x}=\alpha_3^3\circ ev_K (\overline{s})$ then $\overline{\pi}_3^{-1}(\left\lbrace\overline{x}\right\rbrace)$ is the homotopy fiber of $\overline{\pi}_3$. By integrating $\Phi^*\omega$ along this homotopy fiber (or equivalently by taking the pushforward under
$\overline{\pi}_3$) and by performing the pullback by $\alpha_3^3\circ ev_K$ then (see also Eq. (\ref{fvolic}))
      \begin{equation}
      f_{D_{21},K}(\overline{s}) = \left(\left(\alpha_3^3\circ ev_K \right)^*\left( \overline{\pi}_3 \right)_*\Phi^*\overline{\omega}\right)_{\overline{s}}\left(\partial_{\overline{s}}\right)
      \end{equation}  
is a $3$-form in $C(3,{\bf S}^1)$. Thus
      \begin{alignat*}{7}
      \overline{I}_{D_{21}} = \int\limits_{C(3,{\bf S}^1)} f_{D_{21},K}(\overline{s})d\overline{s} 
      &=&& \int\limits_{C(3,{\bf S}^1)}d\overline{s}\int\limits_{\overline{\pi}_3^{-1}(\left\lbrace \overline{x} \right\rbrace)} \Phi^*\overline{\omega}\left(\big[\left(\alpha_3^3\right)_*\dot{K}(\overline{s}\big)]_{\ell},\ldots\right)\\
      &=&& \int\limits_{C(3,{\bf S}^1)}d\overline{s}\int\limits_{{\bf S}^3} \phi^*\overline{\omega}\left(\big[\dot{K}(\overline{s})\big]_{\ell},\ldots\right) + B_{21}, &&
      \numberthis
      \label{Yconfiguration01}
      \end{alignat*}
where $\big[\dot{K}(\overline{s})\big]_{\ell}$ is the lift of the tangent vectors of the knot at each point $\overline{s}=(s_1,s_2,s_3)$ and $B_{12}$ stands for all the boundary terms. These terms exactly cancel with the boundary contribution of another second order diagram (the diagram of figure \ref{laxdibujo}). The non-boundary contribution of (\ref{Yconfiguration01}), $I_{D_{21}}$, is given by the pullback of $\overline{\omega}$ under $\phi$ but now the contribution of the knot is given through the lifts $\big[\dot{K}(\overline{s})\big]_{\ell}$, that is
        \begin{equation*}
        \phi^*\overline{\omega}\left(\big[\dot{K}(\overline{s})\big]_{\ell},\ldots\right)=\phi_{1,4}^*\omega\left(\big[\dot{K}(s_1)\big]_{\ell},\ldots\right)\wedge\phi_{2,4}^*\omega\left(\big[\dot{K}(s_2)\big]_{\ell},\ldots\right)\wedge \phi_{3,4}^*\omega\left(\big[\dot{K}(s_3)\big]_{\ell},\ldots\right),
        \numberthis
        \end{equation*}
where each of these pullbacks has the form 
      \begin{alignat*}{5}
      \phi_{i,4}^*\omega\left(\big[\dot{K}({s}_i)\big]_{\ell},\ldots\right) 
      &=&-& \frac{\varepsilon_{\mu \nu \sigma}}{4\pi} \frac{(x_4- x_i)^{\mu}}{|x_4 - x_i|^3} dx_i^\nu\wedge dx_4^\sigma\left(\big[\dot{K}({s}_i)\big]_{\ell},\ldots\right) \\ \\
      &=&-& \frac{\varepsilon_{\mu \nu \sigma}}{4\pi} \frac{(x_4- K\left(s_i\right))^{\mu}}{|x_4 - K\left(s_i\right)|^3} \dot{K}^\nu\left(s_i\right)dx_4^\sigma \\ \\
      &=& & \Delta_{\nu \sigma} (K\left(s_i\right) - x_4) \dot{K}^{\nu}\left(s_i\right) dx_4^\sigma,
      \numberthis
      \end{alignat*}
with $i=1,2,3$. Substitution of this expression into the non-boundary part of (\ref{Yconfiguration01}) gives 
      \begin{alignat*}{6}
      & & \mathrlap{ \int\limits_{C(3,{\bf S}^1)}d\overline{s}\int\limits_{{\bf S}^3} \phi^*\overline{\omega}\left(\big[\dot{K}(\overline{s})\big]_{\ell},\ldots\right) } && \\
      &=& \mathrlap{ \int\limits_{C(3,{\bf S}^1)} d\overline{s}\int\limits_{{\bf S}^3} \biggl[ \Delta_{\nu_1 \sigma_1} (K(s_1) - x_4) \Delta_{\nu_2 \sigma_2} (K(s_2) - x_4) \Delta_{\nu_3 \sigma_3}(K(s_3) - x_4) } && \\
      & & && \times \dot{K}^{\sigma_1}(s_1) \dot{K}^{\sigma_2}(s_2) \dot{K}^{\sigma_3}(s_3) \biggr] dx_4^{\nu_1}\wedge dx_4^{\nu_2} \wedge dx_4^{\nu_3} \\
      &=& && \int\limits_{C(3,{\bf S}^1)} \dot{K}^{\sigma_1}(s_1)\dot{K}^{\sigma_2}(s_2)\dot{K}^{\sigma_3}(s_3)ds_1\wedge ds_2\wedge ds_3\int\limits_{{\bf S}^3} \varepsilon^{\nu_1 \nu_2 \nu_3} d^3 x_4 \Delta_{\nu_1 \sigma_1} (K(s_1) - x_4) \\
      & & && \times \Delta_{\nu_2 \sigma_2} (K(s_2) - x_4)\Delta_{\nu_3 \sigma_3}(K(s_3) - x_4), \numberthis
      \end{alignat*}
from where
      \begin{align*}
      I_{D_{21}}
      = \int\limits_{C(3,{\bf S}^1)} \dot{K}^{\sigma_1}(s_1)\dot{K}^{\sigma_2}(s_2)\dot{K}^{\sigma_3}(s_3)ds_1\wedge ds_2\wedge ds_3\int\limits_{{\bf S}^3} \varepsilon^{\nu_1 \nu_2 \nu_3} d^3 x_4 \Delta_{\nu_1 \sigma_1} (K(s_1) - x_4) \\
      \times \Delta_{\nu_2 \sigma_2} (K(s_2) - x_4) \Delta_{\nu_3 \sigma_3}(K(s_3) - x_4).
      \label{contribucionD21}      
      \end{align*}
      
In analogy to the self-linking case, this expression can be regarded to match with Eq. (\ref{laycstrabajado}).
      
\begin{figure}[ht]
      \[ \xymatrix{
      & & C\left(4; {\bf S}^1\right) \times \mathcal{K} \ar[dd]^{ev} & & \\
      & & & & \\
      & C\left(4; {\bf S}^1\right) \ar[r]^{ev_K} \ar[dd]^{\alpha_4^1} & C\left(4; {\bf S}^3\right) \ar[dd]^{\alpha_4^3} \ar[rr]^{\phi} & & {\bf S}^2 \times {\bf S}^2 \\
      & & & & \\
      & C\left[4; {\bf S}^1\right] \ar[r]^{\overline{ev}_K} & C\left[4; {\bf S}^3\right] \ar[uurr]_{\Phi} & & \\
      & & & & \\
      \mathcal{K} & & \ar[ll]_{pr} C\left[4;{\bf S}^1\right] \times \mathcal{K} \ar[uu]_{\overline{ev}} & & \\
      } \]
      \caption{Mathematical diagram for $D_{22}$.}
      \label{diagramota04}
      \end{figure}
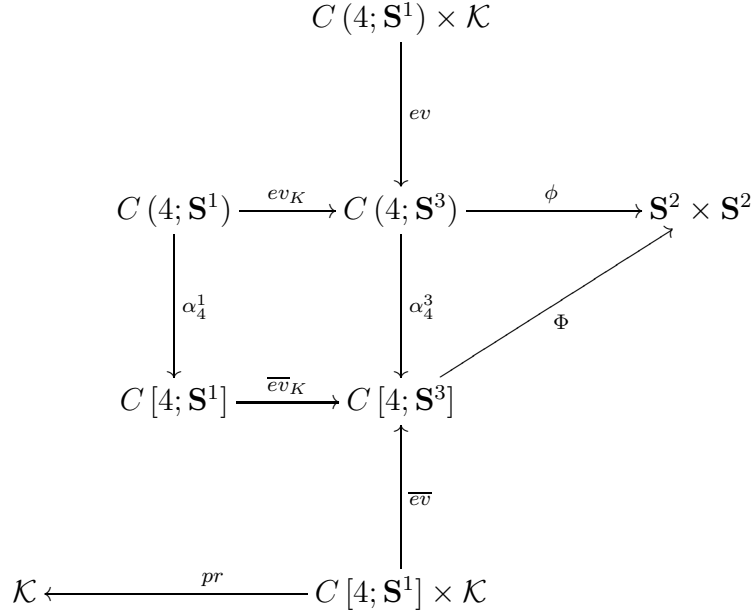

      
Now we proceed to discuss diagram $D_{22}$ by using figure (\ref{diagramota04}).

In this case the $\phi$ map is given by $\phi=\phi_{1,3}\times\phi_{2,4}$ where all the points are on the knot, $\overline{\pi}_4=id$ as in the self-linking case and $\overline{\omega}=\omega\wedge\omega\in\Omega^4\left({\bf S}^2\times {\bf S}^2\right)$. This time the configuration space integral is given by
      \begin{alignat*}{6}
      \overline{I}_{D_{22}} 
      &=&& \int\limits_{C(4,{\bf S}^1)} d\overline{s}\left( \Phi \circ\alpha_4^3\circ ev_K  \right)^* \overline{\omega} + B_{22} \\
      &=&& \int\limits_{C(4,{\bf S}^1)} d\overline{s}\left( \phi \circ ev_K  \right)^* \overline{\omega} + B_{22},
      \numberthis
      \label{configuration22}
      \end{alignat*}
where $B_{22}$ stands for the boundary terms. By using Eq. (\ref{pullbackevkab}) the pullback under $\phi\circ ev_K$ of $\overline{\omega}$ is written as

      \begin{alignat*}{5}
      \left(\phi\circ ev_k\right)^*\overline{\omega}
      &=&& \left( ev_K^*\phi_{1,3}^*\omega\right) & \mathrlap{ \wedge \left( ev_K^*\phi_{2,4}^*\omega \right) } && \\ \\
      &=&& \mathrlap{ \biggl[ \frac{\varepsilon_{\mu_1 \nu_1 \sigma_1}}{4\pi} \frac{(K(s_1)- K(s_3))^{\mu_1}}{|K(s_1) - K(s_3)|^3}\frac{\varepsilon_{\mu_2 \nu_2 \sigma_2}}{4\pi} \frac{(K(s_2)- K(s_4))^{\mu_2}}{|K(s_2) - K(s_4)|^3} \biggr]} \\
      &&& & && \times \dot{K}^{\sigma_1}\left(s_3\right) \dot{K}^{\nu_1}\left(s_1\right) \dot{K}^{\sigma_2}\left(s_4\right)\dot{K}^{\nu_2}\left(s_2\right) ds_1\wedge ds_3\wedge ds_2 \wedge ds_4\\ \\
      &=&& \mathrlap{ \biggl[ \Delta_{\nu_1 \sigma_1}(K(s_1)-K(s_3)) \Delta_{\nu_2 \sigma_2}(K(s_2)-K(s_4)) \biggr]} \\
      &&& & && \times (-1) \dot{K}^{\nu_1}\left(s_1\right) \dot{K}^{\nu_2}\left(s_2\right) \dot{K}^{\sigma_1}\left(s_3\right)\dot{K}^{\sigma_2}\left(s_4\right) ds_1\wedge ds_2\wedge ds_3\wedge ds_4,
      \numberthis
      \end{alignat*}      
and then $I_{D_{22}}$ is written as
      \begin{align*}
      I_{D_{22}}
      =-\int\limits_{C(4,{\bf S}^1)} \biggl[ \Delta_{\nu_1 \sigma_1}(K(s_1)-K(s_3)) \Delta_{\nu_2 \sigma_2}(K(s_2)-K(s_4)) \dot{K}^{\nu_1}\left(s_1\right) \dot{K}^{\nu_2}\left(s_2\right) \\
      \times \dot{K}^{\sigma_1}\left(s_3\right) \dot{K}^{\sigma_2}\left(s_4\right) \biggr] ds_1\wedge ds_2\wedge ds_3\wedge ds_4,
      \numberthis
      \label{contribucionD22}      
      \end{align*}
which can be seen to match with Eq. (\ref{laxcstrabajado}).

\subsection{Third order Vassiliev knot invariant}

The third order Vassiliev invariant has an effective contribution given by diagrams $D_{31}$, $D_{32}$, $D_{33}$ and $D_{34}$ (figures
\ref{dibujo31}, \ref{dibujo32}, \ref{dibujo33} and \ref{dibujo34}, respectively) corresponding to terms (\ref{dibujo31cstrabajado}),
(\ref{dibujo32cstrabajado}), (\ref{dibujo33cstrabajado}) and (\ref{dibujo34cstrabajado}), in that order.

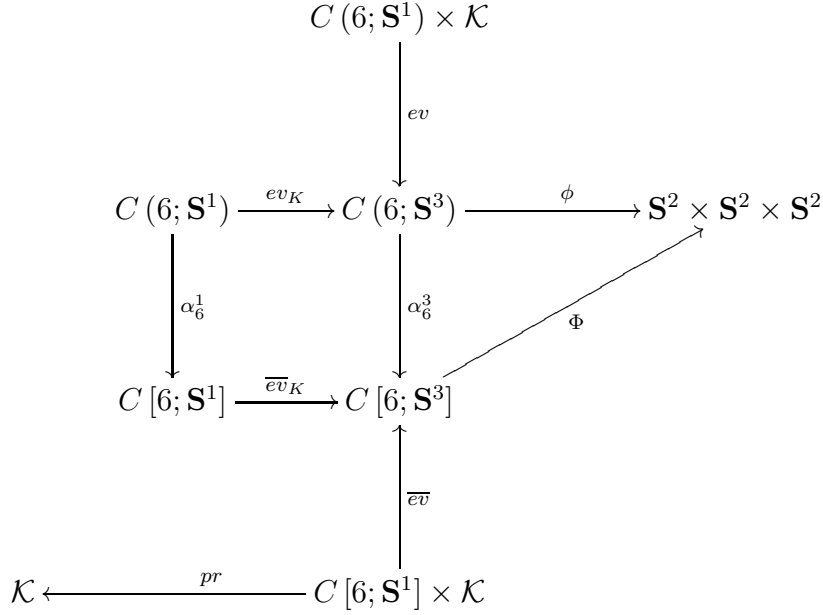
\begin{figure}[ht]
      \[ \xymatrix{
      & & C\left(6; {\bf S}^1\right) \times \mathcal{K} \ar[dd]^{ev} & & \\
      & & & & \\
      & C\left(6; {\bf S}^1\right) \ar[r]^{ev_K} \ar[dd]^{\alpha_6^1} & C\left(6; {\bf S}^3\right) \ar[dd]^{\alpha_6^3} \ar[rr]^{\phi} & & {\bf S}^2 \times {\bf S}^2 \times {\bf S}^2 \\
      & & & & \\
      & C\left[6; {\bf S}^1\right] \ar[r]^{\overline{ev}_K} & C\left[6; {\bf S}^3\right] \ar[uurr]_{\Phi} & & \\
      & & & & \\
      \mathcal{K} & & \ar[ll]_{pr} C\left[6; {\bf S}^1\right] \times \mathcal{K} \ar[uu]_{\overline{ev}} & & \\
      } \]
      \caption{Mathematical diagram for $D_{34}$.}
      \label{diagramota08}
      \end{figure}

In a completely analogous way to the analysis of the first and second order Vassiliev invariants, mathematical diagrams of figures
\ref{diagramota05} - \ref{diagramota08} are to be used to build the corresponding configuration space expressions. The unitary volume form $\omega\in\Omega^2 \left(S^2\right)$ and the corresponding products for each diagram will be used. 

The first step is to analyse diagram $D_{31}$. In this case the $\phi$ map in figure \ref{diagramota05} is given by the restriction of
      \begin{equation*}
      \phi_{1,6} \times \phi_{2,5} \times \phi_{3,5} \times \phi_{4,6} \times \phi_{5,6} : \prod_{i=1}^6 {\bf S}^3 \longrightarrow \prod_{i=1}^5 {\bf S}^2 \numberthis
      \end{equation*}     
to $C(4+2,{\bf S}^3)$. By taking the pullback of $\overline{\omega}:=\omega^5\in\Omega^{10}\left(\left({\bf S}^2\right)^5\right)$ under the previous map it is obtained
      \begin{alignat*}{6}
      \phi^*\left(\overline{\omega}\right) 
      &=&& \left(\phi_{1,6} \times \phi_{2,5} \times \phi_{3,5} \times \phi_{4,6} \times \phi_{5,6}\right)^*(\overline{\omega}) \\
      &=&&\phi_{1,6}^*\omega\wedge\phi_{2,5}^*\omega\wedge\phi_{3,5}^*\omega\wedge\phi_{4,6}^*\omega\wedge\phi_{5,6}^*\omega,
      \numberthis
      \end{alignat*}
where again each of these pullbacks are given by Eq. (\ref{pullbackevkab}).

By applying Eq. (\ref{fvolic}) to a point $\overline{s}\in C(4,{\bf S}^1)$ and according with figure \ref{diagramota05} we have
      \begin{equation}
      f_{D_{31},K}(\overline{s}) =\left(\left(\alpha_4^3\circ ev_K \right)^*\left(\overline{\pi}_4\right)_*\Phi^*\overline{\omega}\right)_{\overline{s}}\left(\partial_{\overline{s}}\right),
      \end{equation}  

      \begin{figure}[ht]
      \begin{subfigure}[b]{0.5\textwidth}
      \centering
		  \begin{tikzpicture}
		  \draw[thick,dashed] (30:2) -- (0:1) node[pos=-0.3] {$s_1$};
		  \draw[thick,dashed] (150:2) -- (180:1) node[pos=-0.3] {$s_2$};
		  \draw[thick,dashed] (210:2) -- (180:1) node[pos=-0.3] {$s_3$} node[pos=1.3] {$x_5$};
		  \draw[thick,dashed] (330:2) -- (0:1) node[pos=-0.3] {$s_4$} node[pos=1.3] {$x_6$};
		  \draw[thick,dashed] (180:1) -- (0:1);
		  \filldraw (30:2) circle (2pt);
		  \filldraw (150:2) circle (2pt);
		  \filldraw (210:2) circle (2pt);
		  \filldraw (-30:2) circle (2pt);
		  \filldraw (0:1) circle (2pt);
		  \filldraw (180:1) circle (2pt);
		  \draw [->] [domain=-90:0] plot ({2*cos(\x)}, {2*sin(\x)});
		  \draw [->] [domain=0:90] plot ({2*cos(\x)}, {2*sin(\x)});
		  \draw [->] [domain=90:180] plot ({2*cos(\x)}, {2*sin(\x)});
		  \draw [->] [domain=180:270] plot ({2*cos(\x)}, {2*sin(\x)});
		  \draw[draw=none] (72:2) -- (288:2) node[pos=-0.1] {} node[pos=1.1] {};
		  \end{tikzpicture}
      \caption{Feynman diagram $D_{31}$.}
      \label{dibujo31}
      \end{subfigure}%
      \begin{subfigure}[b]{0.5\textwidth}
      \centering
		  \begin{tikzpicture}
		  \draw[thick,dashed] (144:2) -- (0:0) node[pos=-0.3] {$s_3$};
		  \draw[thick,dashed] (216:2) -- (0:0) node[pos=-0.3] {$s_4$};
			\draw[name path=a, opacity=0] (72:2) -- (288:2);
			\draw[name path=b, opacity=0] (0:0) -- (0:2);
			\path[name intersections={of = a and b, by = ab}];
		  \draw[thick,dashed] (0:0) -- (0:2) node[pos=-0.3] {$x_6$} node[pos=1.3] {$s_1$};
		  \filldraw[white] (ab) circle (4pt);
		  \draw[thick,dashed] (72:2) -- (288:2) node[pos=-0.1] {$s_2$} node[pos=1.1] {$s_5$};
		  \filldraw (0:2) circle (2pt);
		  \filldraw (72:2) circle (2pt);
		  \filldraw (144:2) circle (2pt);
		  \filldraw (216:2) circle (2pt);
		  \filldraw (288:2) circle (2pt);
		  \filldraw (0:0) circle (2pt);
		  \draw [->] [domain=-36:36] plot ({2*cos(\x)}, {2*sin(\x)});
		  \draw [->] [domain=36:108] plot ({2*cos(\x)}, {2*sin(\x)});
		  \draw [->] [domain=108:180] plot ({2*cos(\x)}, {2*sin(\x)});
		  \draw [->] [domain=180:252] plot ({2*cos(\x)}, {2*sin(\x)});
		  \draw [->] [domain=252:324] plot ({2*cos(\x)}, {2*sin(\x)});
		  \end{tikzpicture} 
      \caption{Feynman diagram $D_{32}$.}
      \label{dibujo32}
      \end{subfigure}%
      \\
      \begin{subfigure}[b]{0.5\textwidth}
      \centering
		  \begin{tikzpicture}
		  \draw[thick,dashed] (180:2) -- (0:2) node[pos=-0.1] {$s_4$} node[pos=1.1] {$s_1$};
		  \filldraw[white] (0:1) circle (4pt);
		  \filldraw[white] (180:1) circle (4pt);
		  \draw[thick,dashed] (120:2) -- (240:2) node[pos=-0.1] {$s_3$} node[pos=1.1] {$s_5$};
		  \draw[thick,dashed] (60:2) -- (300:2) node[pos=-0.1] {$s_2$} node[pos=1.1] {$s_6$};
		  \filldraw (0:2) circle (2pt);
		  \filldraw (60:2) circle (2pt);
		  \filldraw (120:2) circle (2pt);
		  \filldraw (180:2) circle (2pt);
		  \filldraw (240:2) circle (2pt);
		  \filldraw (300:2) circle (2pt);
		  \draw [->] [domain=-30:30] plot ({2*cos(\x)}, {2*sin(\x)});
		  \draw [->] [domain=30:90] plot ({2*cos(\x)}, {2*sin(\x)});
		  \draw [->] [domain=90:150] plot ({2*cos(\x)}, {2*sin(\x)});
		  \draw [->] [domain=150:210] plot ({2*cos(\x)}, {2*sin(\x)});
		  \draw [->] [domain=210:270] plot ({2*cos(\x)}, {2*sin(\x)});
		  \draw [->] [domain=270:330] plot ({2*cos(\x)}, {2*sin(\x)});
		  \end{tikzpicture}  
      \caption{Feynman diagram $D_{33}$.}
      \label{dibujo33}
      \end{subfigure}%
      \begin{subfigure}[b]{0.5\textwidth}
      \centering
		  \begin{tikzpicture}
		  \draw[thick,dashed] (120:2) -- (300:2) node[pos=-0.1] {$s_3$} node[pos=1.1] {$s_6$};
		  \draw[thick,dashed] (60:2) -- (240:2) node[pos=-0.1] {$s_2$} node[pos=1.1] {$s_5$};
		  \filldraw[white] (0:0) circle (4pt);
		  \draw[thick,dashed] (180:2) -- (0:2) node[pos=-0.1] {$s_4$} node[pos=1.1] {$s_1$};
		  \filldraw (0:2) circle (2pt);
		  \filldraw (60:2) circle (2pt);
		  \filldraw (120:2) circle (2pt);
		  \filldraw (180:2) circle (2pt);
		  \filldraw (240:2) circle (2pt);
		  \filldraw (300:2) circle (2pt);
		  \draw [->] [domain=-30:30] plot ({2*cos(\x)}, {2*sin(\x)});
		  \draw [->] [domain=30:90] plot ({2*cos(\x)}, {2*sin(\x)});
		  \draw [->] [domain=90:150] plot ({2*cos(\x)}, {2*sin(\x)});
		  \draw [->] [domain=150:210] plot ({2*cos(\x)}, {2*sin(\x)});
		  \draw [->] [domain=210:270] plot ({2*cos(\x)}, {2*sin(\x)});
		  \draw [->] [domain=270:330] plot ({2*cos(\x)}, {2*sin(\x)});
		  \end{tikzpicture}  
      \caption{Feynman diagram $D_{34}$.}
      \label{dibujo34}
      \end{subfigure}%
      \caption{Feynman diagrams for the third order Vassiliev invariant.}
      \label{thirdordervassilievdibujo}
      \end{figure}
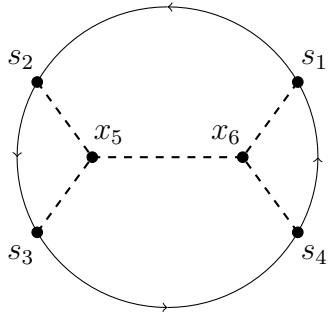
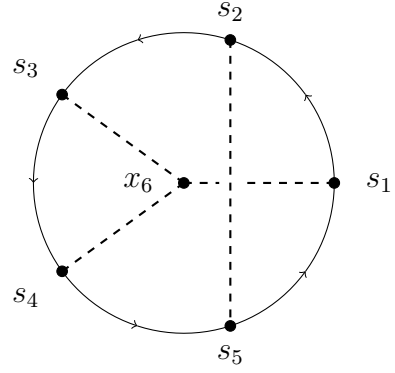
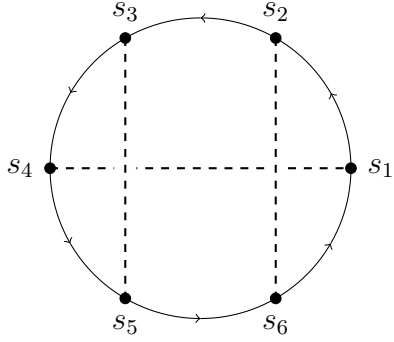
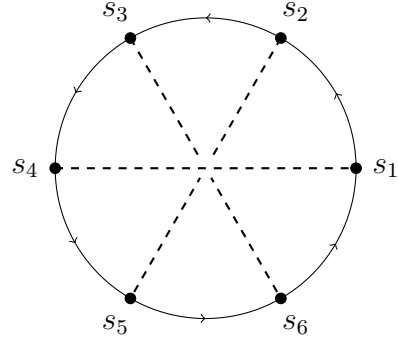
     
      \begin{figure}[ht]
      \[ \xymatrix{
      & & C\left(4; {\bf S}^1\right) \times \mathcal{K} \ar[dd]^{ev} & & & \\
      & & & & & \\
      & C\left(4; {\bf S}^1\right) \ar[r]^{ev_K} \ar[dd]^{\alpha_4^1} & C\left(4; {\bf S}^3\right) \ar[dd]^{\alpha_4^3} & \ar[l]_{\pi_4} C\left(4+2; {\bf S}^3\right) \ar[rr]^{\phi} \ar[dd]^{\alpha_{4+2}^3} & & {\bf S}^2 \times {\bf S}^2 \times {\bf S}^2 \times {\bf S}^2 \times {\bf S}^2 \\
      & & & & & \\
      & C\left[4; {\bf S}^1\right] \ar[r]^{\overline{ev}_K} & C\left[4; {\bf S}^3\right] & \ar[l]_{\overline{\pi}_4} C\left[4+2; {\bf S}^3\right] \ar[uurr]_{\Phi} & & \\
      & & & & & \\
      \mathcal{K} & & \ar[ll]_{pr} C\left[4; {\bf S}^1\right] \times \mathcal{K} \ar[uu]_{\overline{ev}} & \ar[l]_{pr_1} C\left[4, 2; {\bf S}^1, {\bf S}^3\right] \ar[uu]_{pr_2} \pullbackcorner & &
      } \]
      \caption{Mathematical diagram for $D_{31}$.}
      \label{diagramota05}
      \end{figure}
\noindent
which is a $4$-form in $C(4,{\bf S}^1)$. The fiber of $\overline{\pi}_4$ is a space of the same homotopy type than $\mathbb{R}^3 \times\mathbb{R}^3$ and so
      \begin{alignat*}{7}
      \overline{I}_{D_{31}} = \int\limits_{C(4,{\bf S}^1)} f_{D_{31},K}(\overline{s})d\overline{s} 
      &=&& \int\limits_{C(4,{\bf S}^1)}d\overline{s}\int\limits_{\overline{\pi}_4^{-1}(\left\lbrace \overline{x} \right\rbrace)} \Phi^*\overline{\omega}\left(\big[(\alpha_4^3)^*\dot{K}(\overline{s})\big]_{\ell},\ldots\right)\\
      &=&& \int\limits_{C(4,{\bf S}^1)}d\overline{s}\int\limits_{\mathbb{R}^3\times\mathbb{R}^3} \phi^*\overline{\omega}\left(\big[\dot{K}(\overline{s})\big]_{\ell},\ldots\right) + B_{31}, &&
      \numberthis
      \label{31configuration01}
      \end{alignat*}
where $B_{31}$ represents the boundary terms. As in the previous cases the boundary term cancels with the boundary contribution of another third order diagram (the one in figure \ref{dibujo32}).
      
For this diagram the non-boundary contribution of (\ref{31configuration01}) is given by the pullback of $\overline{\omega}$ under $\phi$ evaluated at lifts $\big[\dot{K}(\overline{s})\big]_{\ell}$ of the tangent vectors of the knot at each point in
$\overline{s}=\left(s_1,\ldots,s_4\right)$, that is
      \begin{alignat*}{7}
      \phi^*\overline{\omega}\left(\big[\dot{K}(\overline{s})\big]_{\ell},\ldots\right) & = && \phi_{1,6}^*\omega\left(\big[\dot{K}({s}_1)\big]_{\ell},\ldots\right)&\wedge &\phi_{2,5}^*\omega\left(\big[\dot{K}({s}_2)\big]_{\ell},\ldots\right)\wedge\phi_{3,5}^*\omega\left(\big[\dot{K}({s}_3)\big]_{\ell},\ldots\right)\\
      &&&& \wedge & \phi_{4,6}^*\omega\left(\big[\dot{K}({s}_4)\big]_{\ell},\ldots\right)\wedge\phi_{5,6}^*\omega,
      \numberthis
      \end{alignat*}
where these pullbacks are rewritten as
      \begin{alignat*}{5}
      \phi_{i,j}^*\omega\left(\big[\dot{K}({s}_i)\big]_{\ell},\ldots\right) 
      &=&-& \frac{\varepsilon_{\mu \nu \sigma}}{4\pi} \frac{(x_j- x_i)^{\mu}}{|x_j - x_i|^3} dx_i^\nu\wedge dx_j^\sigma\left(\big[\dot{K}({s}_i)\big]_{\ell},\ldots\right) \\ \\
      &=&-& \frac{\varepsilon_{\mu \nu \sigma}}{4\pi} \frac{(x_j- K\left(s_i\right))^{\mu}}{|x_j - K\left(s_i\right)|^3} \dot{K}^\nu\left(s_i\right)dx_j^\sigma \\ \\
      &=& & \Delta_{\nu \sigma} (K\left(s_i\right) - x_j) \dot{K}^{\nu}\left(s_i\right) dx_j^\sigma,
      \numberthis
      \label{pullbackomegaijenK}
      \end{alignat*}
with $i=1,2,3,4$, $j=5,6$ and $\phi_{5,6}^*\omega$ as in Eq. (\ref{pullbackab}). Substitution of this expression into (\ref{31configuration01}) directly yields to
      \begin{alignat*}{7}
      & & \mathrlap{ \int\limits_{C(4,{\bf S}^1)}d\overline{s}\int\limits_{\mathbb{R}^3\times\mathbb{R}^3} \phi^*\overline{\omega}\left(\big[\dot{K}(\overline{s})\big]_{\ell},\ldots\right) } && \\
      &=& && \int\limits_{C(4,{\bf S}^1)} d\overline{s}\int\limits_{\mathbb{R}^3\times\mathbb{R}^3} \biggl[ \Delta_{\nu_1 \sigma_1} (K(s_1) - x_5) \Delta_{\nu_2 \sigma_2} (K(s_2) - x_5) \Delta_{\nu_3 \sigma_3}(K(s_3) - x_6) \\
      & & && \times \Delta_{\nu_4 \sigma_4}(K(s_4) - x_6) \Delta_{\nu_5 \sigma_5}(x_5 - x_6) \dot{K}^{\nu_1}(s_1) \dot{K}^{\nu_2}(s_2) \dot{K}^{\nu_3}(s_3)\dot{K}^{\nu_4}(s_4) \biggr] \\
      & & && \times dx_5^{\sigma_1}\wedge dx_5^{\sigma_2} \wedge dx_6^{\sigma_3} \wedge dx_6^{\sigma_4} \wedge dx_5^{\nu_5} \wedge dx_6^{\sigma_5},
   \numberthis
      \end{alignat*}   
from where
      \begin{align*}
      I_{D_{31}}
      = \int\limits_{C(4,{\bf S}^1)} \dot{K}^{\nu_1}(s_1) \dot{K}^{\nu_2}(s_2) \dot{K}^{\nu_3}(s_3)\dot{K}^{\nu_4}(s_4)ds_1\wedge ds_2\wedge ds_3\wedge ds_4\int\limits_{\mathbb{R}^3} \varepsilon^{\sigma_1 \sigma_2 \nu_5} d^3 x_5 \\
      \times \int\limits_{\mathbb{R}^3}\varepsilon^{\sigma_3 \sigma_4 \sigma_5} d^3 x_6 \Delta_{\nu_1 \sigma_1} (K(s_1) - x_5) \Delta_{\nu_2 \sigma_2} (K(s_2) - x_5) \\
      \times \Delta_{\nu_3 \sigma_3}(K(s_3) - x_6)\Delta_{\nu_4 \sigma_4}(K(s_4) - x_6) \Delta_{\nu_5 \sigma_5}(x_5 - x_6).
      \numberthis
      \label{contribucionD31}      
      \end{align*}
      It can be regarded that this expression matches with (\ref{dibujo31cstrabajado}) from Chern-Simons theory. 


Now we analyse diagram $D_{32}$ by using the mathematical construction of figure \ref{diagramota06}. Here the $\phi$ map is given by the restriction of

      \begin{figure}[ht]
      \[ \xymatrix{
      & & C\left(5; {\bf S}^1\right) \times \mathcal{K} \ar[dd]^{ev} & & & \\
      & & & & & \\
      & C\left(5; {\bf S}^1\right) \ar[r]^{ev_K} \ar[dd]^{\alpha_5^1} & C\left(5; {\bf S}^3\right) \ar[dd]^{\alpha_5^3} & \ar[l]_{\pi_5} C\left(5+1; {\bf S}^3\right) \ar[rr]^{\phi} \ar[dd]^{\alpha_{5+1}^3} & & {\bf S}^2 \times {\bf S}^2 \times {\bf S}^2 \times {\bf S}^2 \\
      & & & & & \\
      & C\left[5; {\bf S}^1\right] \ar[r]^{\overline{ev}_K} & C\left[5; {\bf S}^3\right] & \ar[l]_{\overline{\pi}_5} C\left[5+1; {\bf S}^3\right] \ar[uurr]_{\Phi} & & \\
      & & & & & \\
      \mathcal{K} & & \ar[ll]_{pr} C\left[5; {\bf S}^1\right] \times \mathcal{K} \ar[uu]_{\overline{ev}} & \ar[l]_{pr_1} C\left[5, 1; {\bf S}^1, {\bf S}^3\right] \ar[uu]_{pr_2} \pullbackcorner & &
      } \]
      \caption{Mathematical diagram for $D_{32}$}
      \label{diagramota06}
      \end{figure}
       
      \begin{equation*}
      \phi_{1,6} \times \phi_{3,6} \times \phi_{4,6} \times \phi_{2,5} : \prod_{i=1}^6 {\bf S}^3 \longrightarrow \prod_{i=1}^4 {\bf S}^2 \numberthis
      \end{equation*}     
to $C(5+1,{\bf S}^3)$. This time the fiber of $\overline{\pi}_5$ is again a space of the same homotopy type than $\mathbb{R}^3$ and so
      \begin{alignat*}{7}
      \overline{I}_{D_{32}} = \int\limits_{C(5,{\bf S}^1)} f_{D_{32},K}(\overline{s})d\overline{s} 
      &=&& \int\limits_{C(5,{\bf S}^1)}d\overline{s}\int\limits_{\overline{\pi}_p^{-1}(\left\lbrace \overline{x} \right\rbrace)} \Phi^*\overline{\omega}\left(\big[(\alpha_p^3)^*\dot{K}(\overline{s})\big]_{\ell},\ldots\right)\\
      &=&& \int\limits_{C(5,{\bf S}^1)}d\overline{s}\int\limits_{\mathbb{R}^3} \phi^*\overline{\omega}\left(\big[\dot{K}(\overline{s})\big]_{\ell},\ldots\right) + B_{32}, &&
      \numberthis
      \label{32configuration01}
      \end{alignat*}    
where $B_{32}$ stands for all the boundary contributions.

In this case the non-boundary part of (\ref{32configuration01}) is expressed as      
      \begin{align*}
      \phi^*\overline{\omega}\left(\big[\dot{K}(\overline{s})\big]_{\ell},\ldots\right) 
      = \phi_{1,6}^*\omega\left(\big[\dot{K}({s}_1)\big]_{\ell},\ldots\right) \wedge \phi_{2,5}^*\omega\left(\dot{K}\left(s_2\right)\dot{K}\left(s_5\right)\right) \wedge \phi_{3,6}^*\omega\left(\big[\dot{K}({s}_3)\big]_{\ell},\ldots\right) & \\
      \wedge \phi_{4,6}^*\omega\left(\big[\dot{K}({s}_4)\big]_{\ell},\ldots\right), &
      \numberthis
      \end{align*}
where each of the pullbaks are given by Eq. (\ref{pullbackomegaijenK}), {\it i.e.}, by
      \begin{alignat*}{5}
      \phi_{i,j}^*\omega\left(\big[\dot{K}({s}_i)\big]_{\ell},\ldots\right) = \Delta_{\nu \sigma} (K\left(s_i\right) - x_j) \dot{K}^{\nu}\left(s_i\right) dx_j^\sigma,
      \numberthis
      \end{alignat*}
with $i=1,3,4$, $j=6$ and where $\phi_{2,5}^*\omega\left(\dot{K}\left(s_2\right),\dot{K}\left(s_5\right)\right)$ is rewritten as
        \begin{alignat*}{7}
        \phi_{2,5}^*\omega\left(\dot{K}\left(s_2\right),\dot{K}\left(s_5\right)\right) & = & \frac{\varepsilon_{\mu \nu \sigma}}{4\pi} \frac{(K(s_2)- K(s_5))^{\mu}}{|K(s_2) - K(s_5)|^3} \dot{K}^\nu\left(s_2\right)\dot{K}^\sigma\left(s_5\right).
        \end{alignat*}
        
Analogously to the case of diagram $D_{31}$, $\overline{I}_{D_{32}}$ is given by
      \begin{alignat*}{7}
      & & \mathrlap{ \int\limits_{C(5,{\bf S}^1)}d\overline{s}\int\limits_{\mathbb{R}^3} \phi^*\overline{\omega}\left(\big[\dot{K}(\overline{s})\big]_{\ell},\ldots\right) } && \\
      &=& && \int\limits_{C(5,{\bf S}^1)} d\overline{s}\int\limits_{\mathbb{R}^3} \biggl[ \Delta_{\nu_1 \sigma_1} (K(s_1) - x_6) \Delta_{\nu_2 \sigma_2} (K(s_2) - K(s_5)) \Delta_{\nu_3 \sigma_3}(K(s_3) - x_6) \Delta_{\nu_4 \sigma_4}(K(s_4) - x_6) \\
      & & && \times \dot{K}^{\nu_1}(s_1) \dot{K}^{\nu_2}(s_2)\dot{K}^{\sigma_2}(s_5) \dot{K}^{\nu_3}(s_3) \dot{K}^{\nu_4}(s_4) \biggr] dx_6^{\sigma_1} \wedge dx_6^{\sigma_3} \wedge dx_6^{\sigma_4},
      \numberthis
      \end{alignat*}
from where
      \begin{align*}
      I_{D_{32}}
      = \int\limits_{C(5,{\bf S}^1)} \dot{K}^{\nu_1}(s_1) \dot{K}^{\nu_2}(s_2) \dot{K}^{\nu_3}(s_3)\dot{K}^{\nu_4}(s_4)ds_1\wedge ds_2\wedge  ds_3\wedge ds_4\wedge ds_5\int\limits_{\mathbb{R}^3} \varepsilon^{\sigma_1 \sigma_3 \sigma_4} d^3 x_6 \\
      \times \Delta_{\nu_1 \sigma_1} (K(s_1) - x_6) \Delta_{\nu_2 \sigma_2}      (K(s_2) - K(s_5)) \Delta_{\nu_3 \sigma_3}(K(s_3) - x_6)\Delta_{\nu_4 \sigma_4}(K(s_4) - x_6).
      \numberthis
      \end{align*}
      Again this expression matches with Eq. (\ref{dibujo32cstrabajado}) which comes from Chern-Simons theory. 
 

Now we continue our analysis with diagram $D_{33}$. This time the map $\phi$ in figure \ref{diagramota07} is given by the restriction of 
      \begin{align}
      \phi_{1,4} \times \phi_{2,6} \times \phi_{3,5} : \prod_{i=1}^6 {\bf S}^3 \longrightarrow \prod_{i=1}^3 {\bf S}^2
      \end{align}     
to $C(6,{\bf S}^3)$. Here all the points belong to the knot, therefore $\overline{\omega}=\omega\times\omega\times\omega\in\Omega^6\left(({\bf S}^2)^3\right)$ and $\overline{\pi}_6=id$.

      \begin{figure}[ht]
      \[ \xymatrix{
      & & C\left(6; {\bf S}^1\right) \times \mathcal{K} \ar[dd]^{ev} & & \\
      & & & & \\
      & C\left(6; {\bf S}^1\right) \ar[r]^{ev_K} \ar[dd]^{\alpha_6^1} & C\left(6; {\bf S}^3\right) \ar[dd]^{\alpha_6^3} \ar[rr]^{\phi} & & {\bf S}^2 \times {\bf S}^2 \times {\bf S}^2 \\
      & & & & \\
      & C\left[6; {\bf S}^1\right] \ar[r]^{\overline{ev}_K} & C\left[6; {\bf S}^3\right] \ar[uurr]_{\Phi} & & \\
      & & & & \\
      \mathcal{K} & & \ar[ll]_{pr} C\left[6; {\bf S}^1\right] \times \mathcal{K} \ar[uu]_{\overline{ev}} & & \\
      } \]
      \caption{Mathematical diagram for $D_{33}$.}
      \label{diagramota07}
      \end{figure}

This time the configuration space integral is given by
      \begin{alignat*}{6}
      \overline{I}_{D,33} 
      &=&& \int\limits_{C(6,{\bf S}^1)} d\overline{s} \left( \Phi \circ\alpha_p^3\circ ev_K  \right)^* \overline{\omega} + B_{33} \\
      &=&& \int\limits_{C(6,{\bf S}^1)} d\overline{s} \left( \phi \circ ev_K  \right)^* \overline{\omega} + B_{33},
      \numberthis
      \label{33configuration01}      
      \end{alignat*}
where $B_{33}$ stands for all boundary terms. By using Eq. (\ref{pullbackevkab}), the pullback of $\overline{\omega}$ under $\phi\circ ev_K$ for this case is written as
      \begin{alignat*}{5}
       && \mathrlap{ \left(\phi\circ ev_k\right)^*\overline{\omega} } && \\ \\
      =&& \mathrlap{ \left(ev_K\right)^*\phi_{1,4}^*\omega \times \left(ev_K\right)^*\phi_{2,6}^*\omega \times \left(ev_K\right)^*\phi_{3,5}^*\omega } && \\ \\
      =&& && \biggl[ \frac{\varepsilon_{\mu_1 \nu_1 \sigma_1}}{4\pi} \frac{(K(s_1)- K(s_4))^{\mu_1}}{|K(s_1) - K(s_4)|^3}\frac{\varepsilon_{\mu_2 \nu_2 \sigma_2}}{4\pi} \frac{(K(s_2)- K(s_6))^{\mu_2}}{|K(s_2) - K(s_6)|^3} \frac{\varepsilon_{\mu_3 \nu_3 \sigma_3}}{4\pi} \frac{(K(s_3)- K(s_5))^{\mu_3}}{|K(s_3) - K(s_5)|^3} \biggr]  \\
       && && \times \dot{K}^{\nu_1}\left(s_1\right) \dot{K}^{\sigma_1}\left(s_4\right) \dot{K}^{\nu_2}\left(s_2\right) \dot{K}^{\sigma_2}\left(s_6\right) \dot{K}^{\nu_3}\left(s_3\right) \dot{K}^{\sigma_3}\left(s_5\right)\\
       && && \times ds_1\wedge ds_4 \wedge ds_2\wedge ds_6\wedge ds_3\wedge ds_5 \\ \\
      =&& \mathrlap{ \biggl[ \Delta_{\nu_1 \sigma_1}(K(s_1)-K(s_4)) \Delta_{\nu_2 \sigma_2}(K(s_2)-K(s_6))\Delta_{\nu_3 \sigma_3}(K(s_3)-K(s_5)) \biggr] } && \\
       && && \times \dot{K}^{\nu_1}\left(s_1\right) \cdot \dot{K}^{\sigma_1}\left(s_4\right) \dot{K}^{\nu_2}\left(s_2\right)\dot{K}^{\sigma_2}\left(s_6\right)\dot{K}^{\nu_3}\left(s_3\right)\dot{K}^{\sigma_3}\left(s_5\right)\\
       && && \times ds_1\wedge ds_2\wedge ds_3\wedge ds_4\wedge ds_5\wedge ds_6,
      \numberthis
      \end{alignat*}  
and then $I_{D_{33}}$ is given by
      \begin{align*}
      I_{D_{33}}
      = \int\limits_{C(6,{\bf S}^1)} \biggl[ \Delta_{\nu_1 \sigma_1}\left(K(s_1)-K(s_4)\right)\Delta_{\nu_2 \sigma_2}\left(K(s_2)-K(s_6)\right)\Delta_{\nu_3 \sigma_3}\left(K(s_3)-K(s_5)\right) & \\
      \times \dot{K}^{\nu_1}(s_1) \dot{K}^{\nu_2}(s_2) \dot{K}^{\nu_3}(s_3)\dot{K}^{\sigma_1}(s_4)\dot{K}^{\sigma_2}(s_6)\dot{K}^{\sigma_3}(s_5)\biggr] & \\
      \times ds_1\wedge ds_2\wedge ds_3\wedge ds_4\wedge ds_5 \wedge ds_6,
      \numberthis
      \end{align*}
which can be seen to match with Eq. (\ref{dibujo33cstrabajado}) from Chern-Simons theory.


The final step is to analyse diagram $D_{34}$. The map $\phi$ in figure \ref{diagramota08} is given now by the restriction of
      
      \begin{equation*}
      \phi_{1,4} \times \phi_{2,5} \times \phi_{3,6} : \prod_{i=1}^6 {\bf S}^3 \longrightarrow \prod_{i=1}^3 {\bf S}^2 \numberthis
      \end{equation*}
to $C(6,{\bf S}^3)$. Again all the points are defined on the knot so there is no integration on internal points. In this case 
$\overline{\omega}=\omega\times\omega\times\omega\in\Omega^6\left(({\bf S}^2)^3\right)$ and the
configuration space integral reads
      \begin{alignat*}{6}
      \overline{I}_{D_{34}} 
      &=&& \int\limits_{C(6,{\bf S}^1)} d\overline{s} \left( \Phi \circ\alpha_p^3\circ ev_K  \right)^* \overline{\omega} + B_{34} \\
      &=&& \int\limits_{C(6,{\bf S}^1)} d\overline{s} \left( \phi \circ ev_K  \right)^* \overline{\omega} + B_{34},
      \label{34configuration01}      
      \numberthis
      \end{alignat*}
where $B_{34}$ again stands for the boundary terms. The same pullback from diagram $D_{33}$ applies to this case and $\phi\circ ev_K$ is rewritten as

      \begin{alignat*}{5}
       && \mathrlap{ \left(\phi\circ ev_k\right)^*\overline{\omega} } && \\ \\
      =&& \mathrlap{ \left( ev_K^* \phi_{1,4}^*\omega \right) \wedge \left( ev_K^* \phi_{2,5}^*\omega \right) \wedge \left( ev_K^*\phi_{3,6}^*\omega \right) } && \\ \\
      =&& && \biggl[ \frac{\varepsilon_{\mu_1 \nu_1 \sigma_1}}{4\pi} \frac{(K(s_1)- K(s_4))^{\mu_1}}{|K(s_1) - K(s_4)|^3}\frac{\varepsilon_{\mu_2 \nu_2 \sigma_2}}{4\pi} \frac{(K(s_2)- K(s_5))^{\mu_2}}{|K(s_2) - K(s_5)|^3}\frac{\varepsilon_{\mu_3 \nu_3 \sigma_3}}{4\pi} \frac{(K(s_3)- K(s_6))^{\mu_3}}{|K(s_3) - K(s_6)|^3} \biggr] \\
       && && \times \dot{K}^{\nu_1}\left(s_1\right) \dot{K}^{\sigma_1}\left(s_4\right) \dot{K}^{\nu_2}\left(s_2\right)\dot{K}^{\sigma_2}\left(s_5\right)\dot{K}^{\nu_3}\left(s_3\right)\dot{K}^{\sigma_3}\left(s_6\right)\\
       && && \times ds_1\wedge ds_4 \wedge ds_2\wedge ds_5\wedge ds_3\wedge ds_6 \\ \\
      =&& \mathrlap{ \biggl[ \Delta_{\nu_1 \sigma_1}(K(s_1)-K(s_4)) \Delta_{\nu_2 \sigma_2}(K(s_2)-K(s_5))\Delta_{\nu_3 \sigma_3}(K(s_3)-K(s_6)) \biggr]} && \\
       && && \times (-1)\dot{K}^{\nu_1}\left(s_1\right) \cdot \dot{K}^{\sigma_1}\left(s_4\right) \dot{K}^{\nu_2}\left(s_2\right)\dot{K}^{\sigma_2}\left(s_5\right)\dot{K}^{\nu_3}\left(s_3\right)\dot{K}^{\sigma_3}\left(s_6\right) \\
       && && \times ds_1\wedge ds_2\wedge ds_3\wedge ds_4\wedge ds_5\wedge ds_6,
      \numberthis
      \end{alignat*}          
from where
      \begin{align*}
      I_{D_{34}}
      = -\int\limits_{C(6,{\bf S}^1)} \biggl[ \Delta_{\nu_1 \sigma_1}\left(K(s_1)-K(s_4)\right)\Delta_{\nu_2 \sigma_2}\left(K(s_2)-K(s_5)\right)\Delta_{\nu_3 \sigma_3}\left(K(s_3)-K(s_6)\right) & \\
      \times \dot{K}^{\nu_1}(s_1) \dot{K}^{\nu_2}(s_2) \dot{K}^{\nu_3}(s_3)\dot{K}^{\sigma_1}(s_4)\dot{K}^{\sigma_2}(s_5)\dot{K}^{\sigma_3}(s_6)\biggr] & \\    
      \times ds_1\wedge ds_2\wedge ds_3\wedge ds_4\wedge ds_5 \wedge ds_6.
      \numberthis
      \end{align*}
This again matches with the corresponding integral in Eq. (\ref{dibujo34cstrabajado}) from Chern-Simons theory.

\subsection{Boundary cancellation}

There are important aspects to consider about the boundary terms appearing in Eqs. (\ref{selflinkinconfiguration01}), (\ref{Yconfiguration01}) and (\ref{configuration22}). In the self-linking case the boundary term in (\ref{selflinkinconfiguration01}) is cancelled via the introduction of a framing term of the knot \cite{Mos04}. The case of the second order Vassiliev invariant needs some more considerations.

In general for a given knot diagram there will be many boundary terms depending on the number and the rates of point collapses \cite{Vol13}. If two points collapse the face is called {\it principal}, if three or more points (but not all) collapse the face is called {\it hidden}. If all points collapse the face is called {\it anomalous} and if one or more collapsing points are considered to be at infinity the face is called 
{\it face at infinity}.

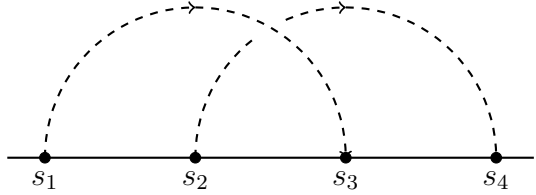
\begin{figure}[ht]
      \centering
		  \begin{tikzpicture}
		  \draw[thick] (-3.5,0) -- (3.5,0);
		  \filldraw (-1,0) circle (2pt);
		  \filldraw (1,0) circle (2pt);
		  \filldraw (-3,0) circle (2pt);
		  \filldraw (3,0) circle (2pt);
		  \draw [thick, dashed, ->] [domain=180:90] plot ({-1+2*cos(\x)}, {2*sin(\x)});
		  \draw [name path=arc1, opacity=0] [domain=90:0] plot ({-1+2*cos(\x)}, {2*sin(\x)});
		  \draw [name path=arc2, opacity=0] [domain=180:90] plot ({1+2*cos(\x)}, {2*sin(\x)});
		  \draw [thick, dashed] [domain=90:0] plot ({1+2*cos(\x)}, {2*sin(\x)});
			\path [name intersections={of = arc1 and arc2, by = arcs}];
		  \draw [thick, dashed, ->] [domain=180:90] plot ({1+2*cos(\x)}, {2*sin(\x)});
			\filldraw [white] (arcs) circle (6pt);
		  \draw [thick, dashed, ->] [domain=90:0] plot ({-1+2*cos(\x)}, {2*sin(\x)});
		  \node at (-3,-0.3) {$s_1$};
		  \node at (-1,-0.3) {$s_2$};
		  \node at (1,-0.3) {$s_3$};
		  \node at (3,-0.3) {$s_4$};
		  \end{tikzpicture}
      \caption{Diagram for the boundary analysis of figure \ref{laxdibujo}.}
      \label{boundarydiagramx}
      \end{figure}

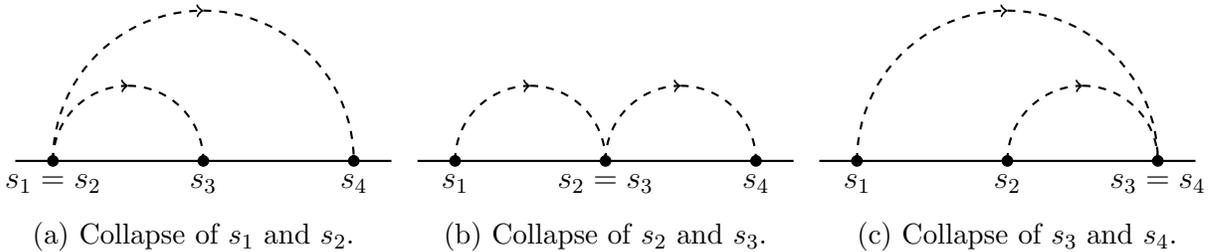
\begin{figure}[ht]
      \begin{subfigure}[b]{0.33\textwidth}
      \centering
		  \begin{tikzpicture}
		  \draw[thick] (-2.5,0) -- (2.5,0);
		  \filldraw (-2,0) circle (2pt);
		  \filldraw (2,0) circle (2pt);
		  \filldraw (0,0) circle (2pt);
		  \draw [thick, dashed, ->] [domain=180:90] plot ({2*cos(\x)}, {2*sin(\x)});
		  \draw [thick, dashed] [domain=90:0] plot ({2*cos(\x)}, {2*sin(\x)});
		  \draw [thick, dashed, ->] [domain=180:90] plot ({-1+1*cos(\x)}, {1*sin(\x)});
		  \draw [thick, dashed] [domain=90:0] plot ({-1+1*cos(\x)}, {1*sin(\x)});
		  \node at (-2,-0.3) {$s_1 = s_2$};
		  \node at (0,-0.3) {$s_3$};
		  \node at (2,-0.3) {$s_4$};
		  \end{tikzpicture}
      \caption{Collapse of $s_1$ and $s_2$.}
      \label{collapse12}
      \end{subfigure}%
      \begin{subfigure}[b]{0.33\textwidth}
      \centering
		  \begin{tikzpicture}
		  \draw[thick] (-2.5,0) -- (2.5,0);
		  \filldraw (-2,0) circle (2pt);
		  \filldraw (2,0) circle (2pt);
		  \filldraw (0,0) circle (2pt);
		  \draw [thick, dashed, ->] [domain=180:90] plot ({-1+1*cos(\x)}, {1*sin(\x)});
		  \draw [thick, dashed] [domain=90:0] plot ({-1+1*cos(\x)}, {1*sin(\x)});
		  \draw [thick, dashed, ->] [domain=180:90] plot ({1+1*cos(\x)}, {1*sin(\x)});
		  \draw [thick, dashed] [domain=90:0] plot ({1+1*cos(\x)}, {1*sin(\x)});
		  \node at (-2,-0.3) {$s_1$};
		  \node at (0,-0.3) {$s_2 = s_3$};
		  \node at (2,-0.3) {$s_4$};
		  \end{tikzpicture}
      \caption{Collapse of $s_2$ and $s_3$.}
      \label{collapse23}
      \end{subfigure}%
      \begin{subfigure}[b]{0.33\textwidth}
      \centering
		  \begin{tikzpicture}
		  \draw[thick] (-2.5,0) -- (2.5,0);
		  \filldraw (-2,0) circle (2pt);
		  \filldraw (2,0) circle (2pt);
		  \filldraw (0,0) circle (2pt);
		  \draw [thick, dashed, ->] [domain=180:90] plot ({2*cos(\x)}, {2*sin(\x)});
		  \draw [thick, dashed] [domain=90:0] plot ({2*cos(\x)}, {2*sin(\x)});
		  \draw [thick, dashed, ->] [domain=180:90] plot ({1+1*cos(\x)}, {1*sin(\x)});
		  \draw [thick, dashed] [domain=90:0] plot ({1+1*cos(\x)}, {1*sin(\x)});
		  \node at (-2,-0.3) {$s_1$};
		  \node at (0,-0.3) {$s_2$};
		  \node at (2,-0.3) {$s_3 = s_4$};
		  \end{tikzpicture}
      \caption{Collapse of $s_3$ and $s_4$.}
      \label{collapse34}
      \end{subfigure}%
      \caption{Principal faces for diagram \ref{boundarydiagramx}.}
      \label{collapsesofboundary}
      \end{figure}

For the case of figure \ref{laxdibujo} the integrals corresponding to hidden and anomalous faces as well as to faces at infinity vanish 
\cite{Vol13}. If such a diagram is sketched as in figure \ref{boundarydiagramx} then the principal faces are those coming from the collapse of
exactly two of those points yielding the configurations shown in figure \ref{collapsesofboundary}. Integrals corresponding to these configurations do not necessarily vanish.

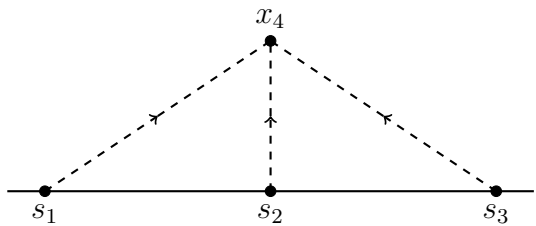
\begin{figure}[ht]
      \centering
		  \begin{tikzpicture}
		  \draw[thick] (-3.5,0) -- (3.5,0);
		  \filldraw (-3,0) circle (2pt);
		  \filldraw (3,0) circle (2pt);
		  \filldraw (0,0) circle (2pt);
		  \filldraw (0,2) circle (2pt);
		  \draw [thick, dashed, ->] (-3,0) -- (-1.5,1);
		  \draw [thick, dashed] (-1.5,1) -- (0,2);
		  \draw [thick, dashed, ->] (0,0) -- (0,1);
		  \draw [thick, dashed] (0,1) -- (0,2);
		  \draw [thick, dashed, ->] (3,0) -- (1.5,1);
		  \draw [thick, dashed] (1.5,1) -- (0,2);
		  \node at (-3,-0.3) {$s_1$};
		  \node at (0,-0.3) {$s_2$};
		  \node at (3,-0.3) {$s_3$};
		  \node at (0,2.3) {$x_4$};
		  \end{tikzpicture}
      \caption{Diagram for the boundary analysis of figure \ref{laydibujo}.}
      \label{boundarydiagramy}
      \end{figure}

The procedure to obtain a topological invariant is then to find another Feynman diagram such that its non-vanishing faces are exactly the same
as the ones in figure \ref{collapsesofboundary} and then subtract them. The choice is just diagram \ref{laydibujo} that is sketched in figure 
\ref{boundarydiagramy}.
      
The integrals corresponding to hidden and anomalous faces, faces at infinity and principal faces coming from the collapse of two points on the 
knot vanish \cite{Vol13}. The remaining non-vanishing boundary contributions are the ones coming from the collapse of the internal point $x_4$ and a point on the knot. In fact the collapse of $x_4$ with $s_1$, $s_2$ and $s_3$ yields, after adjusting orientation of propagators, to the diagram in figure \ref{collapse12}, \ref{collapse23} and \ref{collapse34}, respectively. Thus, subtraction of diagrams \ref{laydibujo} and \ref{laxdibujo} makes all boundary contributions disappear and it is a real topological invariant, this is precisely the second order Vassiliev invariant. It can be shown that for the third order Vassiliev invariant an analogous boundary cancellation occurs between the four diagrams in figure \ref{thirdordervassilievdibujo}, actually this is also true for higher orders \cite{Vol13}. Consequently, the real topological invariant is the sum of contributions of all diagrams in that figure.

\section{Asymptotic Vassiliev invariants from Chern-Simons perturbation theory}

In this section we incorporate a vector field $X$ in the manifold $M$ where Chern-Simons theory is defined.
The key idea is to replace $\dot{K}$ by the vector field $X$, {\it i.e.}, one has the identification 
      \begin{equation}
      \dot{K} \longleftrightarrow X
      \end{equation}
in the sense of section \ref{volumepreserving}. As in the case without flow information the appropriate boundary cancellation will be assumed.

\subsection{First order flow invariant}

As indicated in section \ref{volumepreserving}, $\theta^p$ is the $p$-fold product of the flow generated by the vector field $X$ defined by $\theta^p((x_1,\ldots,x_p),(t_1,\ldots,t_p))=(\theta(x_1,t_1),\ldots,\theta(x_p,t_p))$, where $(x_1,\ldots,x_p)\in C(p,\mathcal{S})$ and $(t_1,\ldots,t_p)\in C(p,{\bf S}^1)$, also $\theta^x(t)=\theta(x,t)$ with $x\in \mathcal{S}$ and $t\in{\bf S}^1$. From now on $\overline{x}$ stands for a $p$-tuple $(x,\ldots,x)$ in $\mathcal{S}^p$.

The first example will be the self-linking expression (\ref{selflinkinconfiguration02}) coming from figure \ref{figuradelselflinking}. For this case, Eq. (\ref{lafdevolicconx}) reads
      \begin{alignat*}{5}
      f_{D_1,X}(\theta^x(t_1),\theta^x(t_2))&=&\left(\left(\alpha_2^3\right)^*\Phi^*\omega\right)_{\left(\theta^x(t_1),\theta^x(t_2)\right)}(X_{\theta^x(t_1)},X_{\theta^x(t_2)}). \numberthis
      \end{alignat*}
      
By using a theorem from Ref. \cite{KV16}, the asymptotic invariant has the form
      \begin{equation}
      \int\limits_{\mathcal{S}}\lambda_{D_1}\mu = \int\limits_{\mathcal{S}\times\mathcal{S}} \lim_{T\rightarrow\infty}\frac{1}{T^2}\left\lbrace\int\limits_0^T \int\limits_0^T f_{D_1,X}(\theta^{x_1}(t_1),\theta^{x_2}(t_2))dt_1\wedge dt_2\right\rbrace \mu_{\Delta},
      \label{asyselflink}
      \end{equation} 
where the part within braces in the integrand can be rewritten as
      \begin{alignat*}{5}
      \int\limits_0^T \int\limits_0^T f_{D_1,X}(\theta^{x_1}(t_1),\theta^{x_2}(t_2))dt_1\wedge dt_2 
      &=&& \int\limits_{C(2,S^1)}\left(\phi\circ\left(\theta^{x_1}\times\theta^{x_2}\right)\right)^*\omega \\
      &=&& \int\limits_{C(2,S^1)}\frac{\varepsilon_{\mu\nu\sigma}}{4\pi}\frac{\left(\theta^{x_1}(t_1)-\theta^{x_2}(t_2)\right)^{\mu}}{|\theta^{x_1}(t_1)-\theta^{x_2}(t_2)|^3}(\dot{\theta}^{x_1})^{\nu}(\dot{\theta}^{x_2})^{\sigma}dt_1\wedge dt_2 \\
      &=&& \text{lk}\left(\theta^{x_1},\theta^{x_2}\right), \numberthis
      \end{alignat*}
and then by substituting this expression into Eq. (\ref{asyselflink}) one gets
      \begin{alignat*}{5}
      \mathcal{H}(X)= \int\limits_{\mathcal{S}}\lambda_{D_1}\mu = \int\limits_{\mathcal{S}\times\mathcal{S}} \lim_{T\rightarrow\infty}\frac{1}{T^2} \text{lk}\left(\theta^{x_1},\theta^{x_2}\right)\mu_{\Delta},
      \numberthis
\label{helicity}      
      \end{alignat*} 
which is the called average asymptotic linking invariant (or Hopf invariant) for the vector field $X$. 

\subsection{Second order flow invariant}

The flow contributions of the diagrams in figure \ref{secondordervassilievdibujo} once the vector field $X$ was introduced are again an 
application of Eq. (\ref{intlambdaeqintf}).

In the case of diagram \ref{laxdibujo} we have to use figure \ref{diagramota04} for which again $\phi=\phi_{1,3}\times\phi_{2,4}$. Then Eq. (\ref{lafdevolicconx}) is written as
      \begin{alignat}{5}
      f_{D_{22},X}(\theta^x(t_1),\ldots,\theta^x(t_4))&=&&\left(\left(\alpha_4^3\right)^*\Phi^*\overline{\omega}\right)_{\left(\theta^x(t_1), \ldots,\theta^x(t_4)\right)}(X_{\theta^x(t_1)},\ldots,X_{\theta^x(t_4)}),
      \end{alignat}
and then the integral (\ref{intasyminv}) takes the form
      \begin{alignat}{5}
      \int_{\mathcal{S}}\lambda_{D_{22}}\mu &=& \int_{\mathcal{S}^4}\lim_{T\rightarrow \infty}\frac{1}{T^4}\left\lbrace \int\limits_0^T \cdots\int\limits_0^T f_{D_{22},X}(\theta^{x_1}(t_1),\ldots,\theta^{x_4}(t_4))dt_1\wedge\cdots\wedge dt_4\right\rbrace\mu_{\Delta}.
\label{contribucionflujoD22}      
      \end{alignat}
Once again, by first making an analysis of the term between braces in the previous equation one gets
	  \begin{alignat}{5}
	  \int\limits_0^T \cdots\int\limits_0^T f_{D_{22},X}(\theta^{x_1}(t_1),\ldots,\theta^{x_4}(t_4))dt_1\wedge\cdots\wedge dt_4 &=&&&& \int\limits_{C(4,{\bf S}^1)}\left(\phi\circ\theta^{4}\right)^*\overline{\omega} + \widetilde{B}_{22}, \numberthis
	  \end{alignat}
where $\widetilde{B}_{22}$ stands for the boundary terms.

The integrand in the first term of the right hand side of the last expression takes the form
	  \begin{alignat*}{7}
	  \left(\phi\circ\theta^4\right)^*\overline{\omega}
	  &=&&(\theta^4)^*(\phi_{1,3}^*\omega\wedge\phi_{2,4}^*\omega) & \\
	  &=&& \mathrlap{ \left[\frac{\varepsilon_{\mu_1\nu_1\sigma_1}}{4\pi}\frac{\left(\theta^{x_1}(t_1)-\theta^{x_3}(t_2)\right)^{\mu_1}}{|\theta^{x_1}(t_1)-\theta^{x_3}(t_2)|^3}\right]\left[\frac{\varepsilon_{\mu_2\nu_2\sigma_2}}{4\pi}\frac{\left(\theta^{x_2}(t_3)-\theta^{x_4}(t_4)\right)^{\mu_2}}{|\theta^{x_2}(t_3)-\theta^{x_4}(t_4)|^3}\right]} \\
	  &&& & \times {(\dot{\theta}^{x_1})}^{\nu_1}{(\dot{\theta}^{x_3})}^{\sigma_1}{(\dot{\theta}^{x_2})}^{\nu_2}{(\dot{\theta}^{x_4})}^{\sigma_2} dt_1\wedge dt_2\wedge dt_3\wedge dt_4 \\
	  &=&& \mathrlap{ \Delta_{\nu_1\sigma_1}(\theta^{x_1}(t_1)-\theta^{x_3}(t_2))\Delta_{\nu_2\sigma_2}(\theta^{x_2}(t_3)-\theta^{x_4}(t_4))} \\
	  &&& & \times X_{\theta^{x_1}(t_1)}^{\nu_1}X_{\theta^{x_3}(t_2)}^{\sigma_1}X_{\theta^{x_2}(t_3)}^{\nu_2}X_{\theta^{x_4}(t_4)}^{\sigma_2} dt_1\wedge dt_2\wedge dt_3\wedge dt_4.
	  \numberthis
	  \end{alignat*}
	  
Then, the asymptotic integral has the form
	  \begin{alignat*}{5}
	  \int_{\mathcal{S}}\lambda_{D_{22}}\mu &=& \int_{\mathcal{S}^4}\lim_{T\rightarrow \infty}\frac{1}{T^4}\biggl\{& \int\limits_{C(4, {\bf S}^1)} \mathrlap{ \Delta_{\nu_1\sigma_1}(\theta^{x_1}(t_1)-\theta^{x_3}(t_2))\Delta_{\nu_2\sigma_2}(\theta^{x_2}(t_3)-\theta^{x_4}(t_4))} \\
	  &&& & \times X_{\theta^{x_1}(t_1)}^{\nu_1}X_{\theta^{x_3}(t_2)}^{\sigma_1}X_{\theta^{x_2}(t_3)}^{\nu_2}X_{\theta^{x_4}(t_4)}^{\sigma_2} dt_1\wedge dt_2\wedge dt_3\wedge dt_4 \bigg\}\mu_{\Delta}.
	  \label{intlambdaD22}
	  \numberthis
	  \end{alignat*}

The diagram \ref{laydibujo} needs different considerations because now there is a point outside the knot and it is necessary to take this
information into account. These considerations imply the use of figure (\ref{diagramota03}) where, analogously to the previous analysis, $\phi$ and $\overline{\omega}$ are given by $\phi=\phi_{1,4}\times\phi_{2,4}\times\phi_{3,4}$ and $\overline{\omega}=\omega\times\omega\times\omega$ then one gets 
	  \begin{alignat*}{7}
 	  & && f_{D_{21},X}(\theta^{3}(\overline{x},(t_1,t_2,t_3)))\\ \\
 	  &=&& \left((\alpha^3_3)^*(\overline{\pi}_3)_*\Phi^*\overline{\omega}\right)_{\left(\theta^{3}(\overline{x},(t_1,t_2,t_3))\right)}\left(X_{\theta^{x}(t_1)},X_{\theta^{x}(t_2)},X_{\theta^{x}(t_3)}\right) \\ \\
 	  &=&& (\alpha^3_3)^*\left(\int\limits_{(\overline{\pi}_3)^{-1}(\theta^3(\overline{x},(t_1,t_2,t_3)))}\Phi^*\overline{\omega}(\big[X_{\theta^{x}(t_1)}\big]_{\ell},\big[X_{\theta^{x}(t_2)}\big]_{\ell},\big[X_{\theta^{x}(t_3)}\big]_{\ell},\ldots)\right) \\ \\
	  &=&& \int\limits_{(\overline{\pi}_3)^{-1}(\alpha^3_3\left(\theta^3(\overline{x},(t_1,t_2,t_3))\right))}\Phi^*\overline{\omega}(\big[(\alpha^3_3)_*\left(X_{\theta^{x}(t_1)}\right)\big]_{\ell},\big[(\alpha^3_3)_*\left(X_{\theta^{x}(t_2)}\right)\big]_{\ell},\big[(\alpha^3_3)_*\left(X_{\theta^{x}(t_3)}\right)\big]_{\ell},\ldots) \\ \\
	  &=&& \int\limits_{\mathbb{R}^3}\Phi^*\overline{\omega}(\big[(\alpha^3_3)_*\left(X_{\theta^{x}(t_1)}\right)\big]_{\ell},\big[(\alpha^3_3)_*\left(X_{\theta^{x}(t_2)}\right)\big]_{\ell},\big[(\alpha^3_3)_*\left(X_{\theta^{x}(t_3)}\right)]_{\ell},\ldots),
	  \numberthis
	  \label{lafvolicYflujo}
	  \end{alignat*}
where $\big[(\alpha^3_3)_*\left(X_{\theta^x(t_i)}\right)\big]_{\ell}$ denotes any lift of the pushforward by $\alpha^3_3$ of the tangent vectors
$X_{\theta^x(t_i)}$ to tangent vectors in $C[3+1,\mathbb{R}^3]$.

Then the average asymptotic term for this diagram can be written as 
	  \begin{alignat}{5}
	  \int_{\mathcal{S}}\lambda_{D_{21}}\mu &=& \int_{\mathcal{S}^3}\lim_{T\rightarrow \infty}\frac{1}{T^3}\left\lbrace \int\limits_0^T \cdots\int\limits_0^T f_{D_{21},X}(\theta^{x_1}(t_1),\theta^{x_2}(t_2),\theta^{x_3}(t_3))dt_1\wedge dt_2\wedge dt_3\right\rbrace\mu_{\Delta}.
      \label{contribucionflujoD21}	  
	  \end{alignat}
	  
By using Eq. (\ref{lafvolicYflujo}) in the last expression it is possible to rewrite the term between braces as
	  \begin{alignat*}{5}
	  \int\limits_0^T\cdots\int\limits_0^T \int\limits_{\mathbb{R}^3}\Phi^*\overline{\omega}(\big[(\alpha^3_3)_*\left(X_{\theta^{x_1}(t_1)}\right)\big]_{\ell},\big[(\alpha^3_3)_*\left(X_{\theta^{x_2}(t_2)}\right)\big]_{\ell},\big[(\alpha^3_3)_*\left(X_{\theta^{x_3}(t_3)}\right)\big]_{\ell},\ldots)dt_1\wedge dt_2\wedge dt_3\\
	  =\int\limits_{C(3,{\bf S}^1)}\int\limits_{\mathbb{R}^3}(\phi\circ(\theta^3\times id_\mathbb{R}^3))^*\overline{\omega}+ \widetilde{B}_{21},
	  \numberthis
	  \end{alignat*}
where $\widetilde{B}_{21}$ stands for the boundary terms.

By using again Eq. (\ref{pullbackab}) the integrand in the first term of the right hand side of the last expression takes the form
	  \begin{alignat*}{5}
	  \left(\phi\circ\left(\theta^3\times id_{\mathbb{R}^3}\right)\right)^*\overline{\omega}
	  &=&& \mathrlap{ \left(\theta^3\times id_{\mathbb{R}^3}\right)^*(\phi_{1,4}^*\omega\wedge\phi_{2,4}^*\omega\wedge\phi_{3,4}^*\omega) } \\
	  &=&& \mathrlap{ \left[\frac{\varepsilon_{\mu_1\nu_1\sigma_1}}{4\pi}\frac{(\theta^{x_1}(t_1)-x_4)^{\mu_1}}{|\theta^{x_1}(t_1)-x_4|^3}{(\dot{\theta}^{x_1})}^{\nu_1}dt_1\wedge dx_4^{\sigma_1}\right]} \\
	  &&&  \wedge \mathrlap{ \left[\frac{\varepsilon_{\mu_2\nu_2\sigma_2}}{4\pi}\frac{(\theta^{x_2}(t_2)-x_4)^{\mu_2}}{|\theta^{x_2}(t_2)-x_4|^3}{(\dot{\theta}^{x_2})}^{\nu_2}dt_2\wedge dx_4^{\sigma_2}\right]} \\
	  &&&  \wedge \mathrlap{ \left[\frac{\varepsilon_{\mu_3\nu_3\sigma_3}}{4\pi}\frac{(\theta^{x_3}(t_3)-x_4)^{\mu_3}}{|\theta^{x_3}(t_3)-x_4|^3}{(\dot{\theta}^{x_3})}^{\nu_3}dt_3\wedge dx_4^{\sigma_3}\right] } \\
	  &=&& & \Delta_{\nu_1\sigma_1}(\theta^{x_1}(t_1)-x_4)\Delta_{\nu_2\sigma_2}(\theta^{x_2}(t_2)-x_4)\Delta_{\nu_3\sigma_3}(\theta^{x_3}(t_3)-x_4)\\
	  &&& & \times X_{\theta^{x_1}(t_1)}^{\nu_1}X_{\theta^{x_2}(t_2)}^{\nu_2}X_{\theta^{x_3}(t_3)}^{\nu_3}dt_1\wedge dx_4^{\sigma_1}\wedge dt_2\wedge dx_4^{\sigma_2}\wedge dt_3\wedge dx_4^{\sigma_3} \\
	  &=&& & \Delta_{\nu_1\sigma_1}(\theta^{x_1}(t_1)-x_4)\Delta_{\nu_2\sigma_2}(\theta^{x_2}(t_2)-x_4)\Delta_{\nu_3\sigma_3}(\theta^{x_3}(t_3)-x_4)\\
	  &&& & \times X_{\theta^{x_1}(t_1)}^{\nu_1}X_{\theta^{x_2}(t_2)}^{\nu_2}X_{\theta^{x_3}(t_3)}^{\nu_3}dt_1\wedge dt_2\wedge dt_3\wedge (-1)^3\varepsilon^{\sigma_1\sigma_2\sigma_3}d^3 x_4.
	  \numberthis
	  \end{alignat*}
	  
Therefore the average asymptotic integral finally reads
	  \begin{alignat*}{5}
	  \int_{\mathcal{S}}\lambda_{D_{21}}\mu &=& \int_{\mathcal{S}^3}\lim_{T\rightarrow \infty}\frac{1}{T^3}\biggl\{& \int\limits_{C(3, {\bf S}^1)} \int\limits_{\mathbb{R}^3}&&\Delta_{\nu_1\sigma_1}(\theta^{x_1}(t_1)-x_4)\Delta_{\nu_2\sigma_2}(\theta^{x_2}(t_2)-x_4)\\
	  &&&&\times &\Delta_{\nu_3\sigma_3}(\theta^{x_3}(t_3)-x_4)X_{\theta^{x_1}(t_1)}^{\nu_1}X_{\theta^{x_2}(t_2)}^{\nu_2}X_{\theta^{x_3}(t_3)}^{\nu_3}\\
	  &&&&\times &(-1)^3\varepsilon^{\sigma_1\sigma_2\sigma_3}dt_1\wedge dt_2\wedge dt_3\wedge d^3 x_4+ \widetilde{B}_{21}\biggr\}\mu_{\Delta}.
	  \label{intlambdaD21}
	  \numberthis
	  \end{alignat*}

The second order asymptotic flow Vassiliev invariant is then the sum of expressions \ref{intlambdaD22} and \ref{intlambdaD21}.

\subsection{Third order flow invariant}

In the case of diagrams (\ref{dibujo31}) and (\ref{dibujo32}) there are inner points and a pushforward process has to be performed for each one.

For diagram (\ref{dibujo31}), the contribution of Eq. (\ref{lafdevolicconx}) is given by
	  \begin{alignat*}{7}
	  &&&f_{D_{31},X}(\theta^{4}(\overline{x},(t_1,\ldots,t_4)))\\
 	  &=&&\left((\alpha^3_4)^*(\overline{\pi}_4)_*\Phi^*\overline{\omega}\right)_{\left(\theta^{4} (\overline{x},(t_1,\ldots,t_4))\right)}\left(X_{\theta^{x}(t_1)},X_{\theta^{x}(t_2)},X_{\theta^{x}(t_3)},X_{\theta^{x}(t_4)}\right) \\
 	  &=&& (\alpha^3_4)^*\left(\int\limits_{(\overline{\pi}_4)^{-1}\left(\theta^4(\overline{x},(t_1,\ldots,t_4))\right)}\Phi^*\overline{\omega}(\big[X_{\theta^{x}(t_1)}\big]_{\ell},\big[X_{\theta^{x}(t_2)}\big]_{\ell},\big[X_{\theta^{x}(t_3)}\big]_{\ell},\big[X_{\theta^{x}(t_4)}\big]_{\ell},\ldots)\right)  \\
	  &=&& \int\limits_{(\overline{\pi}_4)^{-1}\left(\alpha^3_4\left(\theta^4(\overline{x},(t_1,\ldots,t_4))\right)\right)}\Phi^*\overline{\omega}\bigg(\big[(\alpha^3_4)_*\left(X_{\theta^{x}(t_1)}\right)\big]_{\ell},\ldots,\big[(\alpha^3_4)_*\left(X_{\theta^{x}(t_4)}\right)\big]_{\ell},\ldots\bigg) \\
  &=&& \int\limits_{\mathbb{R}^3\times \mathbb{R}^3}\Phi^*\overline{\omega}\bigg(\big[(\alpha^3_4)_*\left(X_{\theta^{x}(t_1)}\right)\big]_{\ell},\ldots,\big[(\alpha^3_4)_*\left(X_{\theta^{x}(t_4)}\right)\big]_{\ell},\ldots\bigg),
	  \numberthis
	  \label{lafvolic31}
	  \end{alignat*}
and then, in the asymptotic integral this takes the form
	  \begin{alignat*}{5}
	  \int_{\mathcal{S}}\lambda_{D_{31}}\mu &=&& \int_{\mathcal{S}^4}\lim_{T\rightarrow \infty}\frac{1}{T^4}\bigg\{&& \int\limits_0^T \cdots\int\limits_0^T f_{D_{31},X}(\theta^{x_1}(t_1),\dots,\theta^{x_4}(t_4))dt_1\cdots\wedge dt_4\bigg\}\mu_{\Delta}\\
	  &=&&\int_{\mathcal{S}^4}\lim_{T\rightarrow \infty}\frac{1}{T^4}\bigg\{&& \int\limits_0^T \cdots\int\limits_0^T \int\limits_{\mathbb{R}^3\times \mathbb{R}^3}\Phi^*\overline{\omega}\bigg(\big[(\alpha^3_4)_*\left(X_{\theta^{x_1}(t_1)}\right)\big]_{\ell},\ldots, \\
	  &&&&&\times \big[(\alpha^3_4)_*\left(X_{\theta^{x_4}(t_4)}\right)\big]_{\ell},\ldots\bigg) dt_1\wedge \cdots\wedge dt_4\bigg\}\mu_{\Delta},
	  \numberthis
	  \label{contribucionflujoD31}	  
	  \end{alignat*}
where once again the integrand form can be separated into its inner and its boundary parts, the former one is given by
	  \begin{alignat*}{5}
	  &&&\phi^*\overline{\omega}\left(\big[\left(\alpha_4^3\right)_*\left(X_{\theta^{x_1}(t_1)}\right)\big]_{\ell},\ldots,\big[\left(\alpha_4^3\right)_*\left(X_{\theta^{x_4}(t_4)}\right)\big]_{\ell},\ldots\right)\\
	  &=&& (\phi_{1,6}^*\omega\wedge\phi_{2,5}^*\omega\wedge\phi_{3,5}^*\omega\wedge\phi_{4,6}^*\omega\wedge\phi_{5,6}^*\omega)\biggl(\big[\left(\alpha_4^3\right)_*\left(X_{\theta^{x_1}(t_1)}\right)\big]_{\ell},\ldots,\big[\left(\alpha_4^3\right)_*\left(X_{\theta^{x_4}(t_4)}\right)\big]_{\ell},\ldots\biggr)  \\ 
	  &=&& \mathrlap{ \left[\frac{\varepsilon_{\mu_1\nu_1\sigma_1}}{4\pi}\frac{(\theta^{x_1}(t_1)-x_6)^{\mu_1}}{|\theta^{x_1}(t_1)-x_6|^3}{(\dot{\theta}^{x_1})}^{\nu_1} dx_6^{\sigma_1}\right]  \wedge
\left[\frac{\varepsilon_{\mu_2\nu_2\sigma_2}}{4\pi}\frac{(\theta^{x_2}(t_2)-x_5)^{\mu_2}}{|\theta^{x_2}(t_2)-x_5|^3}{(\dot{\theta}^{x_2})}^{\nu_2} dx_5^{\sigma_2}\right]	  
	   } \\
	  &&&  \wedge \mathrlap{ 
\left[\frac{\varepsilon_{\mu_3\nu_3\sigma_3}}{4\pi}\frac{(\theta^{x_3}(t_3)-x_5)^{\mu_3}}{|\theta^{x_3}(t_3)-x_5|^3}{(\dot{\theta}^{x_3})}^{\nu_3} dx_5^{\sigma_3}\right]\wedge	  
\left[\frac{\varepsilon_{\mu_4\nu_4\sigma_4}}{4\pi}\frac{(\theta^{x_4}(t_4)-x_6)^{\mu_4}}{|\theta^{x_4}(t_4)-x_6|^3}{(\dot{\theta}^{x_4})}^{\nu_4} dx_6^{\sigma_4}\right]
	   } \\
	  &&&  \wedge \mathrlap{ \left[\frac{\varepsilon_{\mu_5\nu_5\sigma_5}}{4\pi}\frac{(x_5-x_6)^{\mu_5}}{|x_5-x_6|^3}dx_5^{\nu_5}\wedge dx_6^{\sigma_5}\right] } \\
	  &=&& \Delta_{\nu_1\sigma_1}(\theta^{x_1}(t_1)-x_6)\Delta_{\nu_2\sigma_2}(\theta^{x_2}(t_2)-x_5)\Delta_{\nu_3\sigma_3}(\theta^{x_3}(t_3)-x_5)\Delta_{\nu_4\sigma_4}(\theta^{x_4}(t_4)-x_6)\\
	  &&& \times \Delta_{\nu_5\sigma_5}(x_5-x_6) X_{\theta^{x_1}(t_1)}^{\nu_1}X_{\theta^{x_2}(t_2)}^{\nu_2}X_{\theta^{x_3}(t_3)}^{\nu_3}X_{\theta^{x_4}(t_4)}^{\nu_4}\\
	  &&& \times dx_6^{\sigma_1}\wedge dx_5^{\sigma_2}\wedge dx_5^{\sigma_3}\wedge dx_6^{\sigma_4}\wedge dx_5^{\nu_5}\wedge dx_6^{\sigma_5} \\ 
	  &=&& \Delta_{\nu_1\sigma_1}(\theta^{x_1}(t_1)-x_6)\Delta_{\nu_2\sigma_2}(\theta^{x_2}(t_2)-x_5)\Delta_{\nu_3\sigma_3}(\theta^{x_3}(t_3)-x_5)\Delta_{\nu_4\sigma_4}(\theta^{x_4}(t_4)-x_6)\\
	  &&& \times\Delta_{\nu_5,\sigma_5}(x_5-x_6) X_{\theta^{x_1}(t_1)}^{\nu_1}X_{\theta^{x_2}(t_2)}^{\nu_2}X_{\theta^{x_3}(t_3)}^{\nu_3}X_{\theta^{x_4}(t_4)}^{\nu_4} \\
	  &&& \times (-1)^3 \varepsilon^{\sigma_2 \sigma_3 \nu_5}d^3x_5 \wedge \varepsilon^{\sigma_1 \sigma_4 \sigma_5}d^3x_6.
	  \numberthis
	  \end{alignat*}
	  
From Eq. (\ref{intasyminv}), the average asymptotic integral reads
	  \begin{alignat*}{5}
	  \int_{\mathcal{S}}\lambda_{D_{31}}\mu &=& \int_{\mathcal{S}^4}\lim_{T\rightarrow \infty}\frac{1}{T^4}\biggl\{& \int\limits_{C(4,{\bf S}^1)} d\overline{t} \int\limits_{\mathbb{R}^3\times\mathbb{R}^3} \Delta_{\nu_1\sigma_1}(\theta^{x_1}(t_1)-x_6) \Delta_{\nu_2\sigma_2}(\theta^{x_2}(t_2)-x_5)\\
	  &&&\times \Delta_{\nu_3\sigma_3}(\theta^{x_3}(t_3)-x_5) \Delta_{\nu_4\sigma_4}(\theta^{x_4}(t_4)-x_6) \Delta_{\nu_5\sigma_5}(x_5-x_6)\\
	  &&&\times X_{\theta^{x_1}(t_1)}^{\nu_1} X_{\theta^{x_2}(t_2)}^{\nu_2} X_{\theta^{x_3}(t_3)}^{\nu_3} X_{\theta^{x_4}(t_4)}^{\nu_4} (-1)^3 \varepsilon^{\sigma_2 \sigma_3 \nu_5}d^3x_5 \wedge \varepsilon^{\sigma_1 \sigma_4 \sigma_5}d^3x_6\\
	  &&&+\widetilde{B}_{31}\biggr\}\mu_{\Delta}.
      \label{intlambdaD31}	 
	  \numberthis
	  \end{alignat*}
	  
Now we proceed to study diagram (\ref{dibujo32}) in a similar way as in the previous case. We have
	  \begin{alignat*}{7}
 	  & && f_{D_{32},X}(\theta^{5}(\overline{x},(t_1,\ldots,t_5))) \\
 	  &=&& \left((\alpha^3_5)^*(\overline{\pi}_5)_*\Phi^*\overline{\omega}\right)_{\left(\theta^{5}(\overline{x},(t_1,\ldots,t_5))\right)}\left(X_{\theta^x(t_1)},X_{\theta^x(t_2)},X_{\theta^x(t_3)},X_{\theta^x(t_4)},X_{\theta^x(t_5)}\right) \\
 	  &=&& (\alpha^3_5)^*\left(\int\limits_{(\overline{\pi}_5)^{-1}(\theta^5(\overline{x},(t_1,\ldots,t_5)))}\Phi^*\overline{\omega}(\big[X_{\theta^x(t_1)}\big]_{\ell},\big[X_{\theta^x(t_2)}\big]_{\ell},\big[X_{\theta^x(t_3)}\big]_{\ell},\big[X_{\theta^x(t_4)}\big]_{\ell},\big[X_{\theta^x(t_5)}\big]_{\ell},\ldots)\right) \\
	  &=&& \int\limits_{(\overline{\pi}_5)^{-1}(\alpha^3_5\left(\theta^5(\overline{x},(t_1,\ldots,t_5))\right))}\Phi^*\overline{\omega}\biggl(\big[(\alpha^3_5)_*\left(X_{\theta^x(t_1)}\right)\big]_{\ell},\ldots,\big[(\alpha^3_5)_*\left(X_{\theta^x(t_5)}\right)\big]_{\ell},\ldots\biggr) \\ 
	  &=&& \int\limits_{\mathbb{R}^3}\Phi^*\overline{\omega}\biggl(\big[(\alpha^3_5)_*\left(X_{\theta^x(t_1)}\right)\big]_{\ell},\ldots,\big[(\alpha^3_5)_*\left(X_{\theta^x(t_5)}\right)\big]_{\ell},\ldots\biggr),
	  \numberthis
	  \label{lafvolic32}
	   \end{alignat*}
and then the asymptotic integral reads
	  \begin{alignat*}{7}
	  \int_{\mathcal{S}}\lambda_{D_{32}}\mu &=&& \int_{\mathcal{S}^5}\lim_{T\rightarrow \infty}\frac{1}{T^5}\bigg\{&& \int\limits_0^T \cdots\int\limits_0^T f_{D_{32},X}(\theta^{x_1}(t_1),\dots,\theta^{x_5}(t_5))dt_1\cdots\wedge dt_5\bigg\}\mu_{\Delta}\\
	  &=&&\int_{\mathcal{S}^4}\lim_{T\rightarrow \infty}\frac{1}{T^4}\bigg\{&& \int\limits_0^T \cdots\int\limits_0^T \int\limits_{\mathbb{R}^3}\Phi^*\overline{\omega}\biggl(\big[(\alpha^3_5)_*\left(X_{\theta^{x_1}(t_1)}\right)\big]_{\ell},\ldots,\big[(\alpha^3_5)_*\left(X_{\theta^{x_5}(t_5)}\right)\big]_{\ell},\ldots\biggr)\\
	  &&&&&\times dt_1\wedge \cdots\wedge dt_5\bigg\}\mu_{\Delta}.
	  \numberthis
	  \label{contribucionflujoD32}	  
	  \end{alignat*}

The inner part in the integrand can be evaluated and it yields 
	\begin{alignat*}{5}
	   & \phi^*\overline{\omega} \Biggl( \big[\left(\alpha_5^3\right)_*\left(X_{\theta^{x_1}(t_1)}\right)\big]_{\ell},\ldots\big[(\alpha^3_5)_*\left(X_{\theta^{x_5}(t_5)}\right)\big]_{\ell}, \ldots \Biggr) \\
	  =& (\phi_{1,6}^*\omega\wedge\phi_{2,5}^*\omega\wedge\phi_{3,6}^*\omega\wedge\phi_{4,6}^*\omega)\biggl(\big[\left(\alpha_5^3\right)_*\left(X_{\theta^{x_1}(t_1)}\right),\ldots,\big[(\alpha^3_5)_*\left(X_{\theta^{x_5}(t_5)}\right)\big]_{\ell},\ldots\biggr) \\
      =& \left[\frac{\varepsilon_{\mu_1\nu_1\sigma_1}}{4\pi}\frac{(\theta^{x_1}(t_1)-x_6)^{\mu_1}}{|\theta^{x_1}(t_1)-x_6|^3}{(\dot{\theta}^{x_1})}^{\nu_1} dx_6^{\sigma_1}\right]\wedge
	  \left[\frac{\varepsilon_{\mu_2\nu_2\sigma_2}}{4\pi}\frac{(\theta^{x_2}(t_2)-\theta^{x_5}(t_3))^{\mu_2}}{|\theta^{x_2}(t_2)-\theta^{x_5}(t_3)|^3}{(\dot{\theta}^{x_2})}^{\nu_2} {(\dot{\theta}^{x_5})}^{\sigma_2} \right]\\
      & \wedge \left[\frac{\varepsilon_{\mu_3\nu_3\sigma_3}}{4\pi}\frac{(\theta^{x_3}(t_4)-x_6)^{\mu_3}}{|\theta^{x_3}(t_4)-x_6|^3}{(\dot{\theta}^{x_3})}^{\nu_3} dx_6^{\sigma_3}\right]\wedge	  
	  \left[\frac{\varepsilon_{\mu_4\nu_4\sigma_4}}{4\pi}\frac{(\theta^{x_4}(t_5)-x_6)^{\mu_4}}{|\theta^{x_4}(t_5)-x_6|^3}{(\dot{\theta}^{x_6})}^{\nu_4} dx_6^{\sigma_4}\right]  	\\
	  =& \Delta_{\nu_1\sigma_1}(\theta^{x_1}(t_1)-x_6)\Delta_{\nu_2\sigma_2}(\theta^{x_2}(t_2)-\theta^{x_5}(t_3))\Delta_{\nu_3\sigma_3}(\theta^{x_3}(t_4)-x_6)\Delta_{\nu_4\sigma_4}(\theta^{x_4}(t_5)-x_6)\\
	   &\times  X_{\theta^{x_1}(t_1)}^{\nu_1}X_{\theta^{x_2}(t_2)}^{\nu_2}X_{\theta^{x_5}(t_3)}^{\sigma_2}X_{\theta^{x_3}(t_4)}^{\nu_3}X_{\theta^{x_4}(t_5)}^{\nu_4} dx_6^{\sigma_1}\wedge dx_6^{\sigma_3}\wedge dx_6^{\sigma_4} \\
	  =& \Delta_{\nu_1\sigma_1}(\theta^{x_1}(t_1)-x_6)\Delta_{\nu_2\sigma_2}(\theta^{x_2}(t_2)-\theta^{x_5}(t_3))\Delta_{\nu_3\sigma_3}(\theta^{x_3}(t_4)-x_6)\Delta_{\nu_4\sigma_4}(\theta^{x_4}(t_5)-x_6)\\
	   & \times X_{\theta^{x_1}(t_1)}^{\nu_1}X_{\theta^{x_2}(t_2)}^{\nu_2}X_{\theta^{x_5}(t_5)}^{\sigma_2}X_{\theta^{x_3}(t_3)}^{\nu_3}X_{\theta^{x_4}(t_4)}^{\nu_4} \varepsilon^{\sigma_1 \sigma_3 \sigma_4}d^3x_6.
	  \numberthis
	  \end{alignat*}	   
	  
From Eq. (\ref{intasyminv}) again, the average asymptotic integral reads
	  \begin{alignat*}{5}
	  \int_{\mathcal{S}}\lambda_{D_{32}}\mu &=& \int_{\mathcal{S}^5}\lim_{T\rightarrow \infty}\frac{1}{T^5}\biggl\{& \int\limits_{C(5,{\bf S}^1)} d\overline{t} \int\limits_{\mathbb{R}^3} \Delta_{\nu_1\sigma_1}(\theta^{x_1}(t_1)-x_6) \Delta_{\nu_2\sigma_2}(\theta^{x_2}(t_2)-\theta^{x_5}(t_3))\\
	  &&&\times \Delta_{\nu_3\sigma_3}(\theta^{x_3}(t_4)-x_6) \Delta_{\nu_4\sigma_4}(\theta^{x_4}(t_5)-x_6) X_{\theta^{x_1}(t_1)}^{\nu_1} X_{\theta^{x_2}(t_2)}^{\nu_2} X_{\theta^{x_5}(t_3)}^{\sigma_2}\\
	  &&&\times X_{\theta^{x_3}(t_4)}^{\nu_3} X_{\theta^{x_4}(t_5)}^{\nu_4} \varepsilon^{\sigma_1\sigma_3\sigma_4}d^3 x_6+\widetilde{B}_{32}\biggr\}\mu_{\Delta},
      \label{intlambdaD32}	  
	  \numberthis
	  \end{alignat*}
where $\widetilde{B}_{32}$ takes into account all the boundary terms.

For diagrams \ref{dibujo33} and \ref{dibujo34} that just have points on the knot the expressions for Eq. (\ref{intasyminv}) 
(use the notation $X_{\theta^x(\overline{t})}=(X_{\theta^x(t_1)},\ldots,X_{\theta^x(t_n)})$ and consider $\overline{\pi}_p=\text{Id}$) are the same, namely,
	  \begin{alignat*}{7}
 	  f_{D_{33},X}(\theta^{6}(\overline{x}(t_1,\ldots,t_6))) &=&& \left((\alpha^3_6)^*(\overline{\pi}_6)_*\Phi^*\overline{\omega}\right)_{\left(\theta^{6}(\overline{x},(t_1,\ldots,t_6))\right)}\left(X_{\theta^6(\overline{x},(t_1,\ldots,t_6))}\right)\\
 	  &=&&\left(\overline{\pi}_6\right)_*\Phi^*\overline{\omega}_{\alpha^3_6(\theta^6(\overline{x},(t_1,\ldots,t_6)))}\left(X_{\theta^x(\overline{t})}\right)\\
 	  &=&&\phi^*\overline{\omega}_{\theta^6(\overline{x},(t_1,\ldots,t_6))}\left(X_{\theta^x(\overline{t})}\right)+ \widetilde{B}_{33}\\
 	  &=&&(\phi_{1,4}^*\omega\wedge\phi_{2,6}^*\omega\wedge\phi_{3,5}^*\omega)\left(X_{\theta^x(\overline{t})}\right)+\widetilde{B}_{33},
 	  \numberthis
 	  \end{alignat*}
and
 	  \begin{alignat*}{7}
 	  f_{D_{34},X}(\theta^{6}(\overline{x},(t_1,\ldots,t_6))) &=&& \left((\alpha^3_6)^*(\overline{\pi}_6)_*\Phi^*\overline{\omega}\right)_{\left(\theta^{6}(\overline{x}(t_1,\ldots,t_6))\right)}\left(X_{\theta^x(\overline{t})}\right)\\
 	  &=&&\left(\overline{\pi}_6\right)_*\Phi^*\overline{\omega}_{\alpha^3_6(\theta^6(\overline{x},(t_1,\ldots,t_6)))}\left(X_{\theta^x(\overline{t})}\right)\\
 	  &=&&\phi^*\overline{\omega}_{\theta^6(\overline{x},(t_1,\ldots,t_6))}\left(X_{\theta^x(\overline{t})}\right)+ \widetilde{B}_{34}\\
 	  &=&&(\phi_{1,4}^*\omega\wedge\phi_{2,5}^*\omega\wedge\phi_{3,6}^*\omega)\left(X_{\theta^x(\overline{t})}\right)+ \widetilde{B}_{34},
 	  \numberthis
 	  \end{alignat*}
where $\widetilde{B}_{33}$ and $\widetilde{B}_{34}$ are the corresponding contributions of the boundary terms.

After evaluating $f_{D_{33}}$ and $f_{D_{34}}$ in the $6$-fold integral \ref{intlambdaeqintf}, $\phi^*\overline{\omega}_{\theta^6(\overline{t})}\left(X_{\theta^x(\overline{t})}\right)$ is respectively given by
	  \begin{alignat*}{7}
	  &&&\phi^*\overline{\omega}\left(X_{\theta^{x_1}(t_1)},\ldots,X_{\theta^{x_6}(t_6)}\right)\\
	  &=&&(\phi_{1,4}^*\omega\wedge\phi_{2,6}^*\omega\wedge\phi_{3,5}^*\omega)\left(X_{\theta^{x_1}(t_1)},\ldots,X_{\theta^{x_6}(t_6)}\right)\\
 	  &=&&\left[\frac{\varepsilon_{\mu_1\nu_1\sigma_1}}{4\pi}\frac{(\theta^{x_1}(t_1)-\theta^{x_4}(t_2))^{\mu_1}}{|\theta^{x_1}(t_1)-\theta^{x_4}(t_2)|^3}{(\dot{\theta}^{x_1})}^{\nu_1}{(\dot{\theta}^{x_4})}^{\sigma_1}\right]
	  \left[\frac{\varepsilon_{\mu_2\nu_2\sigma_2}}{4\pi}\frac{(\theta^{x_2}(t_3)-\theta^{x_6}(t_4))^{\mu_2}}{|\theta^{x_2}(t_3)-\theta^{x_6}(t_4)|^3}{(\dot{\theta}^{x_2})}^{\nu_2}{(\dot{\theta}^{x_6})}^{\sigma_2} \right] \\
	  &&& \times \left[\frac{\varepsilon_{\mu_3\nu_3\sigma_3}}{4\pi}\frac{(\theta^{x_3}(t_5)-\theta^{x_5}(t_6))^{\mu_3}}{|\theta^{x_3}(t_5)-\theta^{x_5}(t_6)|^3}{(\dot{\theta}^{x_3})}^{\nu_3}{(\dot{\theta}^{x_5})}^{\sigma_3}\right] \\ 
	  &=&& \Delta_{\nu_1\sigma_1}(\theta^{x_1}(t_1)-\theta^{x_4}(t_2))\Delta_{\nu_2\sigma_2}(\theta^{x_2}(t_3)-\theta^{x_6}(t_4))\Delta_{\nu_3\sigma_3}(\theta^{x_3}(t_5)-\theta^{x_5}(t_6))\\
	  &&& \times X_{\theta^{x_1}(t_1)}^{\nu_1}X_{\theta^{x_4}(t_2)}^{\sigma_1}X_{\theta^{x_2}(t_3)}^{\nu_2}X_{\theta^{x_6}(t_4)}^{\sigma_2}X_{\theta^{x_3}(t_5)}^{\nu_3}X_{\theta^{x_5}(t_6)}^{\sigma_3},
	  \numberthis
	  \label{lafvolic33}
	   \end{alignat*}
and
	  \begin{alignat*}{7}
 	  &&& \phi^*\overline{\omega}\left(X_{\theta^{x_1}(t_1)},\ldots,X_{\theta^{x_6}(t_6)}\right)\\
 	  &=&&(\phi_{1,4}^*\omega\wedge\phi_{2,5}^*\omega\wedge\phi_{3,6}^*\omega)\left(X_{\theta^{x_1}(t_1)},\ldots,X_{\theta^{x_6}(t_6)}\right)\\ 
	  &=&& \left[\frac{\varepsilon_{\mu_1\nu_1\sigma_1}}{4\pi}\frac{(\theta^{x_1}(t_1)-\theta^{x_4}(t_2))^{\mu_1}}{|\theta^{x_1}(t_1)-\theta^{x_4}(t_2)|^3}{(\dot{\theta}^{x_1})}^{\nu_1}{(\dot{\theta}^{x_4})}^{\sigma_1}\right]
	  \left[\frac{\varepsilon_{\mu_2\nu_2\sigma_2}}{4\pi}\frac{(\theta^{x_2}(t_3)-\theta^{x_5}(t_4))^{\mu_2}}{|\theta^{x_2}(t_3)-\theta^{x_5}(t_4)|^3}{(\dot{\theta}^{x_2})}^{\nu_2}{(\dot{\theta}^{x_5})}^{\sigma_2} \right] \\
	  &&&  \times \left[\frac{\varepsilon_{\mu_3\nu_3\sigma_3}}{4\pi}\frac{(\theta^{x_3}(t_5)-\theta^{x_6}(t_6))^{\mu_3}}{|\theta^{x_3}(t_5)-\theta^{x_6}(t_6)|^3}{(\dot{\theta}^{x_3})}^{\nu_3}{(\dot{\theta}^{x_6})}^{\sigma_3}\right] \\ 
	  &=&& \Delta_{\nu_1\sigma_1}(\theta^{x_1}(t_1)-\theta^{x_4}(t_2))\Delta_{\nu_2\sigma_2}(\theta^{x_2}(t_3)-\theta^{x_5}(t_4))\Delta_{\nu_3\sigma_3}(\theta^{x_3}(t_5)-\theta^{x_6}(t_6))\\
	  &&&\times  X_{\theta^{x_1}(t_1)}^{\nu_1}X_{\theta^{x_4}(t_2)}^{\sigma_1}X_{\theta^{x_2}(t_3)}^{\nu_2}X_{\theta^{x_5}(t_4)}^{\sigma_2}X_{\theta^{x_3}(t_5)}^{\nu_3}X_{\theta^{x_6}(t_6)}^{\sigma_3}.
	  \numberthis
	  \label{lafvolic34}
	   \end{alignat*}
	   
Thus, the average asymptotic integrals are then given as 
	  \begin{alignat*}{5}
	  \int_{\mathcal{S}}\lambda_{D_{33}}\mu
	  &=&& \int_{\mathcal{S}^6}\lim_{T\rightarrow \infty}\frac{1}{T^6}\biggl\{\int\limits_{C(6,{\bf S}^1)} d\overline{t} \Delta_{\nu_1\sigma_1}(\theta^{x_1}(t_1)-\theta^{x_4}(t_2)) \Delta_{\nu_2\sigma_2}(\theta^{x_2}(t_3)-\theta^{x_6}(t_4)) \\
	  & && \times \Delta_{\nu_3\sigma_3}(\theta^{x_3}(t_5)-\theta^{x_5}(t_6)) X_{\theta^{x_1}(t_1)}^{\nu_1}X_{\theta^{x_4}(t_2)}^{\sigma_1}X_{\theta^{x_2}(t_3)}^{\nu_2}X_{\theta^{x_6}(t_4)}^{\sigma_2}X_{\theta^{x_3}(t_5)}^{\nu_3}X_{\theta^{x_5}(t_6)}^{\sigma_3} \\
	  & && + \widetilde{B}_{33} \biggr\}\mu_{\Delta},
      \label{intlambdaD33}	  
	  \numberthis
	  \end{alignat*}
and
	  \begin{alignat*}{5}
	  \int_{\mathcal{S}}\lambda_{D_{34}}\mu
	  &=&& \int_{\mathcal{S}^6}\lim_{T\rightarrow \infty}\frac{1}{T^6}\biggl\{\int\limits_{C(6,{\bf S}^1)} d\overline{t} \Delta_{\nu_1\sigma_1}(\theta^{x_1}(t_1)-\theta^{x_4}(t_2)) \Delta_{\nu_2\sigma_2}(\theta^{x_2}(t_3)-\theta^{x_5}(t_4))\\
	  & && \times \Delta_{\nu_3\sigma_3}(\theta^{x_3}(t_5)-\theta^{x_6}(t_6)) X_{\theta^{x_1}(t_1)}^{\nu_1}X_{\theta^{x_4}(t_2)}^{\sigma_1}X_{\theta^{x_2}(t_3)}^{\nu_2}X_{\theta^{x_5}(t_4)}^{\sigma_2}X_{\theta^{x_3}(t_5)}^{\nu_3}X_{\theta^{x_6}(t_6)}^{\sigma_3} \\
	  & && + \widetilde{B}_{34}\biggr\}\mu_{\Delta}.
      \label{intlambdaD34}	  
	  \numberthis
	  \end{alignat*}
where as stated at the begining of the section boundary cancellations lead to the average asymptotic Vassiliev invariants  as the sum of \ref{intlambdaD31}, \ref{intlambdaD32}, \ref{intlambdaD33} and \ref{intlambdaD34}.

\section{Final Remarks}
In this article we pursued the implementation of the procedure
followed in Ref. \cite{VV94}, which define the asymptotic Jones-Witten invariants, to find
invariants for flows,  or triplets $(M_3,{\cal F}, \mu)$, in the context of Chern-Simons perturbative theory. In this
situation the invariants of interest are the Vassiliev invariants of knots and links. 
The traditional way of obtaining asymptotic invariants is to give a partial foliation with leaves of certain dimension \cite{SA,RS,S}. One can endow the manifold with a collection of flow boxes and orient the set of flow boxes along the foliation. On the transversal submanifolds to the leaves give a transverse Borel measure of the foliation preserved by the flow. These data gives rise to a {\it geometric current} and it used as an object dual to differential forms defined on the leaves of the foliation. That determines a homology cycle dependent on the flow (or vector field) and an invariant transverse probability measure with respect to the flow. Consequently in order to define topological invariants for flows we integrate differential forms on the transverse measure. For Jones-Witten theory it is precisely the case  \cite{VV94}, where observables are defined as integrals of differential forms on $M_3$ over asymptotic one-dimensional asymptotic cycles.   In the original version of perturbative Chern-Simons theory it was very difficult to write Vassiliev invariants as integrals of certain differential forms.  However the formulation of perturbative Chern-Simons theory using Bott-Taubes integrals on configuration spaces \cite{BT94}, gives rise in a natural way of determining the cohomology of such spaces and in consequence the Vassiliev invariants can be easily rewritten as integrals of certain differential forms on these spaces.  

In the present paper we study in a systematic way the correspondence between Feynman diagrams in perturbative Chern-Simons theory and 
the associated Bott-Taubes integrals. For the Feynman diagrams of order one in $1/k$, we regain the self-linking number (\ref{selflinkinconfiguration01}). For second order in the expansion of $1/k$ there are two relevant contributions to the Vassiliev invariant which come from Eqs. (\ref{Yconfiguration01}) and (\ref{configuration22}). For order three, there are four diagrams which contribute to the Vassiliev invariant. There are four Bott-Taubes integrals given in Eqs. (\ref{31configuration01}),  (\ref{32configuration01}), (\ref{33configuration01}), (\ref{34configuration01}). The analysis of the first order contribution for Feynman diagrams having all the marked points lying on the knot and the discussion of one second order diagram with three points lying on the knot and one outside from it was worked out in Ref.  \cite{Thu99}. In the present paper we obtained the Vassiliev invariant at the second order in perturbation theory constructed from the relevant diagrams $D_{21}$ and $D_{22}$. This was obtained after a proper discussion of the behavior of the boundary terms the Vassiliev invariant.  Moreover, we go further to third order and obtain the corresponding Vassiliev invariant; an analysis of the boundary terms of the Bott-Taubes integrals is also discussed. The problem arising in the computation of the Jones-Witten invariants for flows \cite{VV94} involving the distinction between the Abelian and non-Abelian cases does not appear here. Even if we are discussing 
the non-Abelian case, the Vassiliev invariants are obtained as a perturbative series and then the exponential 
in the Wilson loop operators are expanded leaving all the terms as Lie algebra valued objects.

We have used the previous results and the advantage of writing Vassiliev invariants as Bott-Taubes integrals
in order to introduce flows on the underlying manifold. Thus we were able to incorporate easily the non-singular and non-divergence smooth vector field $X$ on $M_3$ ($\mathbb{R}^3$ or ${\bf S}^3$) to obtain invariants of triplets $(M_3,{\cal F}, \mu)$. This approach was followed in Ref. \cite{KV16} to compute some asymptotic Vassiliev invariants, namely the asymptotic self-linking number was obtained. This invariant was obtained at higher-order with all marked points lying on the knot diagrams. For the first order in $1/k$ the asymptotic Vassiliev invariant corresponds precisely with the asymptotic self-linking number or helicity (\ref{helicity}) obtained in \cite{KV16}. Furthermore, at the second order there are two contributions to the asymptotic Vassiliev invariant, which is given by the sum of  Eqs. (\ref{intlambdaD22}) and  (\ref{contribucionflujoD21}). The boundary terms cancell by the same reason that in the case without flows.  Moreover, the asymptotic third order Vassiliev invariant is given by the superposition of four average integrals in Eqs. (\ref{intlambdaD31}), (\ref{intlambdaD32}), (\ref{intlambdaD33}) and (\ref{intlambdaD34}). From the previous results it is clear that Vassiliev invariants obtained from higher-order diagrams in Chern-Simons theory will be constructed following a similar procedure. An algorithm for the construction of any order diagrams is not given here and is a subject of future work. Also, it would be interesting to generalize these explicit constructions to the case of the two component links described in section \ref{Feynman diagrams for links}. It has to be noticed that the match between amplitudes coming directly from perturbative Chern-Simons theory and those arising from Bott-Taubes integrals in configuration spaces is given in this work up to signature. The reason is that Chern-Simons theory was expressed in Lorentzian signature while Bott-Taubes integration is assuming Euclidean signature.

Finally, it is worth mentioning that knot and link invariants can be categorified in terms of Khovanov homology. In particular, the Jones polynomial was discussed in Ref. \cite{Khovanov:1999qla}. The physical approach in terms of gauge theory and brane theory was studied, for instance, in \cite{Witten:2011zz,WItten:2011pz,Witten:2014xwa,Gukov:2012jx}. As a future work it would be interesting to find the asymptotic version of the categorified Jones polynomial. Moreover, a categorified version of the Vassiliev invariants appeared very recently in Refs. \cite{Ito:2019plv,Yoshida}. One would ask whether this categorified version does admit an asymptotic version and if the formalism of Bott-Taubes integrals will play an important role also in this case.   

\vskip 1truecm
\centerline{\bf Acknowledgments}

It is a pleasure to thank Ernesto Lupercio, Jacob Mostovoy, Roberto Santos-Silva, Alberto Verjovsky, Ricardo Vila-Freyer and Alejandro Xicoténcatl for their interest in the work and many fruitful discussions and suggestions. The work of J. de-la-Cruz-Moreno and E. López. was supported by a CONACyT graduate fellowship. 

\begin{appendices}
\section{Pullback bundle} \label{appendixA}

A pullback (or fibre product) of a pair $(f:X\rightarrow Y,g:Z\rightarrow Y)$ is a subspace of the product $X\times Z$ defined by
	  \begin{equation}
			X\times_Y Z:=\{ (x,z)\in X\times Z\;|\; f(x)=g(z) \}.
	  \end{equation}
	  
Then by considering the projections $\pi_1:X\times Z\rightarrow X$ and $\pi_2:X\times Z\rightarrow Z$ from $X\times Z$ into their first and second coordinates, the restriction of this maps to the fibre product
	  \begin{equation}
	  	pr_1=\pi_1|_{X\times_Y Z}:X\times_Y Z\rightarrow X, \qquad pr_2=\pi_2|_{X\times_Y Z}:X\times_Y Z\rightarrow X,
	  \end{equation}
make diagram in figure \ref{diagramapullback} commutative.

	  \begin{figure}[ht]
		\begin{center}
      		\[ \xymatrix{
      		Y & & Z  \ar[ll]^{g} \\
        	& & \\
      		X \ar[uu]^{f} & & X\times_Y Z \ar[ll]^{pr_1}\ar[uu]_{pr_2} \pullbackcorner \\
      		} \]
      	\caption{Pullback of a pair $(f,g)$.}
      	\label{diagramapullback}
	  \end{center}	
	  \end{figure}

In the case that $f:X\rightarrow Y$ is a kind of bundle and $g:Z\rightarrow Y$ is a morphism between the spaces, then the fibre product is usually denoted by $g^*X$ and $pr_2:g^*X\rightarrow Z$ is called {\it the pullback bundle of the bundle $f$ over Z} \cite{BottandTu}.
Since this is a commutative diagram, the icon inside the diagram is a usual notation to identify the fibre product $X\times_Y Z$ and the corner of this icon indicates the direction of all the arrows in the diagram.

\section{Integration along the fibres} \label{appendixB}

Let $\pi:E\rightarrow B$ be a smooth fiber bundle with homotopy compact fiber $F_b:=\pi^{-1}\left(\left\lbrace b\right\rbrace\right)\simeq F$ with dim$(F)=n$. Let $\omega\in\Omega^k\left(E\right)$. There is a map $\pi_*:\Omega^k\left(E\right)\rightarrow \Omega^{k-n}\left(B\right)$
called {\it integration along the fiber of $\pi$} given by 
      \begin{equation}
      \left(\pi_*\omega\right)_b\left(V_b^1,V_b^2,\ldots,V_b^{k-n}\right)=\int\limits_{F_b}i^*\omega_\pi,
      \end{equation}
where $\omega_\pi$ is an $n$-form in the total space $E$ whose pullback through the inclusion map \\ $i:F_b\hookrightarrow E$ is now an $n$-form in the fiber $F_b$ which is given, for a point $p\in\pi^{-1}(\{b\})$, by 
      \begin{equation}
      \left(i^*(\omega_\pi)\right)_p(W_1,\ldots,W_{n}):=\omega\left(W_1,\ldots,W_{n},\big[{V_b^1}\big]_{\ell},\ldots,\big[{V_b^{k-n}}\big]_{\ell}\right),
      \end{equation}
with $\big[{V_b^i}\big]_{\ell}\in T_pE$ any lift of the tangent vector $V_b^i\in T_bB$ and $\{W_1,\ldots,W_n\}$ a set of vectors tangent to $F_b$ at the point $p$.

To ensure that this definition is independent on the choice of the specific lift consider two different lifts $V$ and $V'$ of $V_b^i$ over the point $p\in F_b$. Since both of them are lifts then
      \begin{alignat*}{5}
      d\pi_p\left(V-V'\right)&=&V_b^i-V_b^i=0, \numberthis
      \end{alignat*} 
thus $V-V'\in \text{Ker}\left(d\pi_p\right)=T_a\pi^{-1}(\{b\})$. Now the set $\left\lbrace W_1,\ldots,W_n,\big[{V}\big]_{\ell}-\big[{V}'\big]_{\ell}\right\rbrace$ 
with $n+1$ different tangent vectors on $\pi^{-1}(\{b\})$ (whose dimension is $n$) has to be linearly dependent and since $\omega$ is an alternating
tensor then
      \begin{equation}
      \omega\left(W_1,\ldots,W_{n-k},\big[{V_b^1}]_{\ell},\ldots,\big[V_b^i]_{\ell}-\big[{V}']_{\ell},\ldots,\big[{V_b^{k-n}}]_{\ell}\right)=0.
      \end{equation}
      The previous equation asserts that $\pi_*\omega$ is independent of the choice of the lift of the tangent vectors \cite{BottandTu}.

\section{Gauss map pullback} \label{appendixC}

Explicit calculation will be done here for points $s_1$ and $x_4$ in diagram of figure \ref{laydibujo}. Note that a generalization for any two pair of points in any diagram is straightforward.

The volume form in ${\bf S}^2$ can be taken as \cite{Thu99}
      \begin{equation}
      \omega = \frac{\varepsilon_{\mu \nu \sigma}}{8\pi} \frac{x^{\mu} dx^{\nu} \wedge dx^{\sigma}}{|\overline{x}|^3}.
      \label{omegareducidoappendix}
      \end{equation}

It will be useful to write $\omega$ explicitly in the coordinate system $\{ x, y, z \}$ in $\mathbb{R}^3$ as
      \begin{align*}
      \omega =& \frac{1}{4\pi} \left[ \frac{x dy \wedge dz - y dx \wedge dz + z dx \wedge dy}{\left( x^2 + y^2 + z^2 \right)^{3/2}} \right] \\ \\
	     =& \omega_{12}((x,y,z)) dx \wedge dy + \omega_{23}((x,y,z)) dy \wedge dz + \omega_{13}((x,y,z)) dx \wedge dz, \numberthis
      \label{omegaexplicitoappendix}
      \end{align*}
where the coefficient functions are given by
      \begin{subequations}
      \begin{align*}
      \omega_{12} : {\bf S}^2 \longrightarrow \mathbb{R} & \\
      \omega_{12} ((x,y,z)) &= \frac{1}{4\pi} \left[ \frac{z}{\left( x^2 + y^2 + z^2 \right)^{3/2}} \right], \numberthis \\ \
      \omega_{23} : {\bf S}^2 \longrightarrow \mathbb{R} & \\
      \omega_{23} ((x,y,z)) &= \frac{1}{4\pi} \left[ \frac{x}{\left( x^2 + y^2 + z^2 \right)^{3/2}} \right], \numberthis \\ \
      \omega_{13} : {\bf S}^2 \longrightarrow \mathbb{R} & \\
      \omega_{13} ((x,y,z)) &= \frac{1}{4\pi} \left[ \frac{-y}{\left( x^2 + y^2 + z^2 \right)^{3/2}} \right]. \numberthis
      \end{align*}
      \end{subequations}
      
Remember that the Gauss map 
      \begin{equation}
      \phi : C(3+1, {\bf S}^3) \longrightarrow {\bf S}^2 \times {\bf S}^2 \times {\bf S}^2
      \label{gaussphiappendix}
      \end{equation}
factors for this diagram as
      \begin{equation}
      \phi = \phi_{1,4} \times \phi_{2,4} \times \phi_{3,4},
      \end{equation}
where the indices refer to the coordinate system $\{ x_1, x_2, x_3, x_4 \}$ on $C(3+1, {\bf S}^3)$ seen as a subset of 
${\bf S}^3 \times {\bf S}^3 \times {\bf S}^3 \times {\bf S}^3$.
Each of these coordinates has three indices (for example $x_1$ represents the coordinates $\{ x_1^1, x_1^2, x_1^3 \}$ in the first ${\bf S}^3$ factor)
thus the coordinate system on $C(3+1, {\bf S}^3)$ is really taken to be 
$\{ x_1^1, x_1^2, x_1^3, x_2^1, x_2^2, x_2^3, x_3^1, x_3^2, x_3^3, x_4^1, x_4^2, x_4^3 \}$.

The factors of $\phi$ are explicitly given by
      \begin{subequations}
      \begin{align*}
      \phi_{1,4} : C(3+1, {\bf S}^3) \longrightarrow {\bf S}^2 & \\
      \phi_{1,4} ((x_1, x_2, x_3, x_4)) &= \frac{x_4 - x_1}{|x_4 - x_1|}
      = \frac{(x_4^1 - x_1^1, x_4^2 - x_1^2, x_4^3 - x_1^3)}{\left[ (x_4^1 - x_1^1)^2 + (x_4^2 - x_1^2)^2 + (x_4^3 - x_1^3)^2 \right]^{1/2}},
      \label{phi14explicitappendix} \numberthis \\ \\
      \phi_{2,4} : C(3+1, {\bf S}^3) \longrightarrow {\bf S}^2 & \\
      \phi_{2,4} ((x_1, x_2, x_3, x_4)) &= \frac{x_4 - x_2}{|x_4 - x_2|}
      = \frac{(x_4^1 - x_2^1, x_4^2 - x_2^2, x_4^3 - x_2^3)}{\left[ (x_4^1 - x_2^1)^2 + (x_4^2 - x_2^2)^2 + (x_4^3 - x_2^3)^2 \right]^{1/2}}, \numberthis \\ \\
      \phi_{3,4} : C(3+1,{\bf S}^3) \longrightarrow {\bf S}^2 & \\
      \phi_{3,4} ((x_1, x_2, x_3, x_4)) &= \frac{x_4 - x_3}{|x_4 - x_3|}
      = \frac{(x_4^1 - x_3^1, x_4^2 - x_3^2, x_4^3 - x_3^3)}{\left[ (x_4^1 - x_3^1)^2 + (x_4^2 - x_3^2)^2 + (x_4^3 - x_3^3)^2 \right]^{1/2}}. \numberthis
      \end{align*}
      \end{subequations}
      
In what follows the function $\phi_{1,4}$ is studied in detail. First write
      \begin{equation}
      \phi_{1,4} ((x_1, x_2, x_3, x_4)) = (\phi_x ((x_1, x_4)), \phi_y ((x_1, x_4)), \phi_z ((x_1, x_4))),
      \end{equation}
where (see (\ref{phi14explicitappendix}))
      \begin{subequations}
      \begin{align*}
      \phi_x ((x_1, x_4)) &= \frac{x_4^1 - x_1^1}{\left[ (x_4^1 - x_1^1)^2 + (x_4^2 - x_1^2)^2 + (x_4^3 - x_1^3)^2 \right]^{1/2}}, \label{phixexplicitappendix} \numberthis \\ \\
      \phi_y ((x_1, x_4)) &= \frac{x_4^2 - x_1^2}{\left[ (x_4^1 - x_1^1)^2 + (x_4^2 - x_1^2)^2 + (x_4^3 - x_1^3)^2 \right]^{1/2}}, \label{phiyexplicitappendix} \numberthis \\ \\
      \phi_z ((x_1, x_4)) &= \frac{x_4^3 - x_1^3}{\left[ (x_4^1 - x_1^1)^2 + (x_4^2 - x_1^2)^2 + (x_4^3 - x_1^3)^2 \right]^{1/2}}. \label{phizexplicitappendix} \numberthis
      \end{align*}
      \end{subequations}
   
We assume that (\ref{omegaexplicitoappendix}) is the volume form in the first ${\bf S}^2$ factor of the codomain in (\ref{gaussphiappendix}); then its pullback to $C(3+1, {\bf S}^3)$ under $\phi_{1,4}$ is given by
      \begin{align*}
      \phi_{1,4}^* \omega = \omega_{12}(\phi_{1,4}((x_1, x_2, x_3, x_4))) d\phi_x \wedge d\phi_y + \omega_{23}(\phi_{1,4}((x_1, x_2, x_3, x_4))) d\phi_y \wedge d\phi_z & \\ \\
      + \omega_{13}(\phi_{1,4}((x_1, x_2, x_3, x_4))) d\phi_x \wedge d\phi_z. & \numberthis
      \end{align*}
      
By defining
      \begin{equation}
      \Theta \equiv (x_4^1 - x_1^1)^2 + (x_4^2 - x_1^2)^2 + (x_4^3 - x_1^3)^2,
      \end{equation}
the above equation reads
      \begin{alignat*}{4}
      \phi_{1,4}^* \omega &=& & \omega_{12} \left(  \frac{x_4^1 - x_1^1}{\Theta^{1/2}}, \frac{x_4^2 - x_1^2}{\Theta^{1/2}}, \frac{x_4^3 - x_1^3}{\Theta^{1/2}} \right)  d\phi_x \wedge d\phi_y \\ \\
			 & &+& \omega_{23} \left( \frac{x_4^1 - x_1^1}{\Theta^{1/2}}, \frac{x_4^2 - x_1^2}{\Theta^{1/2}}, \frac{x_4^3 - x_1^3}{\Theta^{1/2}}  \right) d\phi_y \wedge d\phi_z \\ \\       
			 & &+& \omega_{13} \left( \frac{x_4^1 - x_1^1}{\Theta^{1/2}}, \frac{x_4^2 - x_1^2}{\Theta^{1/2}}, \frac{x_4^3 - x_1^3}{\Theta^{1/2}} \right) d\phi_x \wedge d\phi_z \\ \\
      &=& & \frac{1}{4\pi} \left[ \frac{x_4^3 - x_1^3}{\Theta^{1/2}} \right] d\phi_x \wedge d\phi_y
      + \frac{1}{4\pi} \left[ \frac{x_4^1 - x_1^1}{\Theta^{1/2}} \right] d\phi_y \wedge d\phi_z
      - \frac{1}{4\pi} \left[ \frac{x_4^2 - x_1^2}{\Theta^{1/2}} \right] d\phi_x \wedge d\phi_z. \numberthis
      \label{pullback14appendix}
      \end{alignat*}

The next step to write $\phi_{1,4}^* \omega$ explicitly is to analyse the forms $d\phi_x \wedge d\phi_y$, $d\phi_y \wedge d\phi_z$ and
$d\phi_x \wedge d\phi_z$ with $\phi_x$, $\phi_y$ 
and $\phi_z$ defined from (\ref{phixexplicitappendix}) to (\ref{phizexplicitappendix}). Due to the fact that these functions do not depend on coordinates with subindices $2$ and $3$ the following simplifications apply
      \begin{subequations}
      \begin{alignat*}{4}
      d\phi_x 
	      &=& & \frac{\partial \phi_x}{\partial x_1^1} dx_1^1 + \frac{\partial \phi_x}{\partial x_1^2} dx_1^2 + \frac{\partial \phi_x}{\partial x_1^3} dx_1^3  
	      + \frac{\partial \phi_x}{\partial x_4^1} dx_4^1 + \frac{\partial \phi_x}{\partial x_4^2} dx_4^2 + \frac{\partial \phi_x}{\partial x_4^3} dx_4^3, \numberthis \\ \\
      d\phi_y &=& & \frac{\partial \phi_y}{\partial x_1^1} dx_1^1 + \frac{\partial \phi_y}{\partial x_1^2} dx_1^2 + \frac{\partial \phi_y}{\partial x_1^3} dx_1^3  
	      + \frac{\partial \phi_y}{\partial x_4^1} dx_4^1 + \frac{\partial \phi_y}{\partial x_4^2} dx_4^2 + \frac{\partial \phi_y}{\partial x_4^3} dx_4^3, \numberthis \\ \\     
      d\phi_z &=& & \frac{\partial \phi_z}{\partial x_1^1} dx_1^1 + \frac{\partial \phi_z}{\partial x_1^2} dx_1^2 + \frac{\partial \phi_z}{\partial x_1^3} dx_1^3  
	      + \frac{\partial \phi_z}{\partial x_4^1} dx_4^1 + \frac{\partial \phi_z}{\partial x_4^2} dx_4^2 + \frac{\partial \phi_z}{\partial x_4^3} dx_4^3. \numberthis
      \end{alignat*}
      \end{subequations}
      
It is clear from the above equations that the forms $d\phi_x \wedge d\phi_y$, $d\phi_y \wedge d\phi_z$ and $d\phi_x \wedge d\phi_z$ have many mixed terms. In what follows the interest will be concentrated in coordinates $x_4^1$ and $x_4^2$, {\it i.e.}, just the part $dx_4^1 \wedge dx_4^2$ of $\phi_{1,4}^* \omega$ will be analysed. The notation for this part will be $\left[ \phi_{1,4}^* \omega \right]_{4,4}^{1,2}$. Thus from (\ref{pullback14appendix}) we have
      \begin{alignat*}{4}
      \left[ \phi_{1,4}^* \omega \right]_{4,4}^{1,2} &=& & \frac{1}{4\pi} \left[ \frac{x_4^3 - x_1^3}{\Theta^{1/2}} \right] \left[ d\phi_x \wedge d\phi_y \right]_{4,4}^{1,2}
						    + \frac{1}{4\pi} \left[ \frac{x_4^1 - x_1^1}{\Theta^{1/2}} \right] \left[ d\phi_y \wedge d\phi_z \right]_{4,4}^{1,2} \\
						    & &-& \frac{1}{4\pi} \left[ \frac{x_4^2 - x_1^2}{\Theta^{1/2}} \right] \left[ d\phi_x \wedge d\phi_z \right]_{4,4}^{1,2}, \numberthis               
      \label{pullback144412appendix}
      \end{alignat*}
where
      \begin{subequations}
      \begin{alignat*}{4}
      \left[ d\phi_x \wedge d\phi_y \right]_{4,4}^{1,2} &=& & \frac{\partial \phi_x}{\partial x_4^1} \frac{\partial \phi_y}{\partial x_4^2} dx_4^1 \wedge dx_4^2 
			     - \frac{\partial \phi_x}{\partial x_4^2} \frac{\partial \phi_y}{\partial x_4^1} dx_4^1 \wedge dx_4^2 \\ \\
      &=& & \displaystyle \frac{1}{\Theta^3} \left[\Theta^2 - \Theta (x_4^1 - x_1^1)^2 - \Theta (x_4^2 - x_1^2)^2 \right] dx_4^1 \wedge dx_4^2, \numberthis \\ \\
      \left[ d\phi_y \wedge d\phi_z \right]_{4,4}^{1,2} &=& & \frac{\partial \phi_y}{\partial x_4^1} \frac{\partial \phi_z}{\partial x_4^2} dx_4^1 \wedge dx_4^2 
			     - \frac{\partial \phi_y}{\partial x_4^2} \frac{\partial \phi_z}{\partial x_4^1} dx_4^1 \wedge dx_4^2 \\ \\
      &=& & \displaystyle \frac{1}{\Theta^3} \left[ \Theta (x_4^1 - x_1^1)(x_4^3 - x_1^3) \right] dx_4^1 \wedge dx_4^2, \numberthis \\ \\
      \left[ d\phi_x \wedge d\phi_z \right]_{4,4}^{1,2} &=& & \frac{\partial \phi_x}{\partial x_4^1} \frac{\partial \phi_z}{\partial x_4^2} dx_4^1 \wedge dx_4^2 
			     - \frac{\partial \phi_x}{\partial x_4^2} \frac{\partial \phi_z}{\partial x_4^1} dx_4^1 \wedge dx_4^2 \\ \\
      &=& & \displaystyle \frac{1}{\Theta^3} \left[ - \Theta (x_4^2 - x_1^2)(x_4^3 - x_1^3) \right] dx_4^1 \wedge dx_4^2. \numberthis
      \end{alignat*}
      \end{subequations}
      
By substituting these expressions in (\ref{pullback144412appendix}) it is straightforward to find
      \begin{equation}
      \left[ \phi_{1,4}^* \omega \right]_{4,4}^{1,2} 
      = \frac{\varepsilon_{3 \nu \sigma}}{4\pi} \frac{x_4^3 - x_1^3}{|x_4 - x_1|^3} \left( \frac{1}{2} dx_4^{\nu} \wedge dx_4^{\sigma} \right),
      \label{pullback144412completoappendix}
      \end{equation}
with $\nu, \sigma = 1, 2, 3$, which is to be compared with (\ref{pullbackpreab}).

\end{appendices}
\newpage

\end{document}